# Mapping vacancy and bonding electron distributions around aluminium nanovoids

Philip N. H. Nakashima[1]*, Yu-Tsun Shao[2,3], Zezhong Zhang[1]†, Andrew E. Smith[4], Tianyu Liu[5], Nikhil V. Medhekar[1], Joanne Etheridge[4,6], Laure Bourgeois[1,6] & Jian-Min Zuo[1,6-8]

[1]Department of Materials Science & Engineering, Monash University, Victoria 3800, Australia.
[2]Mork Family Department of Chemical Engineering & Materials Science, University of Southern California, Los Angeles, CA 90089, USA.
[3]Core Center of Excellence in Nano Imaging, University of Southern California, Los Angeles, CA 90089, USA.
[4]School of Physics & Astronomy, Monash University, Victoria 3800, Australia.
[5]Institute of Multidisciplinary Research for Advanced Materials, Tohoku University, Sendai 980-8577, Japan.
[6]Monash Centre for Electron Microscopy, Monash University, Victoria 3800, Australia.
[7]Department of Materials Science & Engineering, University of Illinois at Urbana-Champaign, Urbana, IL 61801, USA.
[8]Materials Research Laboratory, University of Illinois at Urbana-Champaign, Urbana, IL 61801, USA.

*corresponding author email:  philip.nakashima@monash.edu
†present address:  AI for Science Institute, Beijing, People's Republic of China.

## Abstract

All materials have defects and many contain nanostructures, both of which disrupt chemical bonding—the basis of materials properties.  No experimental measurements of bonding electron distributions associated with defects and nanostructures have ever been possible.  We present a method enabling such measurements and interrogate nanovoids surrounded by vacancies—the most fundamental of nanostructures and defects—in aluminium.  We measure the volume of a vacancy with 3% uncertainty and map vacancy concentrations surrounding nanovoids with nanometre resolution in three dimensions where previously only two-dimensional mapping was possible.  We discover that radiation-damaged voids can "heal".  Our bonding measurements are depth-resolved, vacancy-sensitive, and agree with density functional theory.  This work opens bonding electron density measurements to inhomogeneous nanostructured multi-phased materials so that the electronic origins of phenomena such as strengthening, weakening, interface functionality, solute diffusion and phase transformations within them may be revealed.

Summary

Interatomic bonding is the foundation of chemistry and the primary basis of materials properties. All techniques for measuring bonding electron distributions have been limited to homogeneous, single-phased crystals[1-18]; however, many useful materials are inhomogeneous, containing nanostructures and defects that are fundamental to their properties. Here we develop a method of quantitative convergent-beam electron diffraction (QCBED) to probe bonding surrounding nanoscale voids (nanovoids) encapsulated in aluminium. We measure the local vacancy concentration, lattice contraction and associated bonding charge density with nanometre resolution in three dimensions and reveal how these vary spatially from the boundaries of nanovoids into the bulk. These accurate and reproducible depth-resolved measurements agree with our density functional theory (DFT) modelling[19,20]. We measure changes in a nanovoid and its neighbouring vacancy concentrations under controlled irradiation and discover that radiation-damaged voids can "heal". This work opens bonding electron density measurements to inhomogeneous materials to reveal fundamental information about defect and interface-modified electron densities and the electronic origin of fundamental phenomena such as strengthening and weakening, the function of interfaces, solute diffusion, and phase transformations in nanostructured materials.

## Main text

All materials have atomic-scale defects and many "real" and useful materials contain nanostructures that impart specialised properties for specific applications. Chemical bonding is at the heart of all materials properties so experiments for measuring bonds in "real" materials should not only be sensitive to small changes in electron density due to defects, but also highly resolved to map charge densities around nanostructures.

Despite the fundamental importance of chemical bonding and the ubiquity of nanostructured and multi-phased materials, all measurements of electron distributions in chemical bonds have been limited to single-phased crystals. Some methods target regions of perfect crystal[2,3,5-9,11-18], while others selectively filter out volumes of perfect crystal using Bragg diffraction in the "ideally imperfect" crystal models of X-ray diffraction[1,4,8,10,15,18]. Recent work on single-phase alloy solid solutions accommodated variations in solute atom distributions via a local disorder parameter[21]; however, the measurement of bonding electron densities around nanostructures of a secondary phase within a host matrix remains elusive.

Here, we introduce a technique for accurate and precise depth-resolved bonding electron density measurements local to a chosen nanovolume. We apply this to the study of chemical bonding around nanovoids in high-purity aluminium, which are cuboctahedral[22-25] (Fig. 1a,b and Extended Data Fig. 1), have strong effects on properties[26-28], and are interesting plasmonic nanostructures[25,29]. Nanovoids are key agents in catalysis, energy materials, metals exposed to extreme environments and high-entropy materials. Nanovoids are surrounded by a cloud of atomic vacancies, enabling us to also study the impact of vacancies on chemical bonding. As the most fundamental defects in materials, vacancies are critical mediators of phase transformations in solids[30-32]. Being able to probe chemical bonding around a nanovoid in a crystal is a milestone in advancing chemical bonding studies into inhomogeneous, multi-phased and nanostructured "real" materials.

## A way to measure bonding around nanostructures

To probe chemical bonds around nanostructures, we focus conical beams of electrons into nanoscale probes to sample the specimen volumes of interest. Instead of sharp diffraction spots due to parallel illumination, convergent-beam electron diffraction (CBED) patterns contain discs (Extended Data Fig. 2) in which the intensity distributions as a function of angle are acutely sensitive to the charge density in the diffracting crystal planes, among other parameters, such as thickness.

To measure the bonding charge density, we use quantitative CBED (QCBED), where a simulated CBED pattern is fitted to an experimental one by refining the parameters to which the intensities are most sensitive, including the lowest spatial frequency Fourier coefficients (structure factors) of the crystal potential, which are the most sensitive to chemical bonding[2,3,5-9,11-18]. Their conversion to electron distribution via the Mott-Bethe formula[2,8,15,18,33,34] is straightforward.

Most applications of QCBED have so far applied the Bloch-wave theory[35] in a manner requiring structural periodicity and homogeneity in 3 dimensions[2,3,5-9,11-18]. Though stacked Bloch-wave methods[36-38] allow some structural aperiodicity in the electron beam direction, they have not been used in experimental determinations of bonding electron densities.

The multislice theory of Cowley and Moodie[39] describes electron scattering by propagating electron waves through arbitrarily thin slices of crystal potential. Structural periodicity in the slicing direction is unnecessary. Therefore, QCBED with a multislice engine is suited to analysing CBED patterns from disordered crystals and the spatial selectivity of nanometre-sized CBED probes means that multislice-based QCBED can be applied to nanostructured materials that can be locally approximated as layered. To take advantage of this, we developed a multislice-based QCBED program, *QCBEDMS* (Supplementary Information). We summarise it in application to aluminium nanovoids below.

## *QCBEDMS* applied to aluminium nanovoids

We applied the *QCBEDMS* method to cuboctahedral nanovoids 10 to 20 nm in size in heat-treated 99.9999+% Al (Methods, Fig. 1 and Extended Data Figs. 1 and 3). We used electron probe diameters of 3–5 nm which were smaller than any of the void facets and defined the lateral resolution of our experiments.

The knock-on damage threshold for bulk aluminium is ~180 keV[40], but only a few tens of keV at free surfaces[24,25,41]. Therefore, vacancy generation by atom sputtering from void entrance faces and atom deposition onto void exit faces is unavoidable during CBED, as are increased background vacancy concentrations.

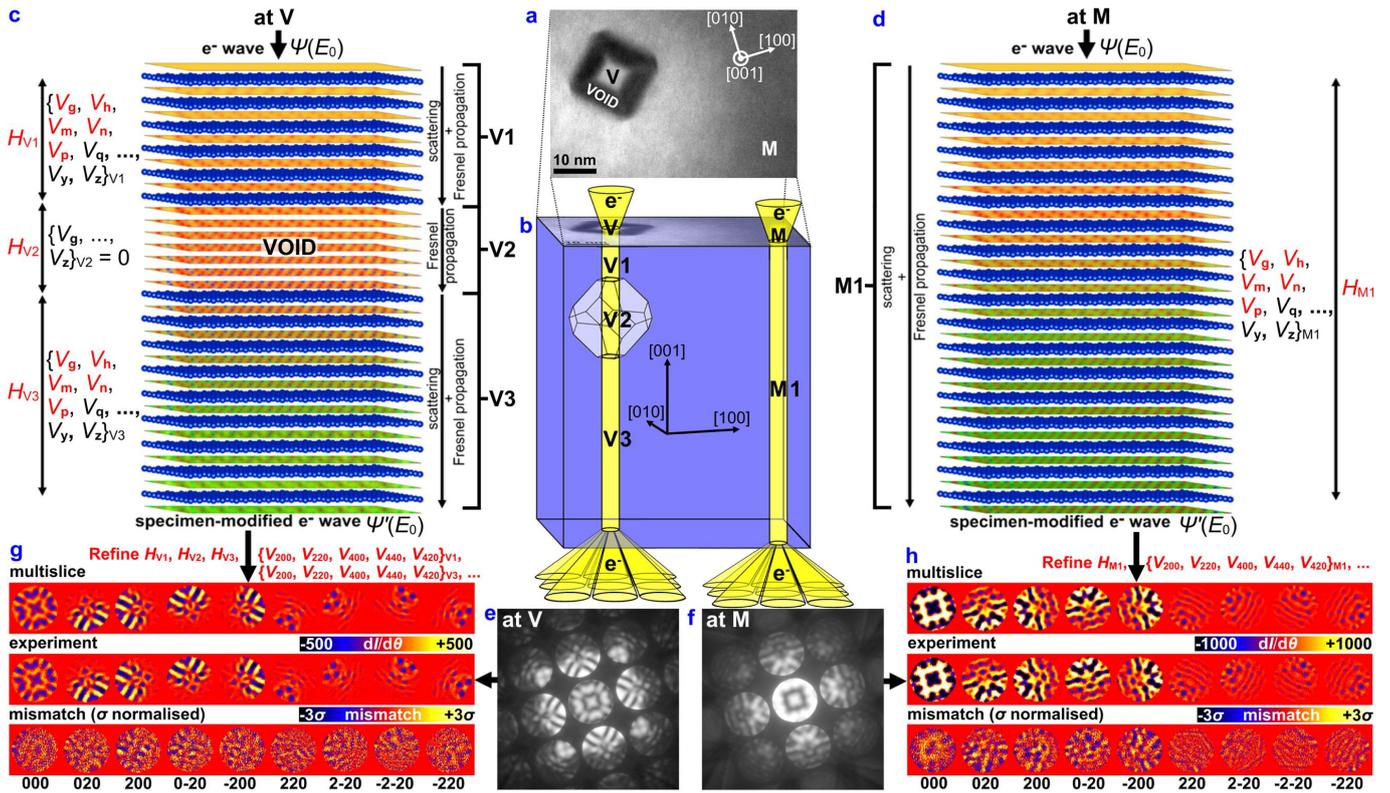

**Fig 1:** *QCBEDMS* in summary – QCBED through a void in aluminium and through the neighbouring continuous matrix. **a,** A void and the surrounding matrix are imaged, the void showing strong diffraction contrast (Methods). **b,** A convergent electron beam incident at location V (**a,b**) passes through three layers, i.e. aluminium-void-aluminium, labelled V1, V2, and V3 respectively, while at location M (**a,b**), it passes through continuous aluminium, labelled M1. The convergence and divergence angles of the incident and exiting diffracted cones are exaggerated ~$10^2$ times (**b**) to highlight that the electron beams are not parallel. The probed volumes within the specimen are drawn as columns (**b**) to approximate the tiny actual convergence angles (6 mrad in the present example – see Extended Data Fig. 2b). **c,** The multislice[39] model for electron scattering at V permits the scattering volume to be partitioned into blocks: V1 with thickness $H_{V1}$ and structure factors $\{V_g, V_h, ......, V_z\}_{V1}$, V2 with thickness $H_{V2}$ and structure factors $\{V_g, V_h, ......, V_z\}_{V2}$, and V3 with thickness $H_{V3}$ and structure factors $\{V_g, V_h, ......, V_z\}_{V3}$. **d,** At location M, the multislice model has only one segment: M1 with thickness $H_{M1}$ and structure factors $\{V_g, V_h, ......, V_z\}_{M1}$. **e,f,** CBED patterns from a 5 nm probe at locations V and M respectively (**a,b** respectively) are shown after taking the cube root of the intensities to improve contrast. **g,h,** QCBED pattern matching compares the calculated CBED intensities with the experimental ones after angular differentiation to remove the diffuse background due to inelastic scattering[12]. The match is optimised by refining the parameters to which the pattern intensities are most sensitive, labelled in red in **c,d,g** and **h**. In this figure, only thicknesses and structure factors are emphasised. The present experimental example involved 160.30 ± 0.06 keV electrons incident along [001], without energy filtering.

To minimise these effects, we used a low-intensity probe from a LaB$_6$ cathode and 160.30±0.06 keV electrons (Supplementary Information) for some of the data collection in the [001] direction (Methods and Extended Data Fig. 3a-d). Additional data were collected with a higher intensity probe in a field-emission gun transmission electron microscope (FEGTEM) and 202.4±0.2 keV electrons (Supplementary Information), with digital scan control to minimise the localised dose[42], in both the [001] and [111] directions (Methods and Extended Data Fig. 3e-m).

We obtained 25 CBED patterns through 8 different voids (voids A-H). The electron beam inevitably causes damage, so we regard the CBED patterns as coming from 25 different "void states" (VS1-VS25). We collected 20 CBED patterns from the continuous matrix near each void and refer to these as "matrix states" (MS1-MS20). See the Methods and Extended Data Fig. 3.

To model electron diffraction using the multislice theory[39], we divided each probed volume of the specimen into slices. Additionally, we grouped the slices into blocks, within each of which, unique sets of parameters were applied. For example, to describe CBED through a void (location V in Fig. 1a,b) one designates a block of slices for the aluminium matrix above the void, another block for the void, and a third for the aluminium matrix below the void (V1, V2, and V3 respectively in Fig. 1b,c). Away from the nanovoid at location M in Fig. 1a,b, only one block is needed (M1 in Fig. 1b,d) to simulate the CBED pattern.

The effect of a missing block of aluminium, V2 (the void), on the CBED intensity distribution is significant. While 4$mm$ symmetry (expected in the [001] direction) is preserved in both CBED patterns (Fig. 1e,f), the intensity distributions are very different despite the total specimen thickness in each case (as refined by *QCBEDMS*) being nearly the same: $H_{V1}+H_{V2}+H_{V3} = 843.4 \pm 0.5$ Å ≈ $H_{M1}$ = 840.2 ± 0.2 Å, where $H_j$ is the thickness of block j (j = V1, V2, V3 or M1). It is important to note that Fresnel propagation of electron waves through the void introduces additional phase, altering the CBED pattern intensity distribution far beyond a simple combination of diffracted intensities from V1 and V3 alone (Extended Data Fig. 2).

Matching both CBED patterns (Fig. 1e,f) by *QCBEDMS* follows the angular difference method[12] for removing the diffuse inelastic scattering signal that hinders QCBED, giving the middle panels of Fig. 1g,h. The same structure factors, {$V_\mathbf{g}$, $V_\mathbf{h}$, ....., $V_\mathbf{p}$}$_j$, are refined independently for each block of slices containing aluminium (j = V1, V3 and M1), as labelled in Fig. 1c,d,g,h. For V2 (the void), all structure factors are zero. The thicknesses of the blocks, i.e. $H_{V1}$, $H_{V2}$, and $H_{V3}$ in case of the void (Fig. 1c,g), and $H_{M1}$ in case of the continuous matrix (Fig. 1d,h), are also refined. Figure 1 only emphasises thicknesses and structure factors as refined parameters, despite there being others.

### Vacancies and the aluminium matrix

To account for the effects of vacancies on the aluminium lattice surrounding nanovoids in our QCBED analyses, we introduced refinable vacancy concentration ($C_{vac}$) profiles into *QCBEDMS* that modify the average atomic occupancy in each slice. Modifying average occupancies is valid since nanometre QCBED probes average over enough unit cells per slice so that individual site symmetries do not dominate CBED patterns as they do with sub-Angstrom probes in scanning transmission electron microscopy (STEM). This assumption is validated by the 4*mm* symmetry retained in CBED patterns through nanovoids as seen in Fig. 1e and Extended Data Fig. 3.

The refinable $C_{vac}$ profile options in *QCBEDMS* are: (i) constant $C_{vac}$, or (ii) $C_{vac}(d)$ – the Laplace solution to Fick's second law of diffusion[24,43] as a function of distance (*d*) from the void centre (Methods and Extended Data Fig. 4).

Atoms near vacancies and free surfaces (void facets) are less constrained in their thermal motion so we apply effective Debye-Waller parameters (DWPs), evaluated slice-by-slice as a function of the local $C_{vac}$ and the number of atomic layers between the slice and nearest void facet (Methods and Extended Data Fig. 5).

Vacancies in elemental FCC metals like aluminium cause lattice contraction[44], which is parametrisable:

$$\frac{V}{V_0} = 1 - LCF \times C_{vac} \qquad (1)$$

Here, $V$ is the supercell volume for $C_{vac} > 0$ and $V_0$ is the supercell volume when $C_{vac} = 0$ (Fig. 2a). The lattice contraction factor ($LCF$) is refinable in $QCBEDMS$ where the $LCF$ and $C_{vac}$ are coupled with the 2-dimensional lattice parameters and slice thicknesses to satisfy Eq. 1 (Supplementary Information). A value of $LCF = 0$ corresponds to total structural rigidity while $LCF = 1$ implies lattice shrinkage by the volume of the missing atom. While $0 \leq LCF \leq 1$ is expected for FCC metals[44], some materials may have $LCF < 0$, i.e. lattice expansion due to vacancies[45].

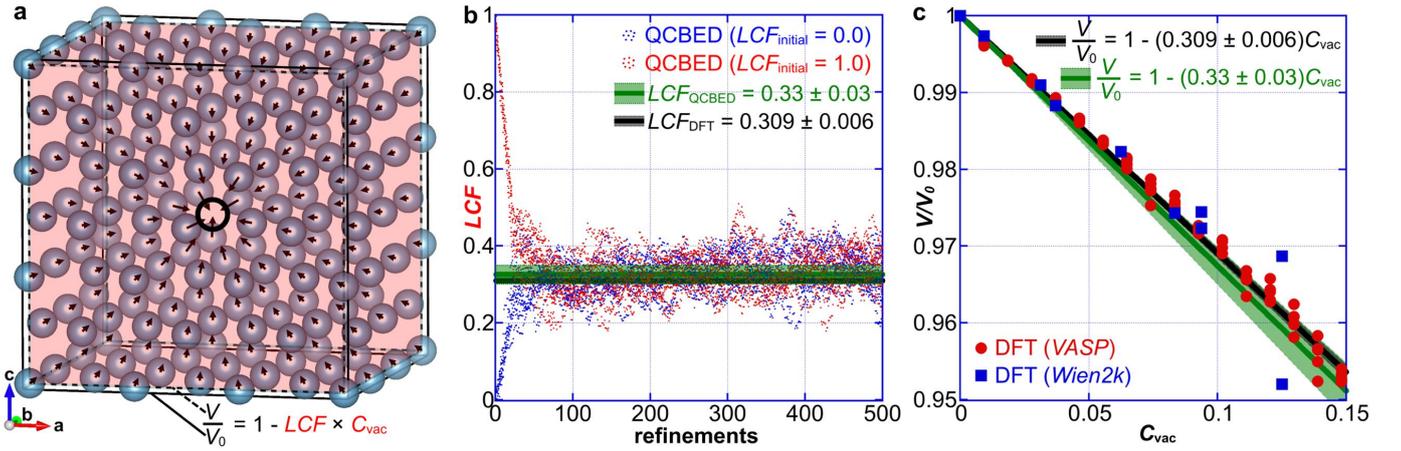

Fig 2: The lattice contraction factor due to vacancies measured by QCBED, $LCF_{QCBED} = 0.33 \pm 0.03$, agrees with two different applications of DFT[19,20] which give $LCF_{DFT} = 0.309 \pm 0.006$. **a,** A 108-site supercell of aluminium containing a vacancy (black circle). The outer cube (solid lines) schematically illustrates the volume, $V_0$, of the supercell for $C_{vac} = 0$. The inner cube (dashed lines and transparent red shading) illustrates the shrinkage to a volume, $V < V_0$, due to the vacancy. Atomic displacement magnitudes and directions are schematically indicated by the black arrows. The $LCF$ relates $V$ and $V_0$ to the vacancy concentration, $C_{vac}$. $LCF$ and $C_{vac}$ are in red as they are refinable in $QCBEDMS$. **b,** Twenty-five CBED patterns taken through eight different voids were each matched 20 times by $QCBEDMS$ for each $LCF$ refinement cycle (totalling 500 refinements per cycle). Ten cycles were executed starting from $LCF = 0$ (blue dots) and another 10 cycles were executed starting from $LCF = 1$ (red dots). The resulting 10,000 refinements converged to $LCF_{QCBED} = 0.33 \pm 0.03$ (95% confidence interval show in green), agreeing with the DFT[19,20] determination of $LCF_{DFT} = 0.309 \pm 0.006$ (95% confidence interval shown in black). **c,** Plots of $V/V_0$ versus $C_{vac}$ from DFT simulations carried out with $VASP$[19] (red dots) and $Wien2k$[20] (blue squares) for a range of different supercells with different numbers of vacancies in a variety of configurations (Methods). The black band shows $LCF_{DFT} = 0.309 \pm 0.006$, resulting from fitting Eq. 1 to the combination of points from $VASP$[19] and $Wien2k$[20], and the green band shows $LCF_{QCBED} = 0.33 \pm 0.03$.

## The refinement of $LCF$ and $C_{vac}$

We refined all void states (VS1-VS25) using $QCBEDMS$ to optimise thicknesses, the 4 parameters defining $C_{vac}(d)$ above and below each void (Methods, Extended Data Figs. 4a-c and 6), the $LCF$, absorption, the incident beam direction, and the structure factors of reflections near the Bragg condition and their coupling vectors (Fig. 1c,g, Extended Data Fig. 6 and Supplementary Information). The refinements were repeated 20 times for all void states in a cycle, giving 500 refinements per cycle (Supplementary Information). Ten cycles were started from the naïve assumption that $LCF = 0$ (blue dots in Fig. 2b) and 10 cycles were started from the other naïve assumption of $LCF = 1$ (red dots in Fig. 2b). In total, 10,000 refinement results are plotted in Fig. 2b.

The convergence to $LCF_{QCBED} = 0.33 \pm 0.03$ was rapid. The green band shows the 95% confidence interval (Fig. 2b,c). Extended Data Fig. 6a gives $LCF$ values determined individually for each void state. Our QCBED results agree with DFT determinations of the $LCF$ due to vacancies by *VASP*[19] and *Wien2k*[20] (Methods) also plotted in Fig. 2b and c.

Table S1 (Supplementary Information) compares the present $LCF$ results with vacancy volumes ($V_v$) and vacancy relaxation volumes ($V_r$) reported in the literature as fractions of the Al atomic volume ($V_a$). Note: $LCF = 1 - V_v/V_a = V_r$. Our work gives:

$(V_v/V_a)_{QCBED} = 0.67 \pm 0.03$, and $(V_v/V_a)_{DFT} = 0.691 \pm 0.006$.

The literature review (not exhaustive) gives experimental ("lit. exp'ts") and theoretical ("lit. theory") averages of:

$(V_v/V_a)_{\text{lit. exp'ts}} = 0.7 \pm 0.2$, and $(V_v/V_a)_{\text{lit. theory}} = 0.8 \pm 0.2$.

The agreement between the present results is an order of magnitude better than the spread in the literature.

We now consider void A, through and adjacent to which, CBED patterns were collected using a 5 nm, 160.3 keV electron probe from a $LaB_6$ source (Methods and Extended Data Fig. 3a-d). Figure 3a gives approximate dose rates and times spanning VS1–VS10, including the period of data collection for MS1-MS5 between VS5 and VS6 when the $C_{vac}(d)$ plots within the probed volumes for the void states show that void A was in a steady state.

Given the void's steady state from VS5 to VS6, we averaged the refined $C_{vac}(d)$ from VS5 and VS6 and applied Fick's second law[24,43] to produce the extended $C_{vac}(d)$ plot and graph of $C_{vac}$ versus $d$ (yellow lines) in Fig. 3b. Our $C_{vac}(d)$ map agrees with separate, Fick-independent refinements of $C_{vac}$ at M1–M5 (orange crosses). These and all other matrix-state QCBED optimised the same parameters as per void-state QCBED, except: (i) the $LCF$ was fixed at 0.33 (the void-state results), and (ii) a constant $C_{vac}$ was refined, ignoring Fick's law (Fig. 1d,h, Extended Data Figs. 4d,e and 7, Methods and Supplementary Information).

Recent beautiful work demonstrated 2D nanoscale vacancy concentration mapping[46]. We note that while Fig. 3b shows $C_{vac}(d)$ in cross-section, it is 3-dimensional.

Figure 3a and Extended Data Fig. 8 (Methods), show that electron-damaged voids can "heal" in periods of low or no dose (i.e. VS1 and VS10 are nearly identical and the number of vacancies in the void and surrounding matrix remained almost constant from state to state). Similar "healing" phenomena have been observed in other materials containing mobile species[47].

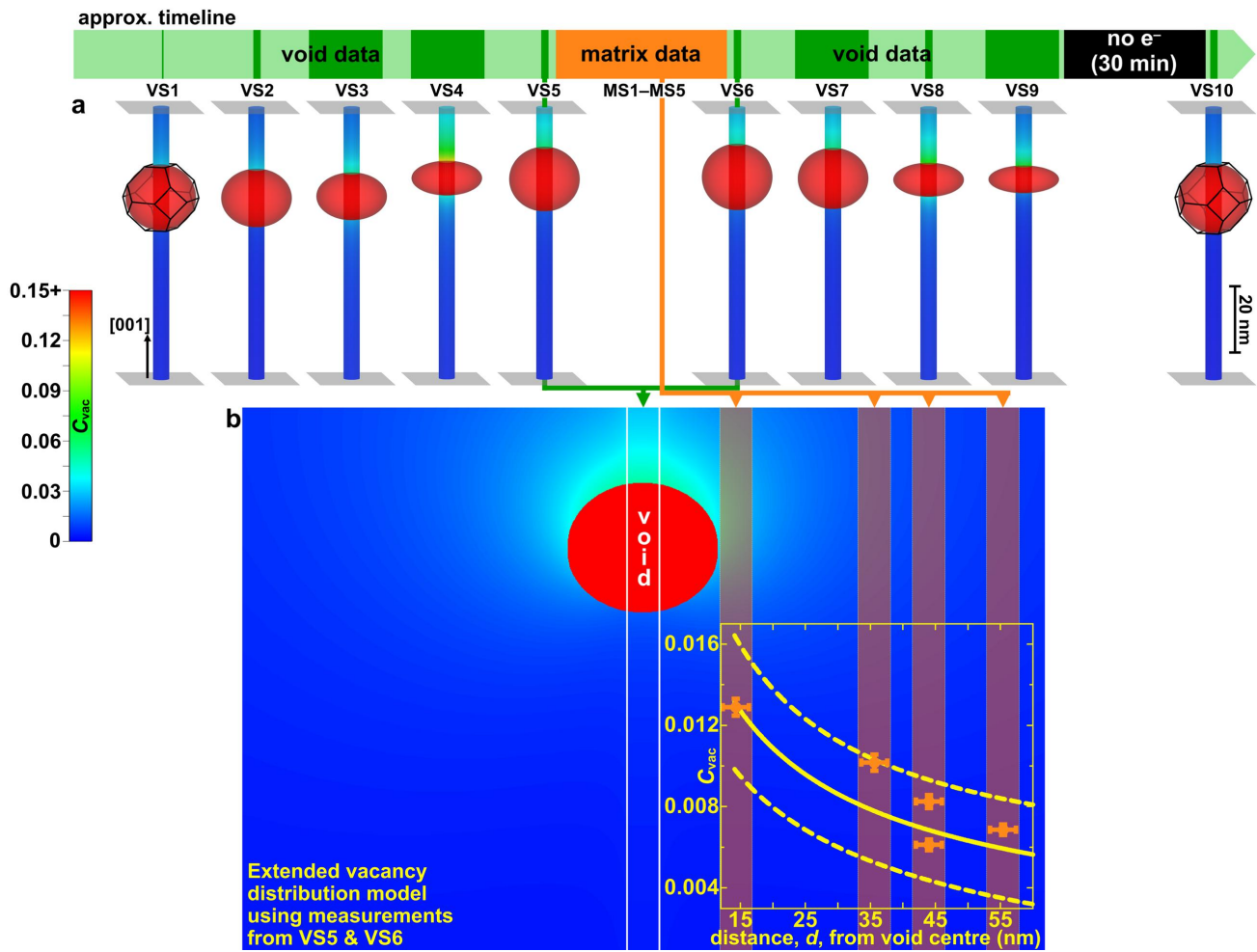

**Fig. 3: A void damaged by the electron beam can "heal" during low-dose periods, and independent matrix-only $C_{vac}$ measurements at different distances, $d$, from the void in a steady state are consistent with the through-void $C_{vac}(d)$ measurements and Fick's second law of diffusion[24,43]. a,** Evolution of void A as an approximate function of time and electron dose. Light green in the approximate timeline corresponds to the void and surroundings being imaged (mild dose rate). Dark green segments correspond to CBED through the void (maximum dose rate) and represent 2, 10 and 100 second exposures (see Extended Data Fig. 3c for approximate dosages) and are exaggerated with respect to the rest of the timeline. Orange shows when CBED in the surrounding matrix was performed. Black shows when the electron beam was blanked (zero dose). Illustrations of the probed volumes, colour coded with the measured $C_{vac}(d)$, for VS1-VS10 (also in Extended Data Fig. 6b) are accompanied by ellipsoids approximating the measured shape and depth of the void in each state (all drawn to scale). Cuboctahedral void nets are shown for VS1 and VS10. **b,** The average of the VS5 and VS6 $C_{vac}(d)$ and void dimension measurements are used to extrapolate $C_{vac}(d)$ using Fick's second law of diffusion[24,43] to generate the extended $C_{vac}(d)$ map in cross-section through the void (Methods and Extended Data Fig. 8). Independent $C_{vac}$ measurements (assuming $C_{vac}$ is constant within the probe) at the indicated distances, $d$, from the void centre (orange columns corresponding to 5 nm probes) in the continuous matrix (MS1-MS5), bounded in time by VS5 and VS6, are plotted (orange crosses) on a graph of $C_{vac}$ versus $d$. The yellow lines show the mean (solid) and 95% confidence interval (dashed) of $C_{vac}(d)$ generated from the VS5 and VS6 results, averaged over 5 nm probes.

## Debye-Waller parameters – static disorder and thermal displacements

In addition to increased thermal displacements of atoms in their neighbourhood, nanovoids and vacancies introduce static disorder contributions to DWPs that are difficult to assess locally. The reproducibility, accuracy, and self-consistency of our *LCF* and $C_{vac}$ results suggest that though our method fixes fractional atomic coordinates, it may be accommodating static disorder via these parameters.

As mentioned earlier, *QCBEDMS* applies effective DWPs as a function of slice proximity to free surfaces and the local $C_{vac}$ in order to deal with less constrained atomic thermal motion caused by these features (Methods and Extended Data Fig. 5). For all QCBED refinements, the reference DWP for aluminium[48] at 295 K ($B_{Al}$) was fixed at $B_{Al} = 0.86$ Å$^2$.

## The refinement of structure factors

Table 1 summarises all refined structure factors (converted from $V_{hkl}$ in Volts to $f_{hkl}$ in e$^-$/atom via the Mott-Bethe formula[33,34]).

Table 1: A summary of measured structure factors after conversion to $f_{hkl}$ (e$^-$/atom) by the Mott-Bethe formula[33,34], above, below and near aluminium nanovoids (V1, V3 and M1 respectively – see Fig. 1b–d and Extended Data Figs. 6 and 7), suggests that static atomic displacements, structural disorder and less constrained thermal motion under the influence of void surfaces and vacancies have been accurately accounted for by *QCBEDMS*. Uncertainties corresponding to the 95% confidence interval are given in parentheses for the last significant figure. Values for V1 and V3 are averaged over VS1-VS25 (Extended Data Fig. 6) and those for M1 are averaged over MS1-MS20 (Extended Data Fig. 7). The IAM values[49] and previous measurements in aluminium[13] ("bench") are also presented. All values correspond to $T = 295$ K ($B_{Al} = 0.86$ Å$^2$)[48].

| (e$^-$/atom) | $f_{200}$ | $f_{220}$ | $f_{400}$ | $f_{420}$ | $f_{422}$ | $f_{440}$ |
|---|---|---|---|---|---|---|
| V1 | 7.942(2) | 6.606(2) | 4.682(1) | 3.993(2) | 3.424(4) | 2.580(1) |
| V3 | 7.938(2) | 6.606(1) | 4.681(1) | 3.991(2) | 3.426(4) | 2.581(1) |
| M1 | 7.931(3) | 6.604(4) | 4.685(3) | 3.993(2) | 3.418(6) | 2.580(2) |
| IAM[49] | 8.080 | 6.607 | 4.682 | 3.992 | 3.424 | 2.580 |
| bench[13] | 7.94(1) | 6.61(3) | 4.73(5) | 3.97(1) | 3.65(7) | 2.75(6) |

The precision and agreement of all *QCBEDMS*-measured structure factors, even for higher orders (larger *hkl*), across all regions (V1, V3, and M1) is about an order of magnitude better than previous benchmark measurements for aluminium[13]. The significant improvement is most probably a result of taking the impact of atomic vacancies and void free surfaces on the aluminium lattice into account, while a perfect vacancy-free crystal was assumed in the older work and atomic displacements were treated as purely thermal[13].

Structure factors for 220 and higher orders are known to be equal to those of the independent atom model (IAM – the electron density of an assembly of spherical, neutral, unbonded atoms) for aluminium[13,16,48,49]. All of the present results agree with this within their 95% confidence intervals.

Table 1 and Extended Data Figs. 6 and 7 show how precise the refined structure factors are from state to state across all void and matrix states, across all regions (V1, V3, and M1) and a range of vacancy concentrations. Differences between individual results and the different region averages are generally ~0.1% or less, independent of *hkl*. These are compelling indications that our method has accounted for static and additional dynamic contributions to the DWP via accurate measurements of $C_{vac}$, the *LCF* and the incorporation of free-surface effects in the vicinity of voids (Extended Data Fig. 5).

## Bonding electron densities local to aluminium nanovoids

The accuracy, precision, and self-consistency of all results so far (*LCF*, $C_{vac}$, and $f_{hkl}$) are critical for accurate mapping of bonding electron densities surrounding aluminium nanovoids, which we now explore by presenting the first experimental, depth-resolved bonding electron densities local to an embedded nanostructure.

Bonding electron densities, $\Delta\rho(\mathbf{r})$ (**r** is a real-space vector), are computed by Fourier summing the measured and IAM structure factor differences. Only $f_{200}$ differs from the aluminium IAM[49] in the present work. The only other structure factor of aluminium that differs from the IAM[13,16,48,49], $f_{111}$, could not be measured as there were no void facet orientations permitting this. Mapping $\Delta\rho(\mathbf{r})$ is not hindered in the present work as $f_{111}$ does not contribute to <001> or <111> projections of $\Delta\rho(\mathbf{r})$.

The QCBED results give bonding electron densities projected within each slice of a refined multislice model. Figure 4 presents projected $\Delta\rho(\mathbf{r})$ maps from QCBED at the indicated locations in VS4 and VS24 along <001> (Fig. 4a) and <111> (Fig. 4b) respectively. These correspond to values of $C_{vac}(d)$ (also mapped for each void state) which are practical to model with DFT.

For comparison to our QCBED $\Delta\rho(\mathbf{r})$ maps, we used *Wien2k*[20] and the Doyle and Turner IAM[49] (Methods and Supplementary Information) to generate the maps marked "DFT" in Fig. 4a,b. To make the DFT $\Delta\rho(\mathbf{r})$ plots comparable to those from QCBED, site averaging was necessary (Supplementary Information), given that this is what CBED probes do. This is shown for $0.00926 \leq C_{vac} \leq 0.125$ in Extended Data Fig. 9, which also shows how the projected $\Delta\rho(\mathbf{r})$ maps were produced.

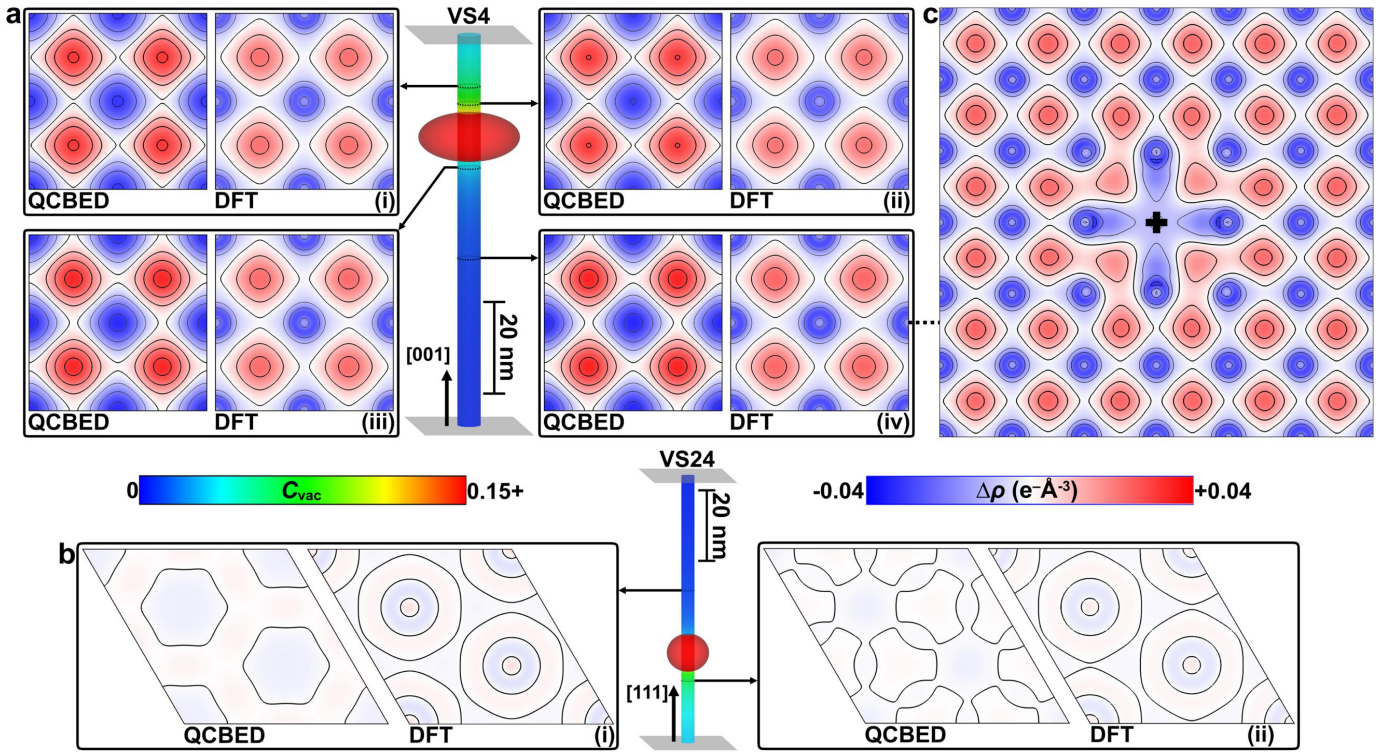

Fig 4: Experimental, depth-resolved bonding electron density plots local to aluminium nanovoids, show the influence of vacancy concentration gradients on the bonding electron density in their neighbourhood. **a,b,** Void states VS4 (**a**) and VS24 (**b**), probed along [001] and [111] respectively, are presented with their $C_{vac}(d)$ plots for the probed volumes, with the voids represented by ellipsoids. Two-dimensional plots of $\Delta\rho(\mathbf{r})$, determined by QCBED and DFT (*Wien2k*[20]) with site averaging, are shown for locations in the bulk where $C_{vac}(d)$ corresponds to integer numbers of vacancies in small DFT supercells (Methods, Extended Data Fig. 9, and Supplementary Information). All DFT-generated plots used effective DWPs ($B_{Al, eff}$) in the determinations of $\Delta\rho(\mathbf{r})$ (Methods and Extended Data Fig. 5). The details for each location are: **a(i):** $C_{vac} = 0.0626$ (DFT model: a di-vacancy in a 32-site supercell), $B_{Al, eff} = 0.942$ Å$^2$; **a(ii):** $C_{vac} = 0.0935$ (DFT model: a tri-vacancy in a 32-site supercell), $B_{Al, eff} = 0.988$ Å$^2$; **a(iii):** $C_{vac} = 0.0315$ (DFT model: a mono-vacancy in a 32-site supercell), $B_{Al, eff} = 0.899$ Å$^2$; **a(iv):** $C_{vac} = 0.00925$ (DFT model: a mono-vacancy in a 108-site supercell), $B_{Al, eff} = 0.871$ Å$^2$; **b(i):** $C_{vac} = 0.00930$ (DFT model: a mono-vacancy in a 108-site supercell), $B_{Al, eff} = 0.871$ Å$^2$; **b(ii):** $C_{vac} = 0.0627$ (DFT model: a di-vacancy in a 32-site supercell), $B_{Al, eff} = 0.942$ Å$^2$. **c,** A DFT-generated, [001]-projected $\Delta\rho(\mathbf{r})$ plot for slices in a 108-site supercell bounding the modelled vacancy (black cross) and corresponding to **a(iv)** ($C_{vac} = 0.00925$). Site averaging was omitted in this case (Extended Data Fig. 10 and Supplementary Information). Contour intervals in all cases are 0.01 e$^-$Å$^{-3}$ with thick lines for $\Delta\rho(\mathbf{r}) \geq 0$ e$^-$Å$^{-3}$ and thin lines for $\Delta\rho(\mathbf{r}) < 0$ e$^-$Å$^{-3}$. This figure used *VESTA*[50].

The QCBED $\Delta\rho(\mathbf{r})$ maps agree very well topologically with the site-averaged and projected maps from DFT for the [001] orientation (Fig. 4a), despite the ~30% difference in magnitude between them (consistent with previous work[13]). The [111] orientation shows a much smaller magnitude for the projected $\Delta\rho(\mathbf{r})$ (Fig. 4b) than for [001] for both QCBED and DFT. This is because pockets of high bonding density in tetrahedral interstices align with pockets of high anti-bonding density around atoms in the [111] direction (Extended Data Fig. 9n), largely cancelling out in projection. The low magnitudes of the projected $\Delta\rho(\mathbf{r})$ in [111] belie any differences in topology.

Most important is the effect of vacancies on $\Delta\rho(\mathbf{r})$. The QCBED maps for [001] (Fig. 4a) show a reduction in amplitude of $\Delta\rho(\mathbf{r})$ as $C_{vac}$ increases (eg. compare Fig. 4a(iv), $C_{vac}$ = 0.00925 with Fig. 4a(ii), $C_{vac}$ = 0.0935). While more subtle in the corresponding DFT plots, a reduction in volume of $|\Delta\rho(\mathbf{r})| \geq 0.02$ e$^-$Å$^{-3}$ in Extended Data Fig. 9 (eg. parts b and r) is more obvious.

Exploring this further, Fig. 4c shows the $\Delta\rho(\mathbf{r})$ plot for [001] slice projections surrounding a single vacancy (black cross) in a 108-site DFT model without site averaging (Extended Data Fig. 10). The bonding bounding a vacancy is much diminished as is the site-centred antibonding density, while there is little transfer of antibonding density to nearest neighbour atoms. The altered topology of $\Delta\rho(\mathbf{r})$ around a vacancy is low in spatial frequency, which explains why QCBED-measured deviations from the IAM are limited to $F_{200}$ (Table 1, Extended Data Figs. 6 and 7).

The present QCBED and DFT $\Delta\rho(\mathbf{r})$ results (Fig. 4 and Extended Data Fig. 9) are consistent with tetrahedrally-centred bonding in aluminium established some years ago[13]. This shows that even high vacancy concentrations do not change the site-averaged morphology of the bonding density but only dilute that which is already very weak because aluminium is a nearly-free electron gas[13,16,48,51].

## Conclusion

We have made bonding electron density measurements local to a nanostructure embedded in a host material by measuring bonding-sensitive structure factors surrounding aluminium nanovoids with uncertainties of ~0.1% or less. This is an order of magnitude more precise than previous benchmark measurements in homogeneous aluminium[13].

For this to be possible, our technique, *QCBEDMS*, had to account for the influence of vacancies which are more abundant around nanovoids. This resulted in measuring local vacancy concentrations and lattice contraction accurately and self-consistently. Our measurements of lattice contraction corresponded to a determination of the volume of an aluminium vacancy with an uncertainty of 3% and agreed with two different applications of DFT[19,20]. Our depth-resolved vacancy concentration measurements achieved a spatial resolution in *three* dimensions of a few nanometres and a precision sufficient to identify subtle changes in vacancy concentrations and nanovoid shape in response to irradiation. This included the observation that beam-damaged nanovoids in aluminium can "heal" during periods of low or no dose.

Modern materials are largely heterogeneous, containing nanostructures that hybridise their properties for specific applications. The demonstrated ability to measure bonding electron densities associated with nanostructures and point defects, and to measure point defect distributions, is critical for understanding the properties of nanostructured materials at the most fundamental level. This is likely to shed new light on fundamental materials phenomena such as strengthening and weakening, the function of interfaces, solute diffusion, and phase transformations, among others.

------------------------------

## Methods

### Specimen preparation

A 50 mm × 50 mm × 1.0 mm sheet of 99.9999+% purity aluminium manufactured by *Puratronic*®, was purchased from Alfa Aesar and used for all specimen production in this work.

All specimens were made by punching 3 mm diameter discs from the aluminium sheet. The first batch (batch 1) was heated to and maintained at 823 K in a nitrate salt bath for 30 min, followed by water quenching to room temperature. The second batch (batch 2) was heated to and maintained at 850 K in air in a tube furnace for 60 min, followed by water quenching to room temperature. Both batches of specimen discs were mechanically ground to between 120 μm and 180 μm in thickness and then twin-jet electro-polished in 67% methanol – 33% $HNO_3$ at 248 K with a potential difference of 13 V.

### Data collection and processing

*Specimen batch 1:*

One void, void A (Figs. 1 and 3, and Extended Data Figs. 3a-d, 6 and 7, and Supplementary Information), was chosen in a specimen from batch 1 and oriented along [001] in a JEOL 2011 TEM at the Monash Centre for Electron Microscopy. The microscope had a conventional thermionic gun with a $LaB_6$ cathode. All TEM images and CBED patterns associated with void A were collected with 160 keV (nominal energy) electrons, on a Gatan™ UltraScan® 1000 CCD camera below the viewing screen without energy filtering.

The void and surrounding area were imaged in bright-field TEM mode in Fig. 1a,b and Extended Data Fig. 3a. The void facets that were not perpendicular to the electron beam appear dark because the objective aperture excluded diffracted beams that had increased intensities in these areas.

Twelve CBED patterns, corresponding to "void states" VS1-VS12, were collected through void A by positioning the focussed electron probe on and perpendicular to the central (002) facet (Fig.

1a,b,e, Extended Data Figs. 3a,c and 6, and Supplementary Information). The acquisition times varied from single frames of 2 seconds exposure to single frames of 10 seconds and 10-frame sums of 10 seconds exposure per frame. These exposures corresponded to doses of $(1.9\pm0.1)\times10^8$ e$^-$, $(9.5\pm0.5)\times10^8$ e$^-$, and $(9.5\pm0.5)\times10^9$ e$^-$ respectively (Extended Data Fig. 3c). These dosages are schematically represented in the approximate timeline of Fig. 3a for VS1-VS10, ignoring that the probe was focussed on the void for short unknown periods before and after pattern acquisition.

At various intervals between CBED data collection through void A, CBED patterns at numerous distances from the void, through the neighbouring matrix were collected, also with varying exposure times and frame averaging as described above. In total, 13 such patterns were collected, but only 7 of these were used in the present work (matrix states MS1-MS7 – Extended Data Fig. 3d, and Supplementary Information) as 6 of the patterns were collected too close to the void so that the probe partially intersected void A, causing strong departures from 4*mm* symmetry in these patterns. Extended Data Fig. 3c,d show the CBED patterns for VS1-VS12 and MS1-MS7, together with dosages, time stamps and collection locations in the bright-field image of void A and its surroundings.

After all CBED patterns had been acquired, the probe in vacuum and through the specimen was imaged to gauge the probe size in the CBED experiments.

CBED patterns without the specimen in the beam path were acquired under the same electron-optical conditions as the CBED data for VS1-VS12 and MS1-MS7. The resulting condenser aperture images were used for instrumental point spread function (PSF) determination[52] as well as dose estimation.

*Specimen batch 2:*
All CBED data from specimens in batch 2 were collected with a JEOL 2200FS field-emission gun TEM (FEGTEM) and an in-column omega-type energy filter at the Frederick Seitz Materials Research Laboratory (FSMRL) at the University of Illinois at Urbana-Champaign (UIUC). A Gatan™ UltraScan® 1000 CCD camera below the viewing screen of the TEM was used to record all data in conjunction with the omega filter and a 10 eV-wide slit centred on the zero-loss peak in the electron energy-loss (EEL) spectrum for 200 keV (nominal energy) incident electrons.

An in-house Digital Micrograph script for collecting CBED patterns at each point of a predefined scan array[42] was used to collect a CBED pattern at each of three points along a line 80 nm in length, ending at the centre of the void under interrogation (Extended Data Fig. 3f-h,j-m). Point 1, at 80 nm from the void, was where final fine alignments were performed and this point of the specimen suffered considerable beam damage. Point 2 was at 40 nm from the void and point 3 was through the void itself (Extended Data Fig. 3f-h,j-m). Each CBED pattern was acquired for an exposure time of 0.1 second. This method of data collection ensured minimal dose to the matrix 40 nm away from the void and through the void itself.

Three-point-scan CBED data were collected along <001> corresponding to VS13 and VS14 and MS8 and MS9 for void B, VS15-VS18 and MS10-MS13 for void C, and VS19 and MS14 for void D. The same approach along <111> yielded VS20 and VS21, and MS15 and MS16 for void E; VS22 and MS17 for void F; VS23 and VS24, and MS18 and MS19 for void G and VS25 and MS20 for void H (Extended Data Fig. 3f-h for <001>, j-m for <111>).

Images of the converged electron probe were obtained to estimate the probe size and CBED patterns without the specimen were acquired under the same conditions as the through-void and matrix-only CBED data to determine the PSF[52] and dose. The latter is stated for each pattern in Extended Data Fig. 3.

*Pre-QCBED data processing:*

All CBED patterns were corrected for the instrumental PSF[52] followed by noise reduction and determination of the variance in each pixel as a function of signal[53]. It has previously been shown that systematic errors in QCBED are introduced by the assumption that noise in CBED patterns is Poisson in form[53].

The method of angular differentials[12] was applied to all the CBED data to remove the diffuse inelastic background that cannot be matched by elastic scattering theory[12,13,16-18], but with a variation on the technique. Instead of averaging them[12,13,16-18], the directional differentials were kept separate – see Extended Data Fig. 11. *QCBEDMS* was programmed to pattern match these individual components, resulting in 4 times as many data points compared to directionally averaged pattern matching. We note that the QCBED examples in Fig. 1 used the present method of directionally resolved angular differences, but the components were averaged post-refinement to simplify the figure (Fig. 1g,h). Extended Data Fig. 11f presents an example of QCBED pattern matching where the experimental and refined calculated angular difference intensities are compared in the directionally-resolved and isotropic (directionally averaged) modes.

### Atomic binding and effective Debye-Waller parameters

An important feature of *QCBEDMS* is its use of effective Debye-Waller parameters, $B_{eff}$, as a function of distance from a free surface and the local vacancy concentration, $C_{vac}$. In the present work, void facets are considered free surfaces while the specimen surfaces are not, because of the capping oxide layer.

In a multislice approach where each slice is a single atomic layer, Extended Data Fig. 5 illustrates how a "binding" function can be self-consistently determined. The "binding" function quantifies how "bound" atoms in the $l^{\,th}$ layer from a free surface are, where the layer has a mean vacancy concentration of $C_{vac}$.

A block segment of a periodic lattice has 1 - $C_{vac}$ of its sites randomly designated as atoms (initial site binding factor = 1) with the rest remaining vacant (initial site binding factor = 0). All sites outside the block are vacant (binding factor = 0).

Each non-zero site in the block has its binding factor re-evaluated by a weighted average over a sphere of radius $r_{max}$ where the weights come from a Lennard-Jones model[54,55] (Extended Data Fig. 5a). The re-evaluation of each non-zero site binding factor is repeated to convergence.

The average site-binding level is determined within each atomic layer in the block within the dotted boundaries (Extended Data Fig. 5b) to eliminate the effects of the other block faces. It is plotted versus the layer number (Extended Data Fig. 5c). In the example of Extended Data Fig. 5, a FCC lattice is modelled with <001> as the slicing direction and $r_{max}$ = 10 voxels. Note that the result is negligibly different if $r_{max}$ = 20 voxels, whilst the evaluation time increases by a factor of 8. The plot in Extended Data Fig. 5c was obtained using 5 different starting blocks with randomly distributed vacancies and $C_{vac}$ = 0.05 in each case (hence the small spread of points in the plot).

Testing different slicing directions and a range of $C_{vac}$, the following binding function was determined for any FCC lattice:

$$binding(l, C_{vac}) = 1 - 1.38455 C_{vac} - (m - nC_{vac})e^{-p(l-1)^{0.9082}}. \quad (2)$$

Only *m*, *n*, and *p* change for different slicing directions, <*uvw*>:

for <001>: *m* = 0.3974, *n* = 0.5985, and *p* = 1.46080, and

for <111>, *m* = 0.3386, *n* = 0.5079, and *p* = 1.7217.

Finally, the effective Debye-Waller parameter, $B_{eff}(l,C_{vac})$ is:

$$B_{eff}(l, C_{vac}) = \frac{B_0}{binding(l,C_{vac})} \quad (3)$$

where $B_0$ is the Debye-Waller parameter for a perfect bulk crystal. This is incorporated into *QCBEDMS* on a slice-by-slice basis where the parameters $C_{vac,0,V1}$, $C_{vac,0,V3}$, $C_{vac,V,V1}$, and $C_{vac,V,V3}$, or $C_{vac,0,M1}$, defined in Extended Data Fig. 4, are refined and determine the value of $C_{vac}$ in Eqs. 2 and 3.

## Applications of Fick's second law of diffusion

The QCBED-refined vacancy distributions in the probed sample volumes above and below each void shown in Fig. 3a for VS1-VS10, Fig. 4a,b for VS4 and VS24 respectively, and Extended Data Fig. 6b for VS1-VS25, are the slice-by-slice *QCBEDMS* output for $H_{V1}$, $H_{V2}$, $H_{V3}$, $C_{vac,0,V1}$, $C_{vac,V,V1}$, $C_{vac,0,V3}$, $C_{vac,V,V3}$, and *LCF* tabulated in Extended Data Fig. 6a. As per Extended Data Fig. 4b,c, $C_{vac,0,V1}$, $C_{vac,V,V1}$, $C_{vac,0,V3}$, and $C_{vac,V,V3}$ parametrise the application of Fick's second law of diffusion within *QCBEDMS*.

The QCBED-refined values of $H_{V1}$, $H_{V2}$, $H_{V3}$, $C_{vac,0,V1}$, $C_{vac,V,V1}$, $C_{vac,0,V3}$, $C_{vac,V,V3}$ (Extended Data Fig. 6a for VS1-VS10), and the lateral void dimensions measured from Fig. 1a and Extended Data Fig. 3a, are used to apply Fick's second law of diffusion with an ellipsoidal approximation of the void shape, to determine the extended plots of $C_{vac}(\mathbf{r})$ in Fig. 3b and Extended Data Fig. 8b.

In Extended Data Fig. 8a, the Laplace solution to Fick's second law of diffusion[24,43] is colour coded for the diagrammatic description immediately below the equation. The green panel describes $R_V/|\mathbf{r}|$ for any point (green dot) with vector $\mathbf{r}$ from the centre of the void having radius $R_V$ along $\mathbf{r}$. This is fully defined by the experimentally measured parameters, $R_{V,xy}$ (from bright-field TEM imaging), and $H_{V2}$ (from QCBED). The parameters $H_{V1}$ and $H_{V3}$ determine the location of the void with respect to the entrance and exit faces of the specimen.

The blue panel describes a Fick-like model of the background vacancy concentration, $C_{vac,0}$. The QCBED-measured $C_{vac,0}$ above ($C_{vac,0,V1}$) and below ($C_{vac,0,V3}$) the void (refer to Extended Data Fig. 4b,c) will be much higher than $C_{vac,0}$ far from the void ($C_{vac,0,\infty}$) due to the electron probe. The probe radius, $r_{pr}$, affects the rate of decay of $C_{vac,0}$ as a function of the minimum distances $d_1$ and $d_3$ of any point (blue dot) from the surface of each probe segment in the matrix, V1 and V3 respectively (dotted lines).

The red panel describes the vacancy concentration at the surface of the void, $C_{vac,V}$, relevant to any point (red dot) with coordinates $(x,y)$. Due to the differences in the refined $C_{vac,V}$ values at the top and bottom of the void ($C_{vac,V,V1}$ and $C_{vac,V,V3}$ respectively), $C_{vac,V}$ is linearly interpolated between $C_{vac,V,V1}$ and $C_{vac,V,V3}$ as a function of the angle, $\theta$, subtended by $(x,y)$ from the vertical through the void centre.

A calculated example of each component of the $C_{vac}(\mathbf{r})$ equation is given below each of the colour-coded panels for the case of VS3 using the QCBED-measured parameters summarised in the panel below Extended Data Fig. 8b (see also Extended Data Fig. 6a). For $C_{vac,0}$ as a function of position, the quenched-in vacancy concentration estimated by the vacancy triangle method[48] and the specimen pre-quench temperature of 823 K was used as $C_{vac,0,\infty}$, i.e. $C_{vac,0,\infty} = 0.00042$.

Combining all three components produces the extended plot of $C_{vac}(\mathbf{r})$ in cross section through VS3 in Extended Data Fig. 8b.

The same approach was used to generate the extended $C_{vac}(\mathbf{r})$ plot in cross section through void A in Fig. 3b, except that the plot in that figure is the average over all permutations of $H_{V1}$, $H_{V2}$, $H_{V3}$, $C_{vac,0,V1}$, $C_{vac,V,V1}$, $C_{vac,0,V3}$, $C_{vac,V,V3}$ and their uncertainties for VS5 and VS6 (see Extended Data Fig. 6a), as well as $R_{V,xy}$ and its associated uncertainty, and three different values of $C_{vac,0,\infty}$. The 3 different values of $C_{vac,0,\infty}$ applied were: (i) the minimum of $C_{vac,0,V1}$ and $C_{vac,0,V3}$ measured by QCBED, (ii) the quenched-in vacancy concentration[48], $C_{vac,0,\infty} = 0.00042$, and (iii) the ambient ($T = 295$ K) equilibrium vacancy concentration[56], $C_{vac,0,\infty} = 9.65 \times 10^{-12}$. This variety of $C_{vac,0,\infty}$ values ensured that the unknown true value would be covered by this range. The probe-averaged plot of $C_{vac}$ as a function of distance from the void centre in Fig. 3b shows the mean (solid line) and 95% Chebyshev confidence interval (dashed lines) obtained from all these parameter value permutations.

The diffusion range ($x$) of vacancies as a function of time ($t$) was determined by $x = \sqrt{4tD_v}$, using an ambient ($T = 295$ K) equilibrium vacancy concentration[56] of $9.65 \times 10^{-12}$ in the vacancy diffusivity ($D_v$) expression[23,24] based on the aluminium self-diffusion coefficient of Volin and Balluffi[23]. The diffusion ranges for different periods of exposure to the focussed electron beam are shown by the white dotted lines in Extended Data Fig. 8b.

The time that void A was exposed to the focussed electron probe before CBED pattern acquisition was estimated to be about 10 seconds. The CBED patterns themselves are averages over the different acquisition times and therefore, the diffusion time applying to each CBED pattern was approximated to be 10 s + half the acquisition time. Using this and $x = \sqrt{4tD_v}$ for the diffusion range, integrating the cross-sectional plots over the relevant diffusion range for each of VS1-VS10 resulted in the plot shown in Extended Data Fig. 8c. This suggests that the number of vacancies in and near void A from state to state remains roughly constant, supporting the conclusion that beam-damaged voids can heal during no or low dose periods (Fig. 3a).

## Applications of DFT

*DFT using VASP:*

All of the red data points in Fig. 2c were generated using the Vienna Ab initio Simulation Package (*VASP*)[19] with the potential constructed under the generalised gradient approximation with the treatment of Perdew, Burke, and Ernzerhof (PBE-GGA)[57] using the projector augmented wave (PAW) method[58,59]. Internal and external structural parameters were fully optimised until Hellmann-Feynman forces were less than 0.01 eV/Å, using the Methfessel-Paxton method[60] with a smearing factor of 0.05 eV. The convergence of the relevant energy differences concerning energy cut-off (500 eV), k-point sampling (~2000/atom), and supercell size was better than 1 meV/atom.

Multiple points at the same vacancy concentrations in Fig. 2c are the result of modelling different vacancy configurations within supercells with the same number of vacancies and atoms.

*DFT using Wien2k:*

All the blue points in Fig. 2c, the 2D bonding electron density plots attributed to DFT in Fig. 4, and all bonding electron density plots in Extended Data Figs. 9 and 10 were generated with *Wien2k_23.2* (ref.[20]). The initialization precision value of prec = 2, as recommended in *Wien2k_23.2* for supercells, was adopted for all calculations. All calculations involved the PBE-GGA[57] exchange correlation functional and at each overall step, the Birch-Murnaghan equation of state[61] determined the calculated lattice parameters. For comparison purposes, the initial computed bulk lattice parameter was calculated with both settings prec = 2 "highly accurate" and prec = 3 "highest precision". For both cases the results to four decimal places were 4.0407 Å. Comparison with Lejaeghere et al.[62] using the earlier *Wien2k_13.1* reveals the recent improvements in the *Wien2k* package.

In particular, the initial computed bulk lattice parameter provided the starting outline for each of the different vacancy supercells where the choice of vacancy sites established the space group symmetry. For each choice of vacancy/atom supercell, initialization with the *Wien2k_23.2* prec = 2 "highly accurate" value enabled automatic, necessarily consistent and efficient, internal parameter choices across all the supercells. Most important of these were the muffin tin radius $R_{MT}$ and the largest $K_{max}$ vector parameter values.

Different vacancy configurations in supercells with the same numbers of atoms and vacancies were modelled, resulting in multiple points lying at the same vacancy concentrations in the graph of Fig. 2c.

Each computation began with a starting loop of *Wien2k* force minimization self-consistent field (SCF) calculations focused on ion position optimisation. Two subsequent levels of fused loop SCF calculations determined a resultant supercell with both optimised charge density and ion positions. Finally, a Birch-Murnaghan process[61] was set up to carry out a set of SCF calculations that started with the resultant supercell with optimised charge density and ion positions. This began with structural files that established incremental volumes that enabled the final series of SCF calculations, each of which were fused charge density and position optimisations.

## Data availability

All CBED data, the *QCBEDMS* code and individual refined results are available upon email request to the corresponding author.

**Acknowledgements:** We thank Emeritus Professor Frederick Mendelsohn FAA FRACP for reviewing several drafts of this manuscript and dispensing valuable advice. The authors acknowledge the use of the instruments and scientific and technical assistance at the Monash Centre for Electron Microscopy, a Node of Microscopy Australia. The authors acknowledge use of transmission electron microscopes in the Frederick Seitz Materials Research Laboratory at the University of Illinois at Urbana-Champaign. This work was supported by Australian Research Council grants FT110100427 & DP210100308. This research was undertaken with the assistance of resources from the National Computational Infrastructure (NCI Australia), an NCRIS enabled capability supported by the Australian Government. Part of this research was carried out on the MonARCH computing facility at Monash University.

**Author contributions:** **P.N.H.N.** conceived the experiments, made the specimens, collected all of the CBED and TEM data at 160 keV, performed all of the QCBED refinements, performed all subsequent analyses, co-wrote the *QCBEDMS* program, wrote *DISTCORR* and all other peripheral programs, developed all of the techniques used in the present work, wrote all parts of the manuscript and Methods and Supplementary Information (with the exception of the *VASP* and *Wien2k* descriptions), and produced all figures and tables and extended data figures and tables. **Y.-T.S.** conducted all TEM and energy-filtered CBED data acquisition at 200 keV and reviewed the manuscript. **Z.Z.** performed the DFT lattice contraction calculations of supercells containing vacancies with *VASP*, provided the text in the Methods section describing this, and reviewed the manuscript. **A.E.S.** performed all DFT calculations of supercells containing vacancies with *Wien2k*, provided the text in the Methods section describing this, and reviewed the manuscript. **T.L.** was involved in the TEM and energy-filtered CBED data acquisition at 200 keV, performed some early trial QCBED refinements, and reviewed the manuscript. **N.V.M.** provided computing resources for all *VASP* calculations and reviewed the manuscript. **J.E.** provided CBED expertise, performed detailed appraisals of the manuscript in numerous drafting stages and provided expertise and advice when it came to the presentation of results and the significance of the key outcomes. **L.B.** provided expertise in the heat treatment of aluminium to produce voids, provided expertise when it came to the morphology of the voids in the specimens, assisted in collecting the CBED and TEM data at 160 keV, and critically reviewed all drafts of the manuscript. **J.-M.Z.** co-authored the *QCBEDMS* program that is critical to the present work, provided access to the TEMs in the Frederick Seitz Materials Research Laboratory at the University of Illinois at Urbana-Champaign, provided CBED expertise, and critically reviewed all drafts of the manuscript.

**Funding:** The collection of data at the Frederick Seitz Materials Research Laboratory at the University of Illinois at Urbana-Champaign was supported by a Monash Engineering Travel Grant (2017). The key QCBED refinements and most of the computation was carried out on a desktop high performance computer funded by a Monash Engineering SEED Grant (2020). This work was supported by Australian Research Council grants FT110100427 & DP210100308.

**Competing interests:** The authors declare no competing interests.

**Supplementary information:** The supplementary material is attached

**Correspondence and requests for materials** should be addressed to Philip Nakashima (philip.naskashima@monash.edu).

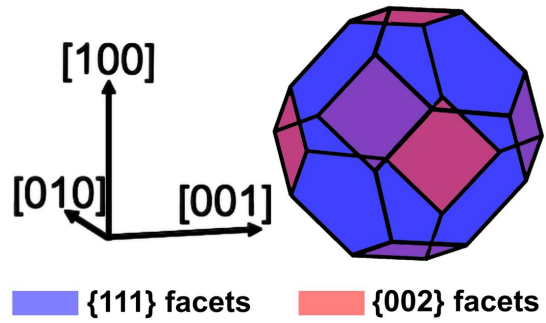

Extended Data Fig. 1: A fully facet-indexed schematic of a typical cuboctahedral void in aluminium.

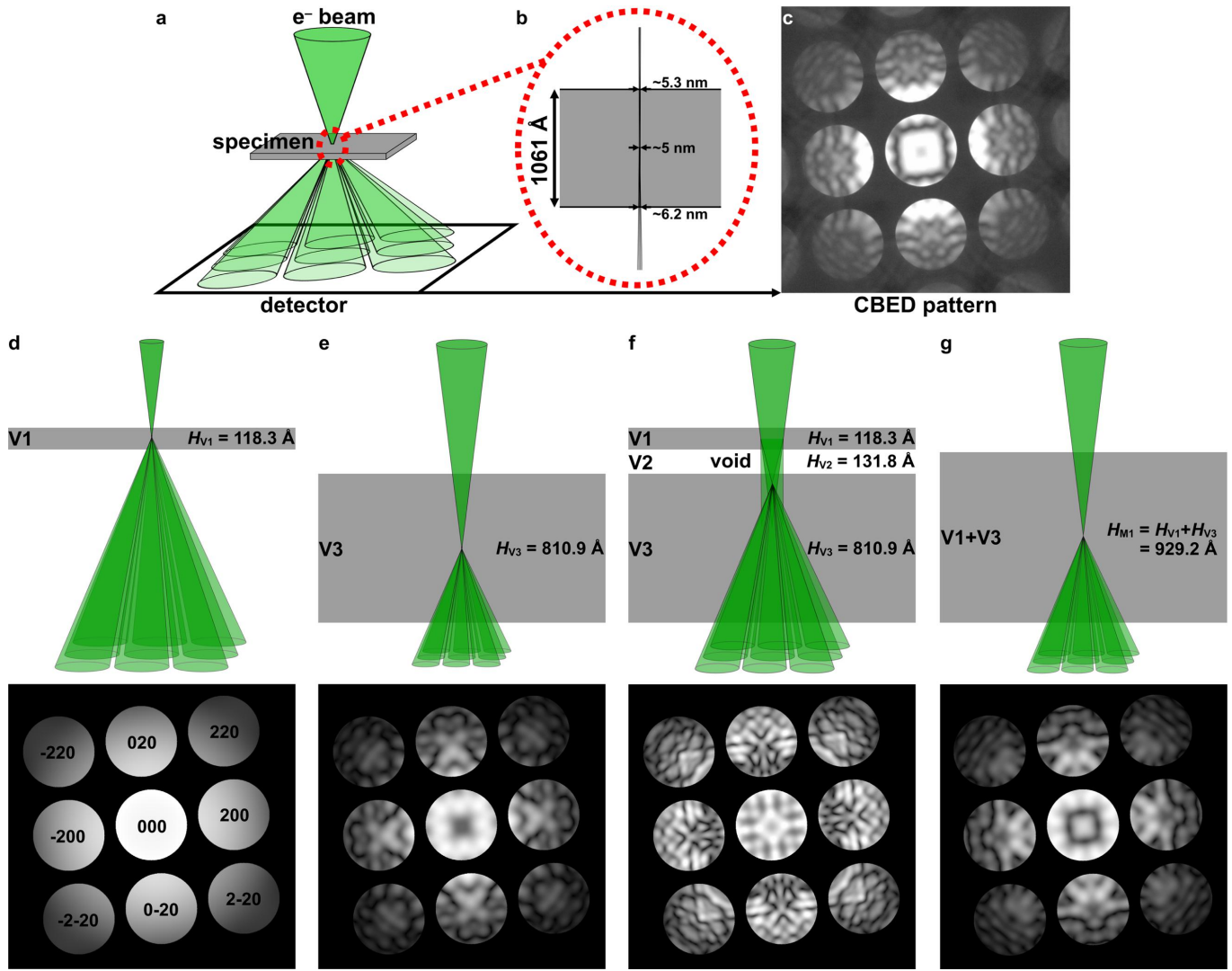

Extended Data Fig. 2: A schematic illustration of CBED and an illustration that the intensity distributions within CBED patterns are very sensitive to specimen morphology. **a,** A convergent electron beam focused onto a thin crystalline specimen produces a diffraction pattern that contains reflection discs. The cone angles are greatly exaggerated in **a** but are drawn to scale in **b**. **b,** The convergence angle in the present example is 0.34° (or 6 mrad). Note that in this schematic (**b**), scattering of the incident beam is drawn from the centre of the specimen and spans only a few diffraction orders. Whilst this is a crude approximation, it still illustrates that treating the probed volume as a column is a workable approximation in the context of the present study. **c,** An example CBED pattern from aluminium along [001] with 202.4 keV electrons corresponding to the thickness in the schematic (**b**). The intensity distributions in the diffraction discs are sensitive to crystal thickness and the electrostatic potential in the diffracting crystal planes. The cube root of the as-captured intensities is displayed (**c**) so that weaker intensity features can be seen. **d,e,f,g,** Calculated CBED patterns from two individual slabs of aluminium (**d,e**), the two slabs sandwiching a void (**f**), and the two slabs together as a single slab without the void (**g**), show that the diffraction pattern through the void (**f**) contains details that cannot be inferred from any of the single-slab CBED patterns (**d,e,g**). The present examples correspond to MS9 (**c**) and VS14 (**f**) (see Extended Data Figs. 3, 6 and 7). Note that only the 9 inner-most reflections (indexed in **d**) were calculated for 202.4 keV electrons along [001]. The cube-root of the intensities was taken to improve contrast in all CBED patterns shown here.

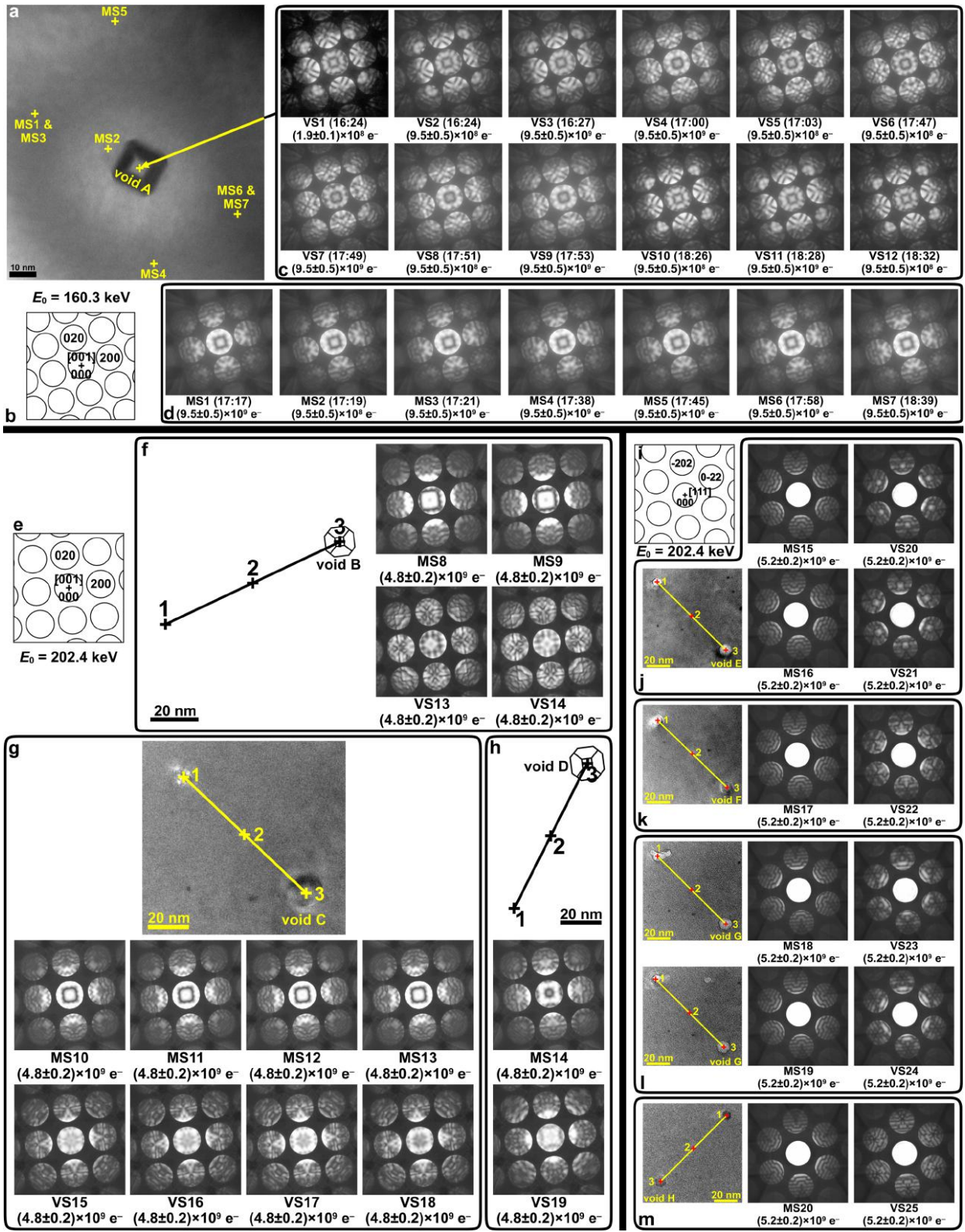

**Extended Data Fig. 3:** A summary of all CBED patterns collected for the present work and their locations with respect to each of the voids examined (voids A-H). **a**, A TEM image of void A marked with all locations where CBED data were collected. **b**, A schematic CBED silhouette showing the location of the [001] zone axis (+) and indices of two reflections with the shortest non-colinear scattering vectors. **c**, All void state (VS1-VS12) and **d**, all matrix state CBED patterns associated with void A including time stamps and dose estimates. Note that for **a–d**, the electron energy was 160.3 keV and no energy filtering was used. **e–m**, CBED patterns for MS8–MS20 and VS13–VS25 were collected with 202.4 keV electrons and ±5 eV energy filtering centred on 0 eV energy loss. **e,i**, Schematic CBED silhouettes showing the location and identity of the zone axes (+) and indices of two reflections with the shortest non-colinear scattering vectors. **f–h,j–m**, The 3-point line scans for data collection are shown (schematically where no image of the area was taken) for voids B–H, together with all of the matrix state (MS8–MS20) and void state (VS13–VS25) CBED patterns (with dose estimates) collected at points 2 and 3 respectively along the scans. Point 1 was where fine alignment was carried out, incurring damage at that point.

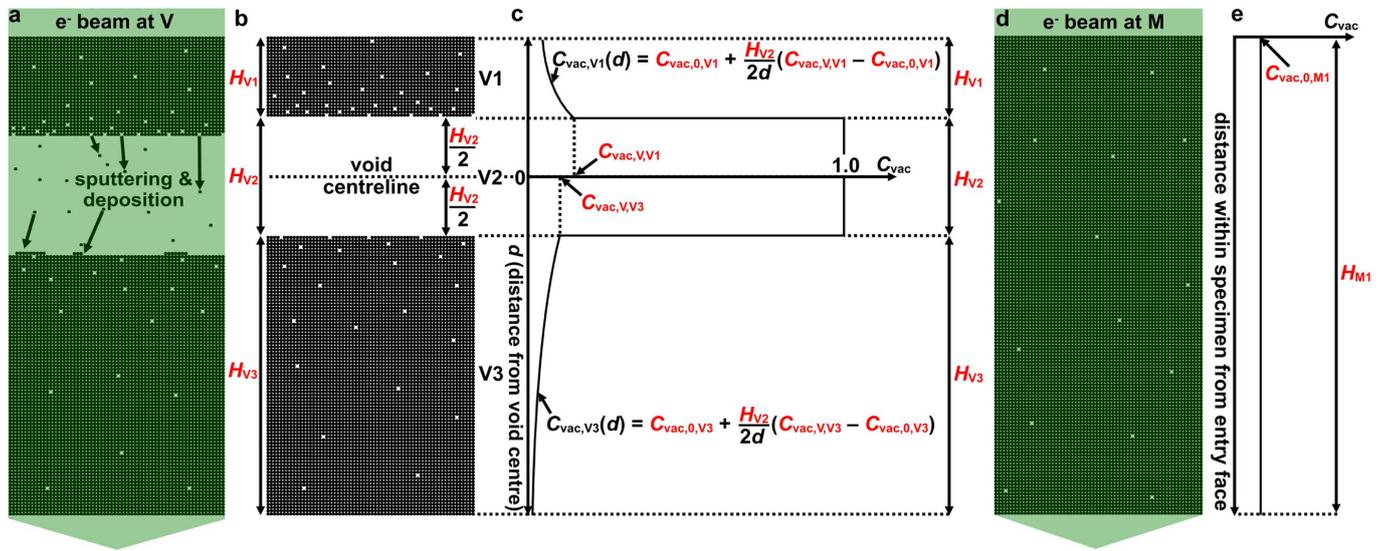

**Extended Data Fig. 4:** Vacancy concentrations as a function of distance from the centre of a void and within the uninterrupted matrix. **a,** A schematic diagram of a section through a void (not to scale) with atoms represented by black dots. The CBED probe will sputter atoms from the top of the void, with evaporated atoms depositing on the bottom surface. **b,** Over prolonged irradiation, a void will migrate upwards in the specimen (i.e. in the direction opposing the electron beam). The three thicknesses defined in Fig. 1c are shown here, and the void centreline is used as the origin of distance measurements, $d$, from the centre of the void. **c,** A schematic graph of vacancy concentration as a function of distance, $d$, from the centre of a void. The equations for vacancy concentrations in the aluminium matrix extending beyond $d = H_{V2}/2$ are obtained from the solution to the Laplace equation describing Fick's second law of diffusion[24,43]. These equations for the upper section (V1), $C_{vac,V1}(d)$, and the lower section (V3), $C_{vac,V1}(d)$, of the matrix surrounding the void, are incorporated into *QCBEDMS*. The equilibrium vacancy concentrations far from the void in sections V1 and V3, $C_{vac,0,V1}$ and $C_{vac,0,V3}$ respectively, and the vacancy concentrations at the interface of matrix sections V1 and V3 with the void, $C_{vac,V,V1}$ and $C_{vac,V,V3}$ respectively, are refinable parameters and coloured red. **d,** A schematic diagram of a section through the matrix being probed. The vacancy concentration, $C_{vac,0,M1}$, which is refinable and therefore coloured red, is assumed to be constant in the scattering volume as illustrated by the schematic graph in **e**, which ranges over the matrix thickness, $H_{M1}$. All thicknesses are refinable and coloured red in this figure.

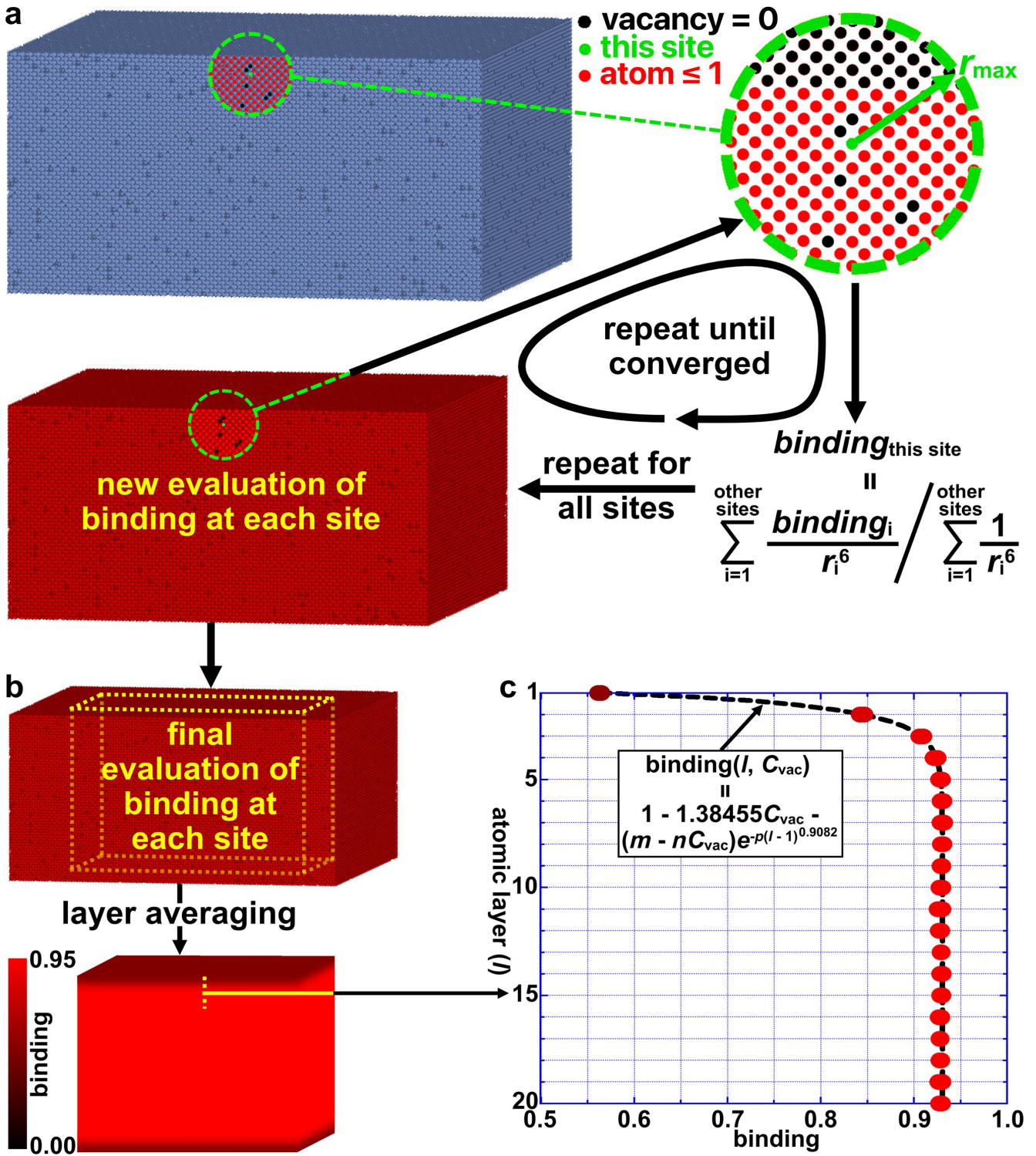

Extended Data Fig. 5: Modelling the degree of atomic binding in layers near a free surface using a Lennard-Jones-type[54,55] approximation, yields an equation for binding as a function of atomic layer number from a free surface, $l$, and vacancy concentration, $C_{vac}$, that can be used to modify the Debye-Waller parameters of atoms in each slice of a multislice model. **a,** A block of atoms and randomly distributed vacancies with concentration $C_{vac}$, is generated with a top facet having a surface normal of $<uvw>$. In the present example, the lattice is FCC, $C_{vac} = 0.05$, and $<001>$ is the surface normal of the top facet. Initially, atomic sites are assigned a binding value of 1 and vacancies are set to 0. For each atom in the block, the level of binding is determined by the weighted average of the site binding values at all other sites within a maximum range ($r_{max}$) where the weights are $1/r_i^6$, and $r_i$ is the distance of another site from the current site under evaluation – adopted from the first term of a Lennard-Jones-type interaction[54,55]. A new array of binding values feeds into the next iteration of the binding revaluation until it converges. **b,** The final binding values are averaged layer-by-layer after truncating the block (dashed yellow lines) to exclude the effects of other block surfaces not desired in the modelling of binding. **c,** A plot of the average binding in each atomic layer from the free surface, $l$, for 5 blocks of randomly distributed vacancies with $C_{vac} = 0.05$, results in a spread of points, albeit small. The binding colour scale in **b** also applies to the points in the graph in **c**. The fitted binding function has been found to apply to any surface normal of a FCC lattice, with only coefficients $m$, $n$ and $p$ changing with $<uvw>$. For example, for $<001>$, $m = 0.3974$, $n = 0.5985$, & $p = 1.46080$.

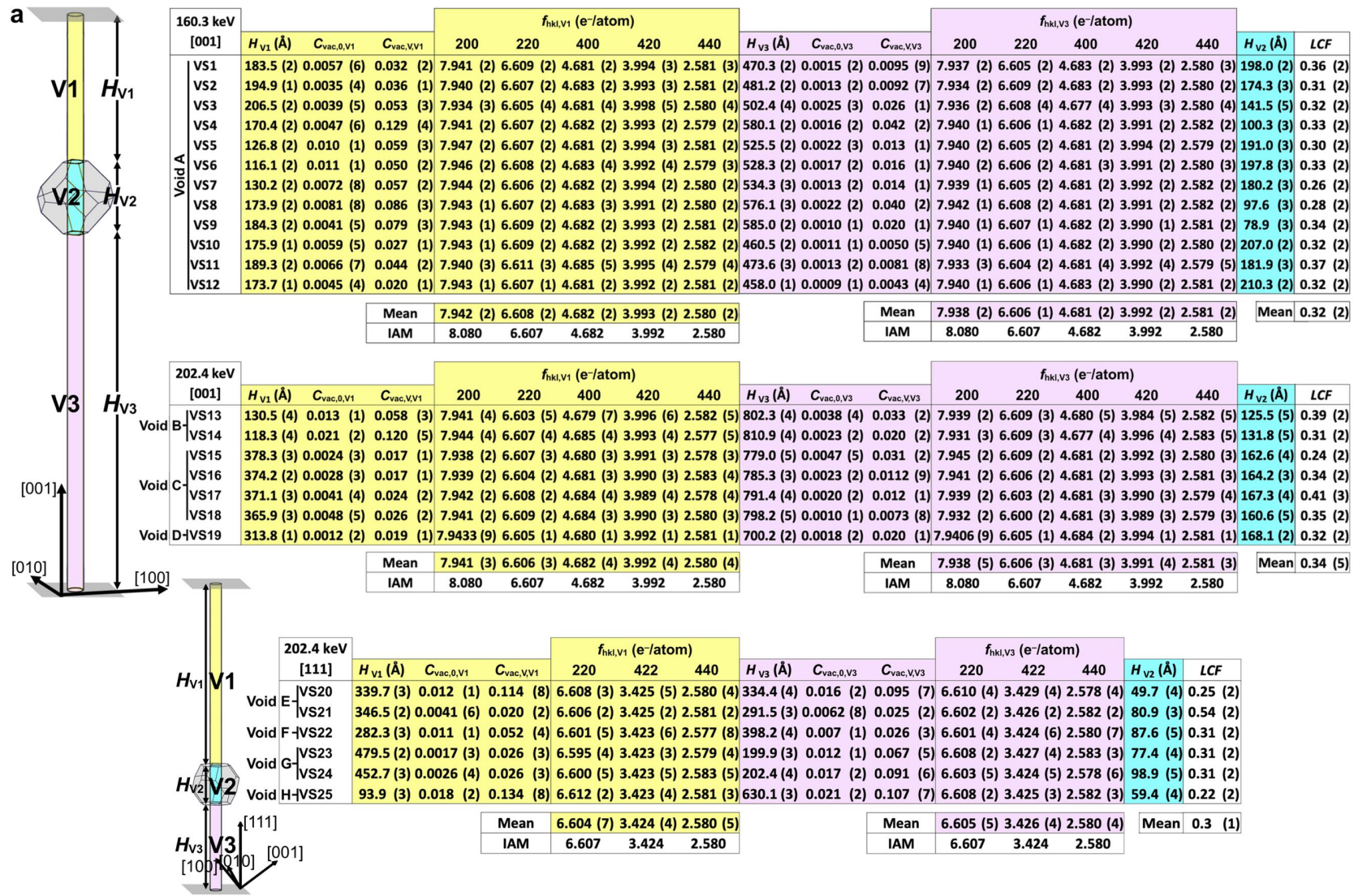

Extended Data Fig. 6a: Refined parameter results (excluding incident beam orientation and phenomenological absorption) for individual void states VS1-VS25 of voids A–H, collected with 160.30 keV electrons without energy filtering, along [001] (void A), and 202.4 keV electrons with ±5 eV energy filtering centred on 0 eV energy loss along [001] (voids B–D) and [111] (voids E–H). The different sections of the probe column through the void schematics and corresponding parts of the tables are colour coded yellow (V1), light blue (V2), and lilac (V3). The $LCF$ is common to V1 and V3 and is reported without colour coding. The reported refined parameters are: the three section thicknesses $H_{V1}$, $H_{V2}$, and $H_{V3}$; the equilibrium vacancy concentrations far from the void in sections V1 and V3, $C_{vac,0,V1}$ and $C_{vac,0,V3}$ respectively; the vacancy concentrations at the interface of matrix sections V1 and V3 with the void, $C_{vac,V,V1}$ and $C_{vac,V,V3}$ respectively; the $LCF$; and the structure factors refined for sections V1 and V3 after conversion to units of e-/atom by the Mott-Bethe formula[33,34]. Values in parentheses indicate the uncertainty in the last significant figure corresponding to the 95% confidence interval.

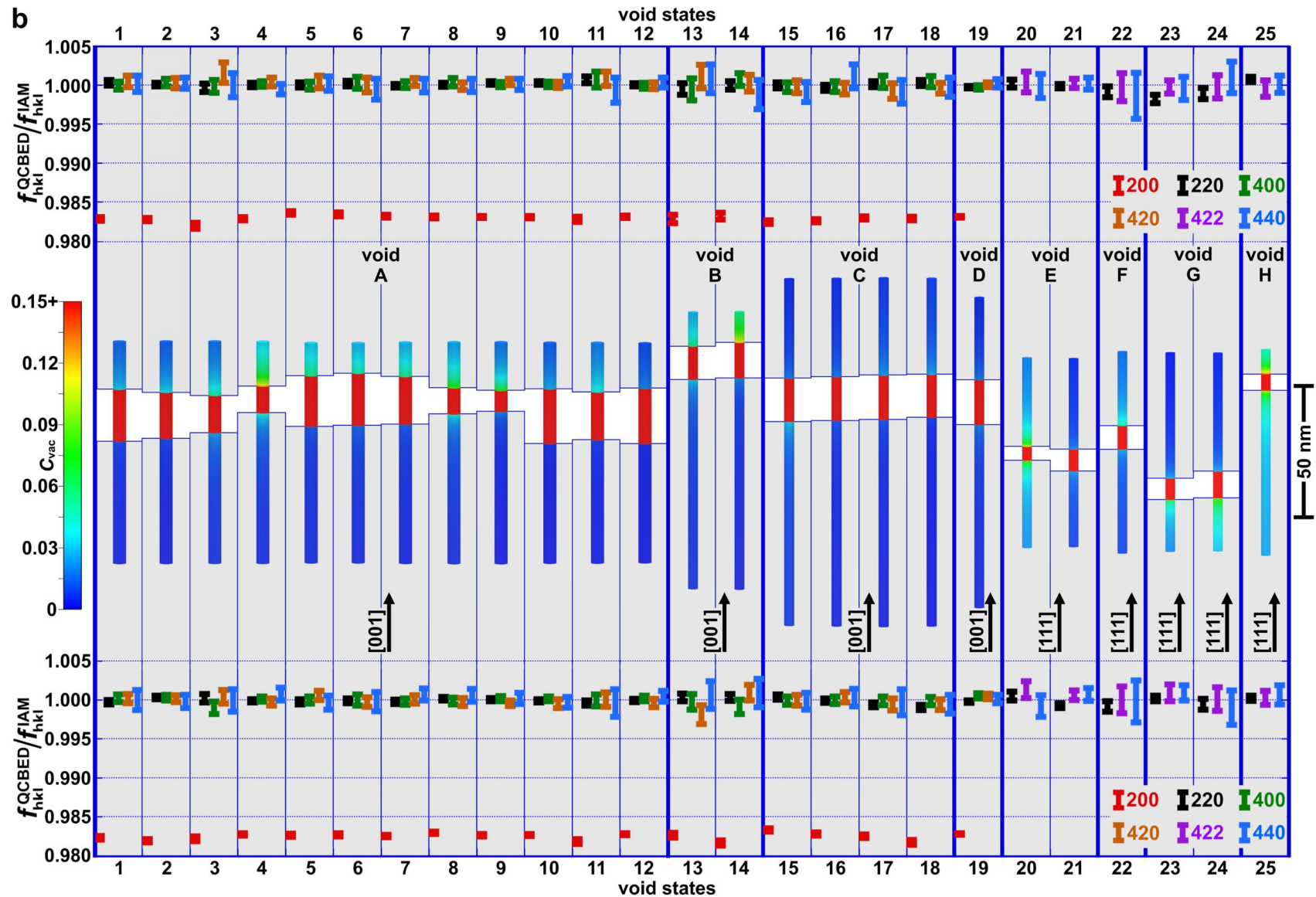

**Extended Data Fig. 6b: Individual, correctly scaled plots of vacancy concentrations as a function of position, $C_{vac}(d)$, within each probe column for every void state (VS1-VS25), and plots of the structure factors in the matrix regions above and below each void, as refined by QCBED.** All structure factors are plotted as ratios of the independent atom model (IAM)[49] values, after conversion from electrostatic potential structure factors, $V_{hkl}$ (Volts), to structure factors of the electron density, $f_{hkl}$ (e⁻/atom), using the Mott-Bethe formula[33,34]. The data for void A (VS1-VS12) were collected with 160.30 ± 0.06 keV electrons, without an energy filter (Extended Data Fig. 3a-d). The data for voids B-H (VS13-VS25) were collected with 202.4 ± 0.2 keV electrons with 0 ± 5 eV energy-loss filtering (Extended Data Fig. 3e-m).

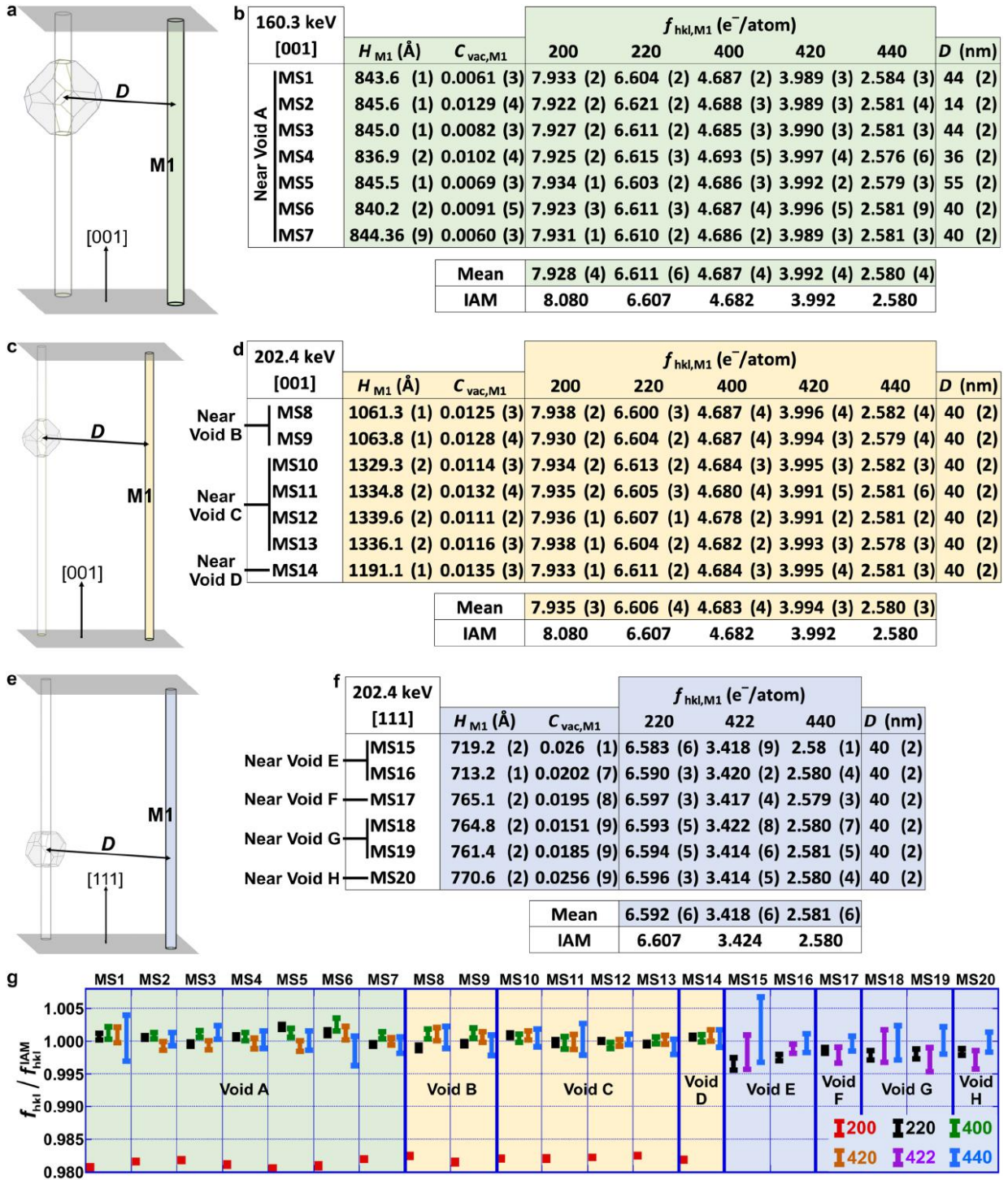

Extended Data Fig. 7: Refined parameter results (excluding incident beam orientation and phenomenological absorption) for individual matrix states MS1-MS20 at the reported distances, $D$, from voids A-H. **a,** A schematic diagram of the uninterrupted matrix-only probe column (green), in the [001] direction, and in relation to void A probed in void states VS1-VS12 (Extended Data Figs. 3a,c and 6). **b,** The table of refined parameters for matrix states MS1-MS7 near void A. The table lists: the refined thickness, $H_{M1}$; the refined constant vacancy concentration, $C_{vac,M1}$, the 5 refined structure factors after conversion to units of e$^-$/atom by the Mott-Bethe formula[33,34], $\{f_{200}, f_{220}, f_{400}, f_{420}, f_{440}\}_{M1}$, and the distance, $D$, of each matrix state from the centre of the void. **c,** A schematic diagram of the uninterrupted matrix-only probe column (yellow), in the [001] direction, and in relation to voids B-D probed in void states VS13-VS19 (Extended Data Figs. 3f-h and 6). **d,** As for **b** but for matrix states MS8-MS14 near voids B-D. **e,** A schematic diagram of the uninterrupted matrix-only probe column (blue), in the [111] direction, and in relation to voids E-H probed in void states VS20-VS25 (Extended Data Figs. 3j-m and 6). **f,** The table of refined parameters for matrix states MS15-MS20 near voids E-H (see Extended Data Fig. 3j-m). The table lists: the refined thickness, $H_{M1}$; the refined constant vacancy concentration, $C_{vac,M1}$, the 3 refined structure factors after conversion to units of e$^-$/atom by the Mott-Bethe formula[33,34], $\{f_{220}, f_{422}, f_{440}\}_{M1}$, and the distance, $D$, of each matrix state from the centre of the void. **b,d,f,** Values in parentheses indicate the uncertainty in the last significant figure with the uncertainty corresponding to the 95% confidence interval. **g,** A plot of all structure factors as ratios of the independent atom model (IAM)[49] values.

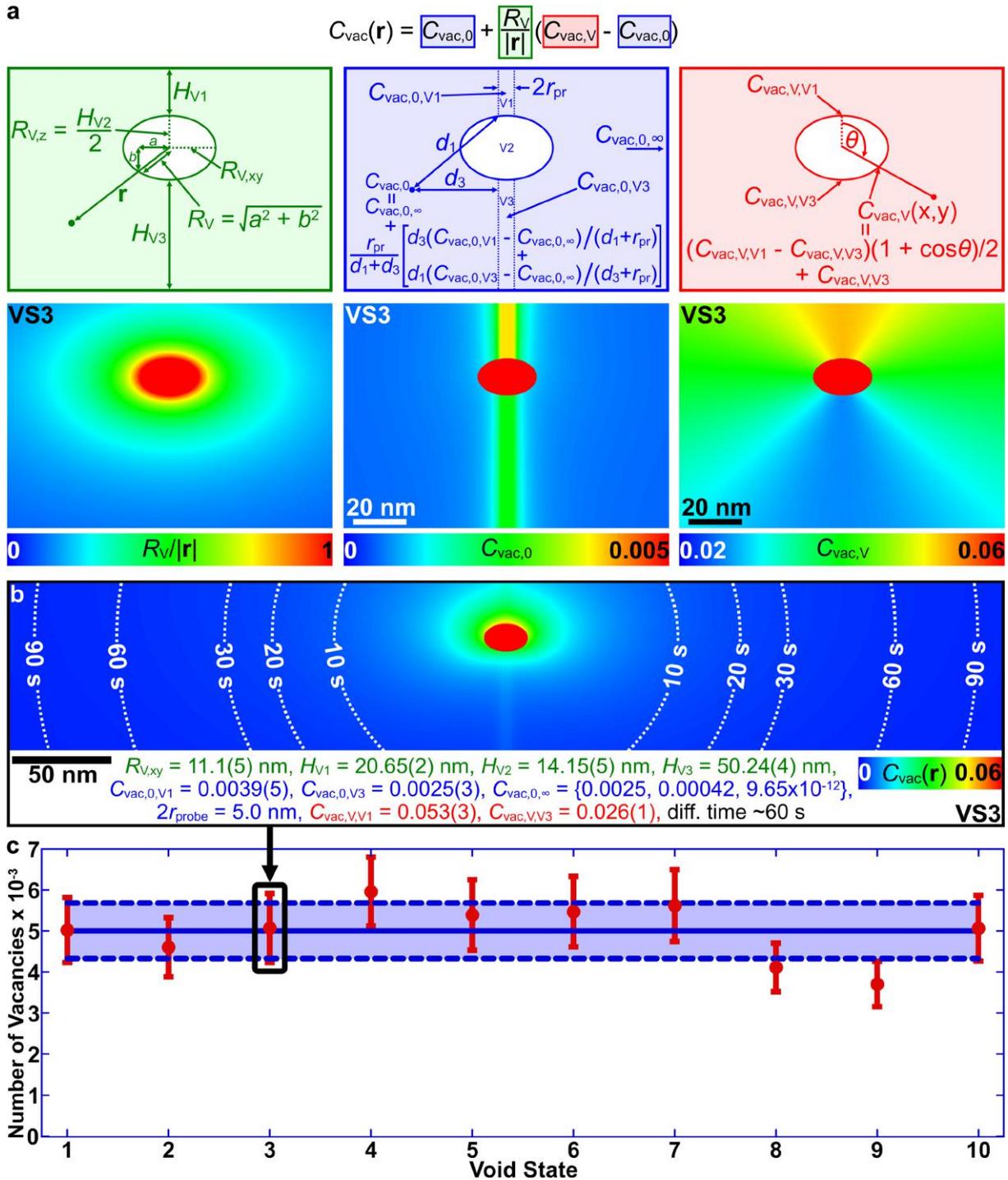

Extended Data Fig. 8: Fick's second law of diffusion[24,43] furnished with QCBED-measured parameters for void states VS1-VS10 (void A) shows that the number of vacancies in the void and its surroundings remains approximately constant from state to state. **a,** The Laplacian solution to Fick's second law[24,43] of diffusion, $C_{vac}(\mathbf{r})$, is stated with the colour-coded components modelled as shown in the corresponding diagrams. A computed example of each component follows, using the parameters for VS3 listed in **b** (see also Extended Data Fig. 6a). **b,** A plot of $C_{vac}(\mathbf{r})$ extending ±250 nm laterally from the void centre was computed for VS3 using the central values of all the parameters listed below the plot. The parameters are coloured to show which component of the $C_{vac}(\mathbf{r})$ model they pertain to. All parameters except for the background vacancy concentration far from the void, $C_{vac,0,\infty}$, were determined experimentally (uncertainty in the last significant figure in parentheses – see Fig. 1a and Extended Data Figs. 3 and 6a). The dotted white lines in the plot of $C_{vac}(\mathbf{r})$ for VS3 show the diffusion ranges of vacancies for different diffusion times during exposure of the void to the focussed CBED probe. **c,** A plot of the total number of vacancies associated with void A in the modelled cross sections of $C_{vac}(\mathbf{r})$ for VS1-VS10. Each $C_{vac}(\mathbf{r})$ model was integrated over the vacancy diffusion range corresponding to the approximate diffusion time relevant to each void state (eg. ~60 s for VS3 as listed in **b**). Error bars span the 95% Chebyshev confidence interval determined from the spread of $C_{vac}(\mathbf{r})$ models computed over all permutations of parameter values and their uncertainties. The example of VS3 is boxed in the plot. The blue region shows the 95% confidence interval (solid line for the mean and dashed lines for the bounds) for the average number of vacancies associated with void A in cross-section from VS1 to VS10.

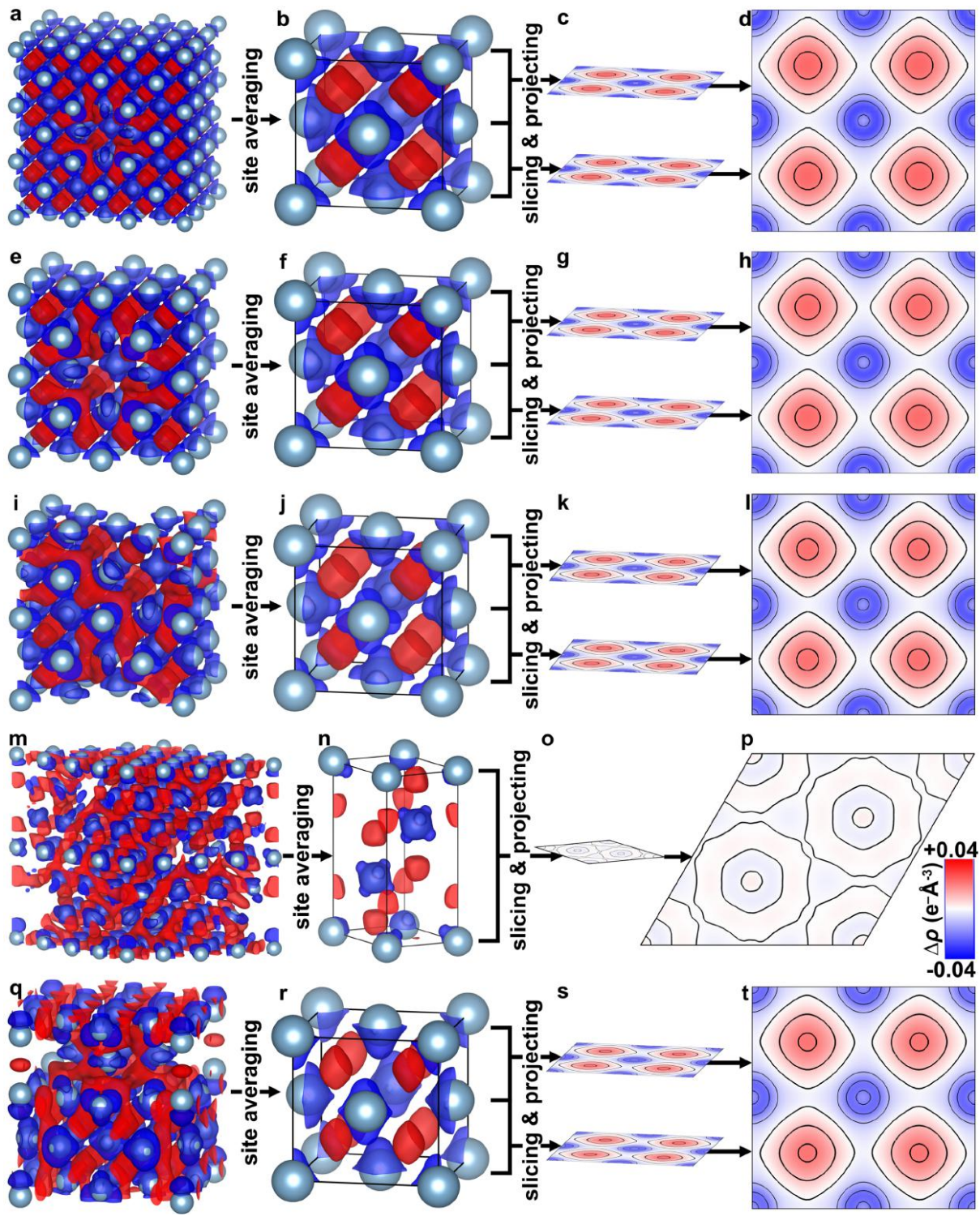

Extended Data Fig. 9: DFT-calculated (*Wien2k*[20]) bonding electron density, $\Delta\rho(\mathbf{r})$, plots for a range of different vacancy concentrations in aluminium, showing the process of obtaining 2-dimensional projections comparable to the $\Delta\rho(\mathbf{r})$ plots obtained from QCBED. **a,** A monovacancy in a 108-site cubic supercell of aluminium (light blue spheres), i.e. $C_{vac}$ = 0.00926, with a plot of $\Delta\rho(\mathbf{r})$ bounded by the −0.02 e⁻·Å⁻³ iso-surface (blue) and the +0.02 e⁻·Å⁻³ iso-surface (red). Note that the applied $B_{Al, eff}$ = 0.871 Å². **b,** The 4-site fcc unit cell after site averaging over the supercell in **a**. **c,** Partitioning the cell in **b** into 2 slices and showing the projected $\Delta\rho(\mathbf{r})$ in each slice. These are identical and are shown in detail in **d**. **e,f,g,h,** The same as for **a,b,c,d**, respectively except for a monovacancy in a 32-site supercell of aluminium, i.e. $C_{vac}$ = 0.03125, and $B_{Al, eff}$ = 0.899 Å². **i,j,k,l,** The same as for **a,b,c,d**, respectively except for two vacancies in a 32-site supercell of aluminium, i.e. $C_{vac}$ = 0.0625, and $B_{Al, eff}$ = 0.942 Å². **m,n,o,p,** The same as for **a,b,c,d**, respectively except for 9 vacancies in a 96-site hexagonal supercell of aluminium, i.e. $C_{vac}$ = 0.09375, and $B_{Al, eff}$ = 0.988 Å². Another exception in this case is that the site averaged cell in **n** is projected into a single slice shown in **o** and **p**. **q,r,s,t,** The same as **a,b,c,d**, respectively except for 4 vacancies in a 32-site supercell of aluminium, i.e. $C_{vac}$ = 0.125, and $B_{Al, eff}$ = 1.040 Å². The colour legend on the far right applies only to the 2-dimensional $\Delta\rho(\mathbf{r})$ plots and does not represent the +0.02 e⁻·Å⁻³ and −0.02 e⁻·Å⁻³ iso-surfaces (red and blue respectively) in the 3D $\Delta\rho(\mathbf{r})$ plots. This figure involved *VESTA*[50].

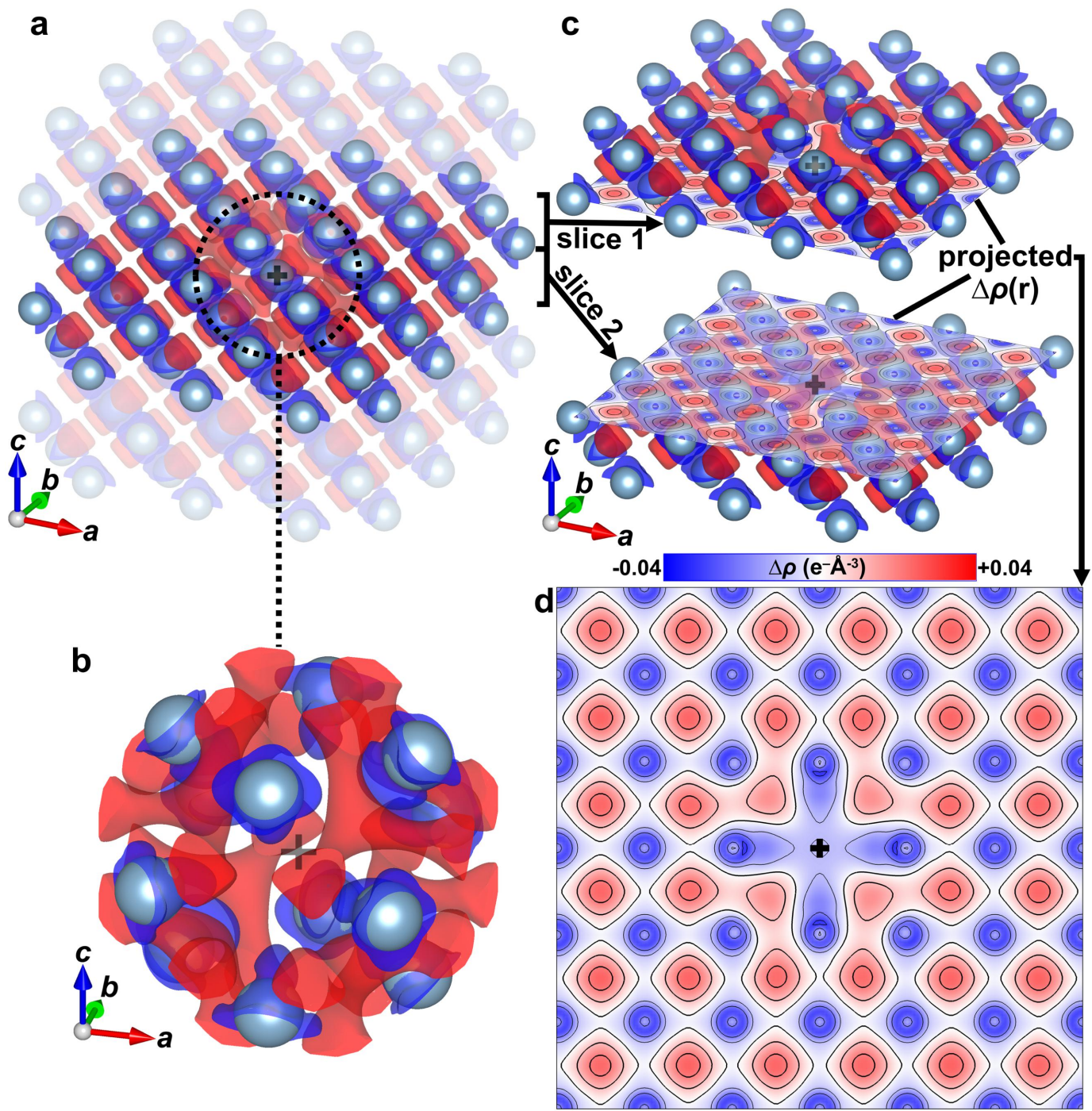

**Extended Data Fig. 10: A closer examination of the bonding electron density surrounding a vacancy. a,** The same 108-site DFT model as for Extended Data Fig. 9a-d is shown except that the vacancy (black cross) is now positioned in the centre of the super cell instead of at one of the faces as in the previous figure. Most of the 3D plot has been faded out to leave only the slices (equivalent in thickness to the multislice slices used in *QCBEDMS* analyses) surrounding the vacancy highlighted. **b,** An extracted portion of the bonding electron density surrounding the vacancy (black cross) shows the significantly altered bonding structure in the immediate neighbourhood of the vacancy. **c,** The two slices bounding the vacancy (black cross in each slice) are separated and projected independently along [001], giving the same plot of $\Delta\rho(\mathbf{r})$, shown also in **d. d,** The [001]-projected 2D $\Delta\rho(\mathbf{r})$ plot that applies to slices 1 and 2 bounding the vacancy (also shown as a black cross in this plot). Unlike in Fig. 4a,b and Extended Data Fig. 9, no site averaging (Supplementary Information) was carried out and this plot is the one shown in Fig. 4c. The colour legend between **c** and **d** applies only to the 2-dimensional $\Delta\rho(\mathbf{r})$ plots and does not represent the +0.02 e$^-$Å$^{-3}$ and −0.02 e$^-$Å$^{-3}$ iso-surfaces (red and blue respectively) in the 3D $\Delta\rho(\mathbf{r})$ plots. This figure involved *VESTA*[50].

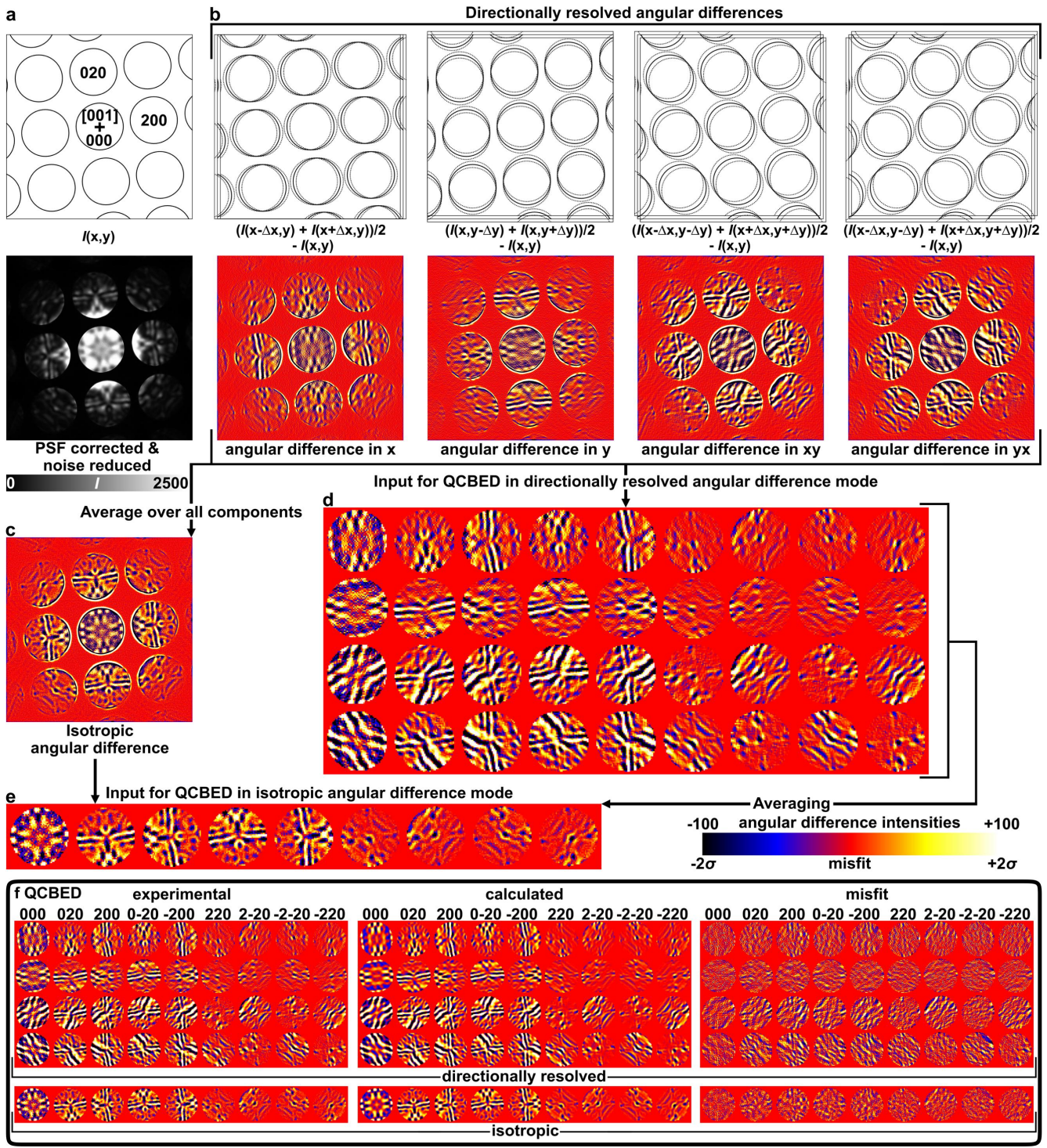

Extended Data Fig. 11: The method of directionally resolved angular differences, used for all QCBED refinements in this work, results in 4 times the number of data points per CBED pattern being matched and therefore, 4 times the number of constraints in the refinement of QCBED parameters. **a,** A schematic outline of the CBED pattern being used as an example is minimally indexed, showing the central beam, 000, and the two shortest non colinear scattering vectors and the position of the [001] zone axis orientation (+). This is accompanied immediately below by the pattern itself, which corresponds to VS15 (which also appears in Extended Data Fig. 3g) and has been PSF corrected[52] and noise reduced[53], and which was collected with 202.4 keV electrons. Here, the intensities are presented on a linear intensity scale. **b,** Schematic illustrations of the individual directional, centrosymmetric angular differences (above) are accompanied by the actual corresponding angular differences (below). In the schematics, dotted and solid discs indicate the shifted and unshifted patterns respectively. **c,** Averaging over the directionally resolved angular difference patterns results in the isotropic angular difference pattern of the previously established angular difference method[12,13,16-18]. **d,** Partitioning each of the directionally resolved angular difference CBED patterns into the reflection arrays used as input for directionally resolved angular difference QCBED, results in the displayed array. **e,** Averaging the 4 rows in **d** is equivalent to partitioning the isotropic angular difference CBED pattern in **c** to give the input for isotropic angular difference QCBED. **f,** An example of pattern matching by *QCBEDMS* using the example of VS15. The colour scale applies to all angular differences shown in this figure and the misfit in numbers of standard uncertainties ($\sigma$) in each pixel.

# Supplementary Information

## The *QCBEDMS* program

*QCBEDMS* is a multislice-based QCBED program, first produced by P.N.H.N. in 2001 and updated regularly over the last 2 decades (by P.N.H.N.) into its present form as used in this work. It was originally assembled from the *Multis* code by J.-M.Z., listed in *Electron Microdiffraction* by Spence and Zuo[2], which was written to calculate diffracted intensity as a function of thickness (i.e. Pendellösung). Code from *RefineCB_5* – a Bloch-wave QCBED program[2,63] written and shared by J.-M.Z. – has also been integrated into *QCBEDMS* for generating the scan over incident beam orientations (to produce rocking curves) and to facilitate pattern-matching refinement of parameters affecting the intensity distributions within the calculated CBED patterns. Results produced from the earliest version of *QCBEDMS* first appeared in 2003 in a paper concerning bonding electron densities in $\alpha$-$Al_2O_3$[64].

Since its first application 20+ years ago, *QCBEDMS* has undergone many modifications, and its capabilities have been expanded significantly. Many offshoot versions of *QCBEDMS* exist that contain bespoke modifications for specific QCBED purposes. There is also a fully parallelised version by T.L., used for bonding electron density measurements in copper[65]. The present work uses a non-parallel version that has evolved in its modifications to be suited to the present interrogation of voids and the vacancy concentrations surrounding them. In this work, 45 independent data sets were refined simultaneously (1 data set per core), so that the parallelism of the refinements came from the large number of single jobs run at the same time on different cores.

The key features of *QCBEDMS* critical to the present work are:

(i) The subdivision of the scattering volume into a user-specified number of regions along the zone axis nearest the incident beam direction. The regions have their own ensemble of slices defined to describe the atomic structure in each region. **For each region**, one must:

(a)  Specify the slice dimensions, i.e. the $a$, $b$ and $\gamma$ lattice parameters and slice thickness.

(b)  Specify the number of atoms and their identity, plus the coordinates of each atom and their occupancy, for each of the slices declared in (a).

(c)  Specify the starting thickness of each region in the scattering volume and the thickness refinement switches for each region. Refinement switches set which parameters are to be refined and which ones are to be held static. In the case of thicknesses, these should always be refined because CBED pattern intensity distributions are always most sensitive to thickness.

(d)  Allocate one of three vacancy concentration profiles: (1) constant concentration, (2) the Laplace solution to Fick's second law of diffusion[24,43] leading up to the boundary of a void, or (3) the Laplace solution to Fick's second law of diffusion[24,43] leading away from the boundary of a void (see Extended Data Fig. 4). For the first case of constant vacancy concentration, only $C_{vac,0}$ is specified whilst in the second case of leading up to the facet of a void, $C_{vac,0,V1}$ and $C_{vac,V,V1}$ are user-specified, and in the third case of leading away from the facet of a void, $C_{vac,0,V3}$ and $C_{vac,V,V3}$ are user-specified. In each case, the associated refinement switches must also be specified. Should one want to assume a perfect crystal without vacancies, one need only specify profile 1 and set $C_{vac,0} = 0$ and switch off the refinement switch.

(e)  Specify the Debye-Waller parameters (DWPs) for each atomic species and the associated refinement switches. This is done on a region-by-region basis because this will allow for differences in temperature from region to region to be accounted for, owing to different thicknesses and compositions. In this work, the DWPs in all regions (V1, V3 for QCBED of the void data, and M1 for the continuous matrix data) were fixed at the $T = 295$ K value ($B_{Al} = 0.86 Å^2$) in the literature[48].

(f) Specify the $m$, $n$, and $p$ coefficients of the binding equation discussed in the Methods (Eq. 2) and Extended Data Fig. 5. Note that the binding equation is used to modify the input DWPs (Methods, Eq. 3) to produce effective ones as a function of atomic layer number from a free surface, $l$, and the vacancy concentration in each slice, $C_{vac}$. For FCC elemental metals, along <001>: $m = 0.3974$, $n = 0.5985$, and $p = 1.46080$; and along <111>: $m = 0.3386$, $n = 0.5079$, and $p = 1.7217$.

(g) Specify the $hkl$ of the structure factors and corresponding phenomenological absorption coefficients to be refined, specify the starting values of the structure factors as fractions of the IAM[49], specify the starting values of the absorption coefficients as fractions of the values calculated by the phenomenological absorption model based on that of Bird and King[66,67], and specify the refinement switches for the structure factors and absorption coefficients.

(ii) The lattice contraction as a function of vacancy concentration, $C_{vac}$, via the lattice contraction factor ($LCF$). This is treated as a universal parameter that applies to all regions that have the same composition and structure. The $LCF$ is defined in the main text (see Eq. 1 and Fig. 2a). As a result of $C_{vac}$ being able to vary from slice to slice (see the main text, Methods, and Extended Data Fig. 4), the implementation of the $LCF$ requires the determination of $V$ according to Eq. 1 (main text) for every slice. Since $V_0$ is proportional to the product of the user-specified lattice constants, $a_0$, $b_0$, and slice thickness $\Delta z_0$, it is trivial to determine the lattice constants and slice thickness in the $j^{th}$ slice as follows:

$$\frac{a_j}{a_0} = \frac{b_j}{b_0} = \frac{\Delta z_j}{\Delta z_0} = \sqrt[3]{1 - LCF \times C_{vac,j}}. \tag{4}$$

This means that in cases where $C_{vac}$ varies from slice to slice within a region, so do the lattice parameters. This also means that the reciprocal lattice vectors change from slice to slice and the incident beam directions spanning the reciprocal lattice within each slice also change from slice to slice. This has all been implemented in *QCBEDMS*. The effect of variable lattice parameters from slice to slice due to variable $C_{vac}$, can be thought of very crudely as being akin to blurring the incident beam direction for a CBED pattern from a vacancy-free crystal with a short-range point spread function.

(iii) A parametrised approximation to Bird and King's phenomenological absorption model[66,67] summarised by the expression:

$$f'_j(Z_j, B_j, E_0, s) = A d_j^{non-local}(Z_j, B_j, E_0, s) + B g_j^{local}(Z_j, B_j, E_0, s) + C l_j^{local}(Z_j, B_j, E_0, s). \tag{5}$$

Here, $f'_j(Z_j, B_j, E_0, s)$ is the absorptive form factor and $Z_j$ and $B_j$ are the atomic number and Debye-Waller parameters for the $j$th atom in the atomic structure respectively. The beam electron energy is given by $E_0$, and $s = \frac{\sin\theta}{\lambda}$. In Eq. 5, $d_j^{non-local}(Z_j, B_j, E_0, s)$ is a Dirac delta function having a maximum at $s = 0$ Å$^{-1}$ and models phenomenological absorption due to "non-local" effects (i.e. those that are introduced by the nature of the experiment). In Eq. 5, $g_j^{local}(Z_j, B_j, E_0, s)$ is a Gaussian and $l_j^{local}(Z_j, B_j, E_0, s)$ is a Lorentzian that both model absorption local to each atom, the latter having a maximum at $s > 0$ Å$^{-1}$. The functional forms of $d_j^{non-local}(Z_j, B_j, E_0, s)$, $g_j^{local}(Z_j, B_j, E_0, s)$, and $l_j^{local}(Z_j, B_j, E_0, s)$ are invariant from element to element and both $g_j^{local}(Z_j, B_j, E_0, s)$ and $l_j^{local}(Z_j, B_j, E_0, s)$ are normalised such that values of the coefficients $A = 0$, $B = 1$, and $C = 1$ result in Eq. 5 reproducing the Bird and King *ATOM* subroutine[66] absorptive form factors to within ±3% for $4 \leq Z_j \leq 98$ over the ranges: $0.05 \text{ Å}^2 \leq B_j \leq 2.00 \text{ Å}^2$, $1 \text{ keV} \leq E_0 \leq 1000 \text{ keV}$, and $0 \text{ Å}^{-1} \leq s \leq 6 \text{ Å}^{-1}$ (Ref.[67]). The reproduction is not as good for hydrogen, helium and lithium, across the ranges given above, with maximum errors of ±17% for hydrogen, ±11% for helium and ±8% for lithium[67].

In *QCBEDMS*, there is the option to refine the coefficients *A*, *B*, and *C* in Eq. 5. This parametrised approach[67] is implemented instead of using the *ATOM* subroutine of Bird and King[66] because CBED data that have not been electron-optically energy filtered can still be used in QCBED via the angular difference method[12]. In fact, this method also removes the portion of the signal due to thermal diffuse scattering (TDS) that cannot be pattern matched and that cannot be removed by energy filtering. What is retained though is the portion of the inelastic signal that mimics the elastic scattering intensity distribution[12,68] from a range of different sources including elastic rescattering of electrons that have lost energy, and plasmon-loss signals. The result is an additional contribution to the angular difference intensity distribution that is not accounted for by the *ATOM* subroutine[66] but can be accommodated by refining parameters *B* and *C* in the parametrised implementation of phenomenological absorption[67] according to Eq. 5.

In general, there is rarely a need to refine the *A* coefficient in Eq. 5 because this serves only to scale all the calculated intensities up and down uniformly and this is automatically done when the calculated intensities are normalised to the same magnitude as the experimental intensities. However, if the incident beam intensity has been measured separately (by collecting a CBED pattern in the absence of the specimen, and under the same conditions as the pattern being matched) the normalisation of the calculated intensities can be fixed to this predetermined value with the aim of refining the *A* coefficient of the non-local term in Eq. 5. This can be interesting for its own reasons – i.e. to gauge the magnitude of the non-local absorption as a function of scattering geometry, the specimen, the use or not of an energy filter, and the TEM itself, for example.

For this work, all refinements were performed with $A = 0$ and the calculated intensities were normalised to minimise the misfit metric. Only $B$ and $C$ were refined, and $B = C = 1$ was used as a starting point in the refinements.

**Key features of *QCBEDMS* that are generally beneficial for any QCBED application are:**

(iv) Implementation of the angular-difference method to allow pattern-matching of CBED patterns collected with or without an electron-optical energy filter[12]. The present work applied a variation where the individual directional differentials have not been summed but are matched simultaneously. We call this the *method of directionally resolved angular differences*. This approach is illustrated in Extended Data Fig. 11 for the example of VS15 and all data sets (VS1–VS25 and MS1–MS20) were refined in the same manner.

(v) A new form for the pattern-mismatch metric. The conventional form of the misfit metric, $\chi^2$, metric is given by:

$$\chi^2 = \frac{1}{\sum_{i=1}^{n} w_i} \sum_{i=1}^{n} \frac{w_i (I_i^{expt.} - c I_i^{calc.})^2}{\sigma_i^2} \tag{6}$$

where $I_i^{expt.}$ is the experimental intensity in the $i^{\text{th}}$ pixel of the CBED data being matched with $I_i^{calc.}$, which is the theoretically calculated intensity in the same pixel. Note that $c$ is the normalisation factor that scales the calculated intensities to the same magnitude as the experimental intensities. This is not to be confused with the scale factor in X-ray diffraction least-squares refinement, even though the form of the equation is very similar. In the case of QCBED, a complicated and detailed intensity distribution as a function of scattering angle is being matched per reflection, as opposed to an integrated single intensity per reflection in X-ray diffraction. In the latter, the scale factor directly scales the resultant structure factor whilst in QCBED, the structure factor is absolute because it changes the form of the intensity distribution that is being matched. Returning to the quantities in Eq. 6, $\sigma_i^2$ is the variance associated with the $i^{\text{th}}$ pixel in the experimental data and $w_i$ is the individual weight applied to the $i^{\text{th}}$ pixel in the comparison. In all the refinements carried out in this work, $w_i = 1$ or $0$ and the weights are only used to mask out intensities that fall outside of the reflection discs being pattern matched.

Very often, the noise in CBED patterns is non-Poisson[53]. Furthermore, when averaging many frames in efforts to increase the signal-to-noise ratio, the linear and constant components of the noise as a function of signal are amplified and can become as prominent as the Poisson component[53]. This means that QCBED of different data sets with different amounts of frame averaging can have very different variances as a function of signal in the denominator of $\chi^2$ (Eq. 6) and this makes $\chi^2$ difficult to compare across QCBED refinements, even when the data were collected using the same instrument and detector, and even if they were collected in the same session.

To overcome the lack of a standard scale for $\chi^2$ when attempting to gauge the true "goodness of fit", *QCBEDMS* employs a self-normalised form of $\chi^2$, i.e. $\chi_{norm}^2$:

$$\chi_{norm}^2 = \frac{\frac{1}{\sum_{i=1}^n w_i}\sum_{i=1}^n \frac{w_i(I_i^{expt.} - cI_i^{calc.})^2}{\sigma_i^2}}{\frac{1}{\sum_{i=1}^n w_i}\sum_{i=1}^n \frac{w_i(I_i^{expt.} - 0)^2}{\sigma_i^2}}, \text{ i.e. } \quad \chi_{norm}^2 = \frac{\sum_{i=1}^n \frac{w_i(I_i^{expt.} - cI_i^{calc.})^2}{\sigma_i^2}}{\sum_{i=1}^n \frac{w_i(I_i^{expt.})^2}{\sigma_i^2}}. \quad (7)$$

This is simply the $\chi^2$ one is familiar with, divided by $\chi^2$ if one compares the experimental data with nothing at all, i.e. $I_i^{calc} = 0$. In this way, the variance still weights the individual data points within the determination of $\chi^2$ and thus $\chi_{norm}^2$; however, the latter is now normalised by both the experimental data and the variance in having what amounts to a "reference" $\chi^2$ in the denominator.

It should also be noted that $0 \leq \chi^2_{norm} \leq 4$. That is, in the best-case scenario where $I_i^{calc.} = I_i^{expt.}$ for all pixels, $i$, $\chi^2_{norm} = 0$. If one does nothing at all, meaning that $I_i^{calc.} = 0$, then $\chi^2_{norm} = 1$. On the other hand, the worst one can do is if the calculated CBED pattern is the exact opposite of the experimental one (and this is possible, though highly unlikely, with the angular or thickness difference methods of QCBED because the differentials necessarily have both positive and negative values[12,69]), i.e. $I_i^{calc.} = -I_i^{expt.}$, then $\chi^2_{norm} = 4$.

(vi) A hybrid simulated annealing with parabolic minimisation algorithm for multidimensional parameter optimisation. Popular multidimensional lowest-$\chi^2$ search algorithms used in QCBED have included the Simplex algorithm[2,63,70], conjugate and analytical gradient methods[3,7,63,70], and simulated annealing[63,70], among others. In *QCBEDMS*, a bespoke algorithm has been written as a standalone subroutine in Fortran 77, called *SIMPARA*. This is short for simulated annealing – parabolic minimisation and is illustrated in Fig. S1.

In Fig. S1, $\chi^2$ is used generically and in its implementation within *QCBEDMS*, *SIMPARA* acts on $\chi^2_{norm}$. The individual perturbation factors, $\{\Delta_1, \Delta_2, \ldots, \Delta_n\}$, are initially the user-estimated uncertainties in the initial parameter values, each multiplied by a different normal pseudo-random number where the normal distribution is centred on 0 and the normal distribution standard deviation is the "temperature", $T$. This gives rise to a parameter perturbation vector, $T\{\Delta_1, \Delta_2, \ldots, \Delta_n\}$. As the refinement of parameters progresses, the individual parameter uncertainties that constitute $\{\Delta_1, \Delta_2, \ldots, \Delta_n\}$ evolve. They are the running standard deviations of the individual parameter perturbations from each cycle that led to an improvement in $\chi^2$. From cycle to cycle, the temperature factor, $T$, is reduced. It is reduced at a faster rate if a cycle did not result in an improvement in $\chi^2$ and a slower "cooling" rate is used if a cycle resulted in a lower $\chi^2$.

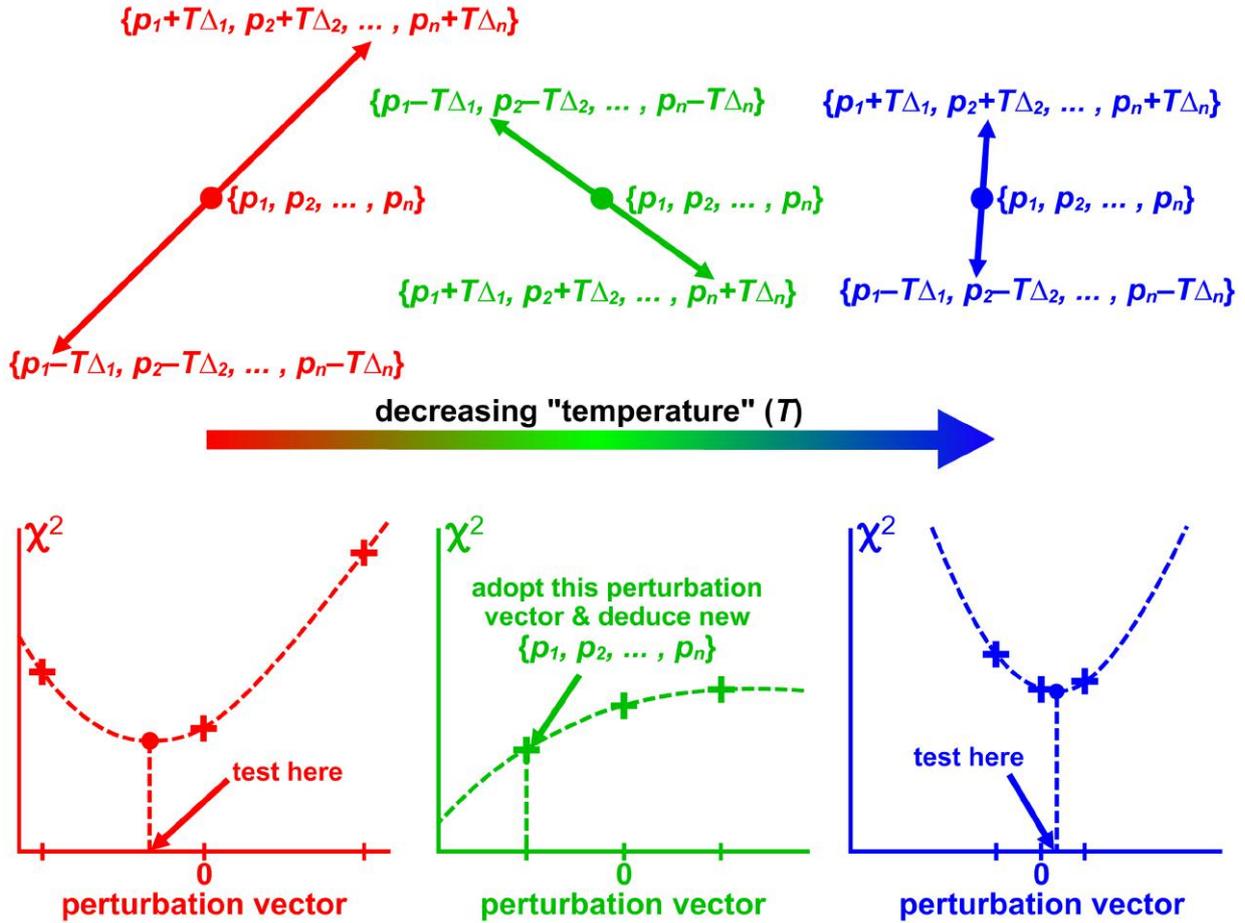

**Fig. S1:** A schematic illustration of the *SIMPARA* (simulated annealing – parabolic minimisation) algorithm. Perturbations $T\{\Delta_1, \Delta_2, ... , \Delta_n\}$ to each of the $n$ parameters $\{p_1, p_2, ... , p_n\}$ are applied in the forward and reverse directions and the $\chi^2$ fit parameter (here $\chi^2$ is used generically to represent any optimisation criterion) is determined for each set of new parameters. In combination with the unperturbed current parameters, the three values of $\chi^2$ permit the fitting of a parabola in the $n$-dimensional parameter space, i.e. a parabola for each parameter. The parabolic minima generate a new perturbation vector which when added to the current set of parameters, gives a new point for testing. Sometimes the minimum is attained from one of the initial perturbations (eg. the situation in green). Whichever set of parameters yields the lowest $\chi^2$ moves forward as the optimal set of parameters $\{p_1, p_2, ... , p_n\}$. After every cycle of this procedure, the "temperature", $T$, is reduced until a minimum temperature is reached or the maximum number of refinement iterations has been reached.

(vii) Cubic spline interpolation of electron and X-ray scattering factors for $0 \leq (\sin\theta)/\lambda \leq 6.0$ Å$^{-1}$ in a new Fortran 77 function called *ATOMICSCATFACT*. The function makes use of the X-ray and electron atomic scattering factors ($f^X(s)$ and $f^e(s)$ respectively, $s = (\sin\theta)/\lambda$) tabulated in the International Tables for Crystallography[71,72] which present the IAM scattering factors from Doyle and Turner[49]. Parametrised functional fits to scattering factors[71-74] are practical when it comes to crystallographic programming but if accuracy is vital, then the best approach is to adopt a direct cubic spline interpolation of the tabulated scattering factors. This is done by *ATOMICSCATFACT*, which can output either electron or X-ray atomic scattering factors over the range $0 \leq s \leq 6.0$ Å$^{-1}$. Where there are gaps in either $f^X(s)$ or $f^e(s)$ for a desired range of $s$ in the tables[49,71,72], the function uses the other type of scattering factor in the desired range for which there is no gap, performs a cubic spline interpolation to obtain the scattering factor at the required value of $s$, and performs a conversion to the desired scattering factor type with the Mott-Bethe formula[33,34]. The *ATOMICSCATFACT* function has replaced parameterised functional approximations to scattering factors in *QCBEDMS*.

## QCBED refinements using *QCBEDMS*

### Electron energy ($E_0$) calibration

Sixty CBED patterns from 99.9999+% pure and unshocked aluminium were used to QCBED-refine the electron energy, $E_0$, at a nominal accelerating voltage of 160 kV as part of a separate study. Ten CBED patterns were collected along [001] and 10 more from near [001] with primary scattering vector $\mathbf{g}_{220}$ bisected by the zone axis. Twenty patterns were collected near [1$\bar{1}$0]: 10 with primary scattering vector $\mathbf{g}_{220}$, and 10 with primary scattering vector $\mathbf{g}_{\bar{2}\bar{2}0}$ bisected by the zone axis. Ten patterns were collected along [$\bar{1}$11] and 10 patterns were collected near [$\bar{1}$11] with primary scattering vector $\mathbf{g}_{422}$ bisected by the zone axis. The probed specimen thicknesses ranged from 415 Å to 1781 Å.

    The electron energy was refined collectively with thickness, relevant structure factors, phenomenological absorption, the DWP and incident beam orientation. The refinements were run in loops that required all 60 QCBED sets to finish before the mean refined $E_0$ was used as new input for the next loop of 60 refinements where all other refined parameters were reset to their original input values. Geometric distortion correction was included for every data set in each refinement loop[75]. Ten such loops were run and the final value of $E_0 = 160.30 \pm 0.06$ keV (95% confidence interval) was determined from the last 5 loops (i.e. $5 \times 60 = 300$ QCBED results), by which stage the value of $E_0$ had converged.

The electron energy, $E_0$, for the CBED data collected with 200 keV (nominal energy) electrons was calibrated using a silicon standard with CBED patterns taken along <411>. An intersecting HOLZ line method was used, returning $E_0 = 202.4 \pm 0.2$ keV.

**The QCBED refinements**

All CBED data were collected along [001] or [111] (Extended Data Fig. 3). For the [001] data, 9 reflections were matched (the 000 disc and 8 surrounding discs with the shortest primary scattering vectors). For the [111] data, 7 reflections were matched (the 000 disc and 6 surrounding discs with the shortest primary scattering vectors).

For all [001] data, the symmetry equivalents of 200 and 220 were near the Bragg condition. Thus, $V_{200}$ and $V_{220}$, and the structure factors of all coupling vectors, i.e. $V_{400}$, $V_{420}$ and $V_{440}$, were refined[16]. For all the [111] data the symmetry equivalents of 220 were near the Bragg condition, so $V_{220}$ plus the structure factors of all coupling vectors, i.e. $V_{422}$ and $V_{440}$, were refined[16].

*Void state refinements*

For all void states (VS1-VS25), the refined parameters were: the three thicknesses $H_{V1}$, $H_{V2}$ and $H_{V3}$ defining the specimen geometry within the probe (Fig. 1c,g), $\{V_{hkl}\}_{V1}$ and $\{V_{hkl}\}_{V3}$ – the relevant sets of structure factors above and below the void respectively (Fig. 1c,g), $C_{vac,0,V1}$ and $C_{vac,V,V1}$ defining $C_{vac}(d)$ above the void, $C_{vac,0,V3}$, and $C_{vac,V,V3}$ defining $C_{vac}(d)$ below the void (Extended Data Fig. 4b,c), the $LCF$, the $B$ and $C$ coefficients of the phenomenological absorption parametrisation[67] (Eq. 5), and the incident orientation.

The QCBED pattern-matching procedure for VS1-VS25 is detailed in Fig. S2 and was repeated 10 times for an initial $LCF = 0$ and 10 times for an initial $LCF = 1$. This produced 10,000 sets of results for VS1-VS25 with 400 sets of results per void state. Extended Data Fig. 6a presents the averages and uncertainties (from the 95% confidence interval), for $H_{V1}$, $H_{V2}$, $H_{V3}$, $C_{vac,0,V1}$, $C_{vac,V,V1}$, $C_{vac,0,V3}$, $C_{vac,V,V3}$, $LCF$, $\{f_{hkl}\}_{V1}$ and $\{f_{hkl}\}_{V3}$ for all void states (where $f_{hkl}$ were converted from $V_{hkl}$ by the Mott-Bethe formula[33,34]). All these results, except the $LCF$, are graphically displayed in Extended Data Fig. 6b.

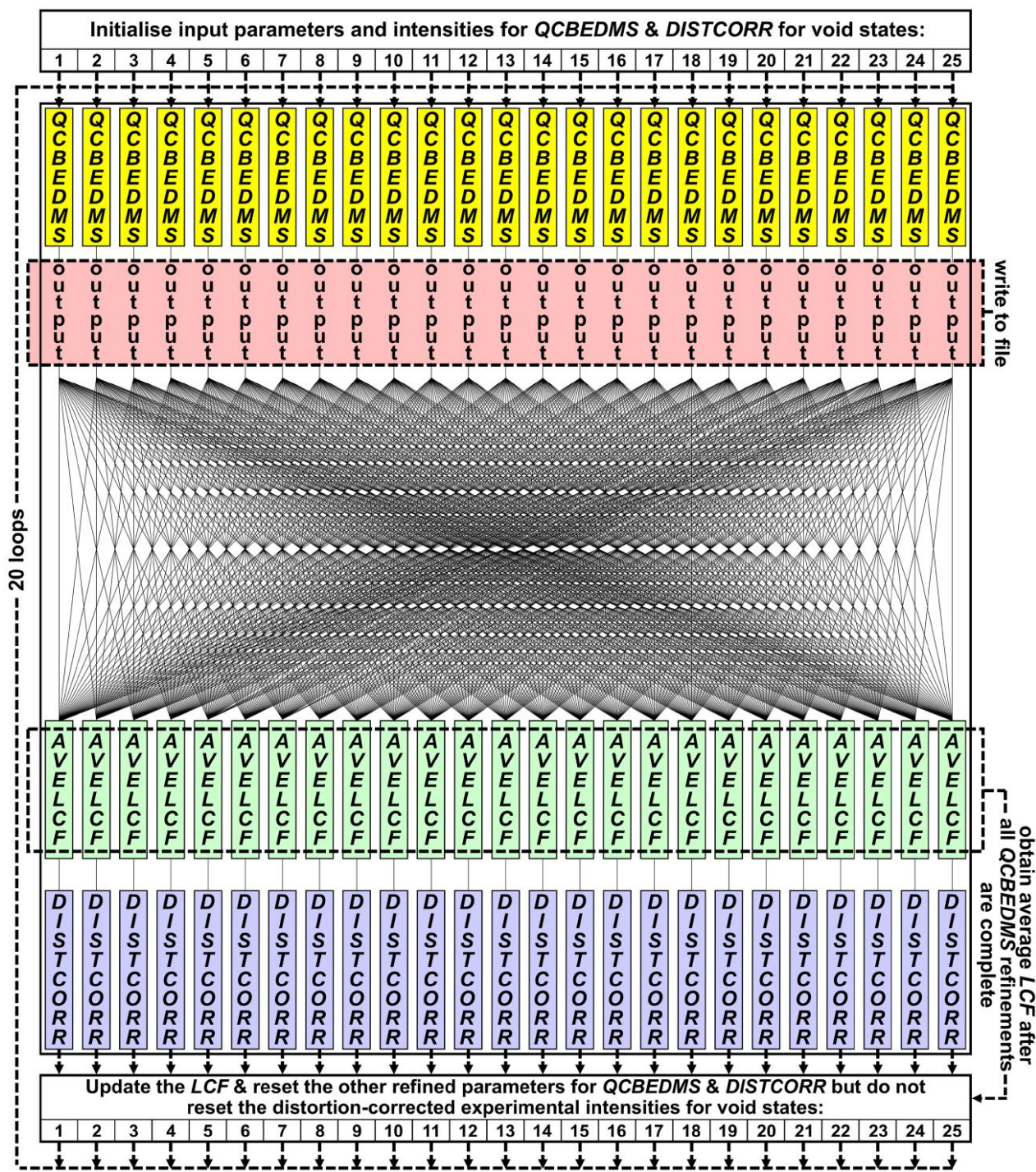

Fig. S2: The architecture of QCBED refinements for VS1-VS25 involves averaging the lattice contraction factor (*LCF*) returned from each void state refinement and feeding the average value back as input into the next refinement loop. The refinement process is initialised with the starting values of the refined parameters and experimental intensities unique to each data set. The first step is a *QCBEDMS* refinement which generates optimised values of the variable parameters for each of VS1-VS25. These results are written to a parameter summary file as each *QCBEDMS* refinement terminates. This is immediately followed by a call to a small program, "*AVELCF*", within each void state refinement. This program obtains the average of all the output *LCF* values from each refinement and if other void states are still running *QCBEDMS*, then the optimal values from the last global refinement loop for the corresponding voids states are used for averaging. The *DISTCORR* program refines geometric distortion corrections[75] and is called by each void state, generating geometrically corrected experimental CBED intensities. A global refinement loop terminates only after all void states have completed the *QCBEDMS–AVELCF–DISTCORR* sequence. The input parameters for *QCBEDMS* and *DISTCORR* are reset to their initial input values except for the *LCF* which is updated for all void state inputs with the average obtained after all void state *QCBEDMS* refinements finished in the current loop. The next refinement loop uses the distortion corrected experimental intensities from the current loop. Twenty loops were performed for the entire refinement procedure illustrated here, returning 500 results (25 void states × 20 refinement loops). For the present work, this procedure was repeated 10 times with an initial *LCF* of 0 and 10 times for an initial *LCF* of 1, giving 10,000 sets of results, i.e. 400 sets of results per void state.

The void state refinements were interlinked by setting the *LCF* starting value for every individual void state to the average refined value from all 25 void states at the end of a refinement loop and before the beginning of the next loop in the procedure illustrated in Fig. S2.

Examples of QCBED pattern matching are given for all void states in Figs. S3–S5 and should be examined in conjunction with the more detailed description of QCBED pattern matching given in Extended Data Fig. 11 and the Methods.

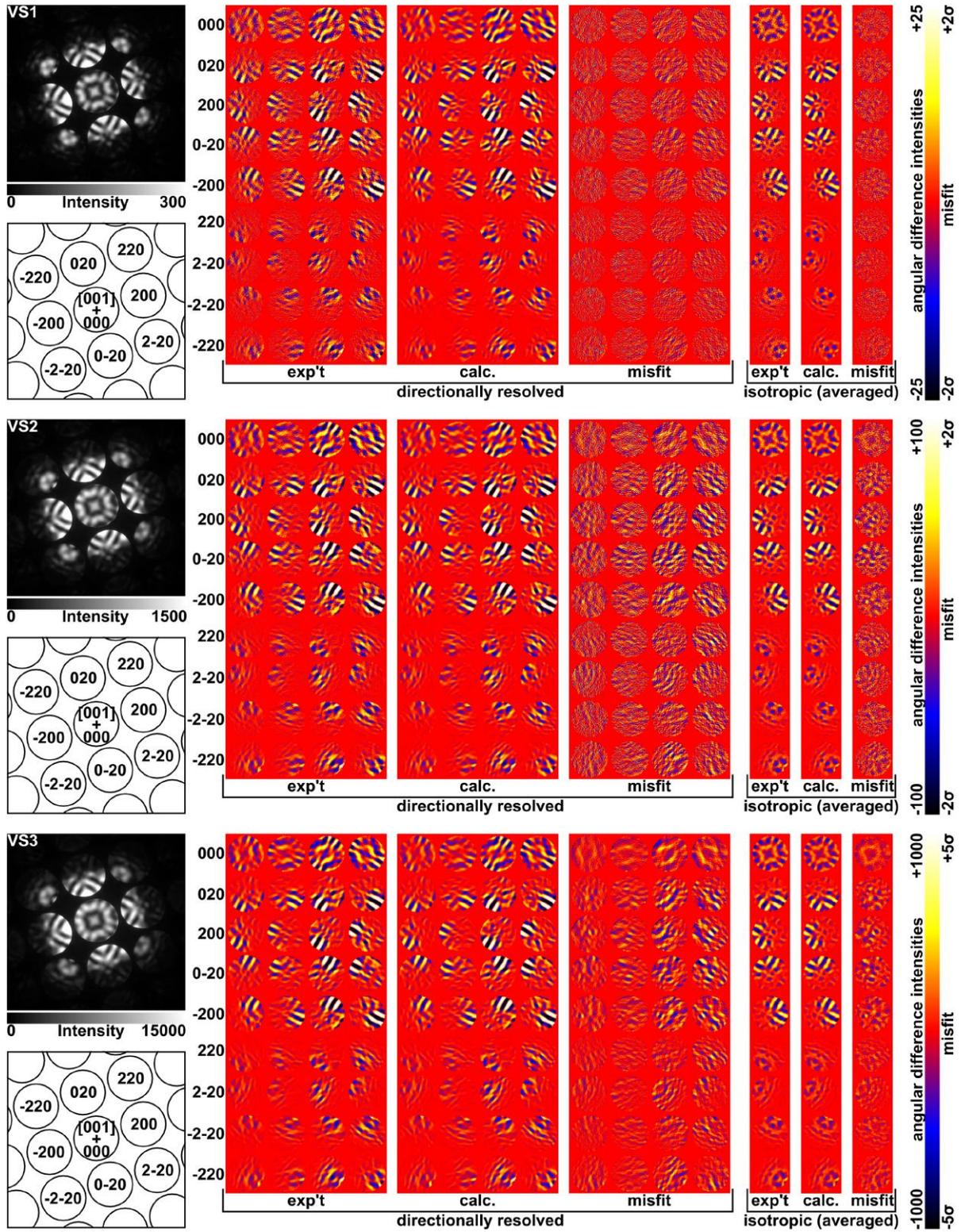

**Fig. S3 (part 1/4):** The CBED patterns and QCBED pattern-matched directionally resolved angular difference intensities for VS1-VS3 are shown together with the isotropic angular difference intensities determined by averaging the directionally resolved components after QCBED was performed. For each void state, the PSF-corrected[52] and noise reduced[53] pattern is shown in linear greyscale together with a matching schematic outline that indexes all the pattern-matched reflections and the location of the zone-axis orientation (+), which in the present cases is [001]. The angular difference intensities (determined as per the method in Extended Data Fig. 11) are all shown in false colour with the colour legend at right applying to all angular difference intensities shown for that void state. Note that the misfit maps are plots of $(I_i^{exp't} - I_i^{calc.})/\sigma_i$, where $I_i^{exp't}$ and $I_i^{calc.}$ are the experimental and calculated angular difference intensities in the $i^{th}$ pixel respectively, and $\sigma_i$ is the standard uncertainty in the $i^{th}$ pixel. The electron energy used for these cases was 160.3 keV. The location from which each CBED pattern was collected is shown in Extended Data Fig. 3. While the present figure presents the output of just one pattern-matching refinement for each void state, the refinements were repeated many times for each CBED pattern (as described above and in Fig. S2 and in the main text). The averages and uncertainties of the key parameters determined from all refinements for each void state are summarised in Extended Data Fig. 6.

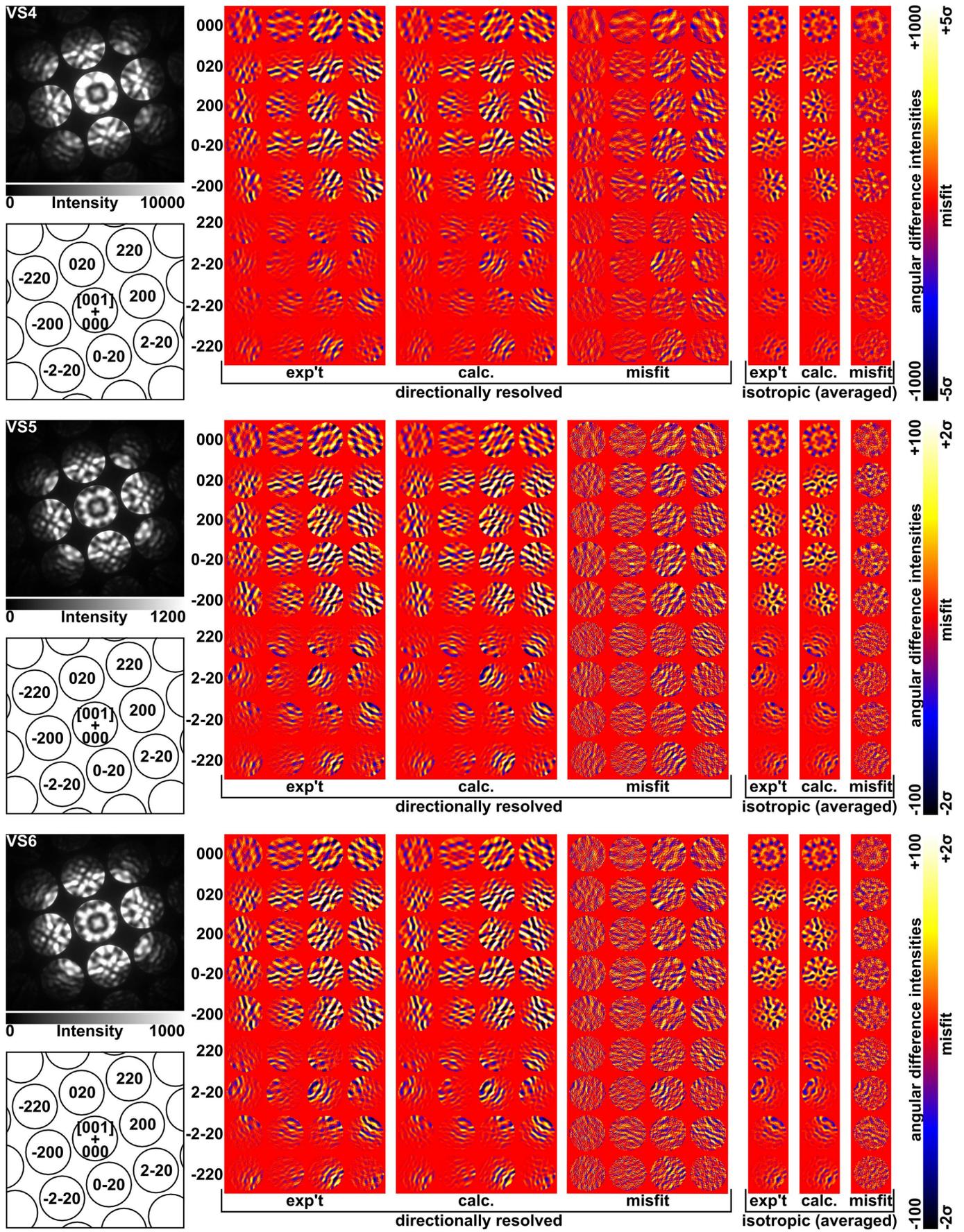

**Fig. S3 (part 2/4):** As per Fig. S3 (part 1/4), except for VS4-VS6.

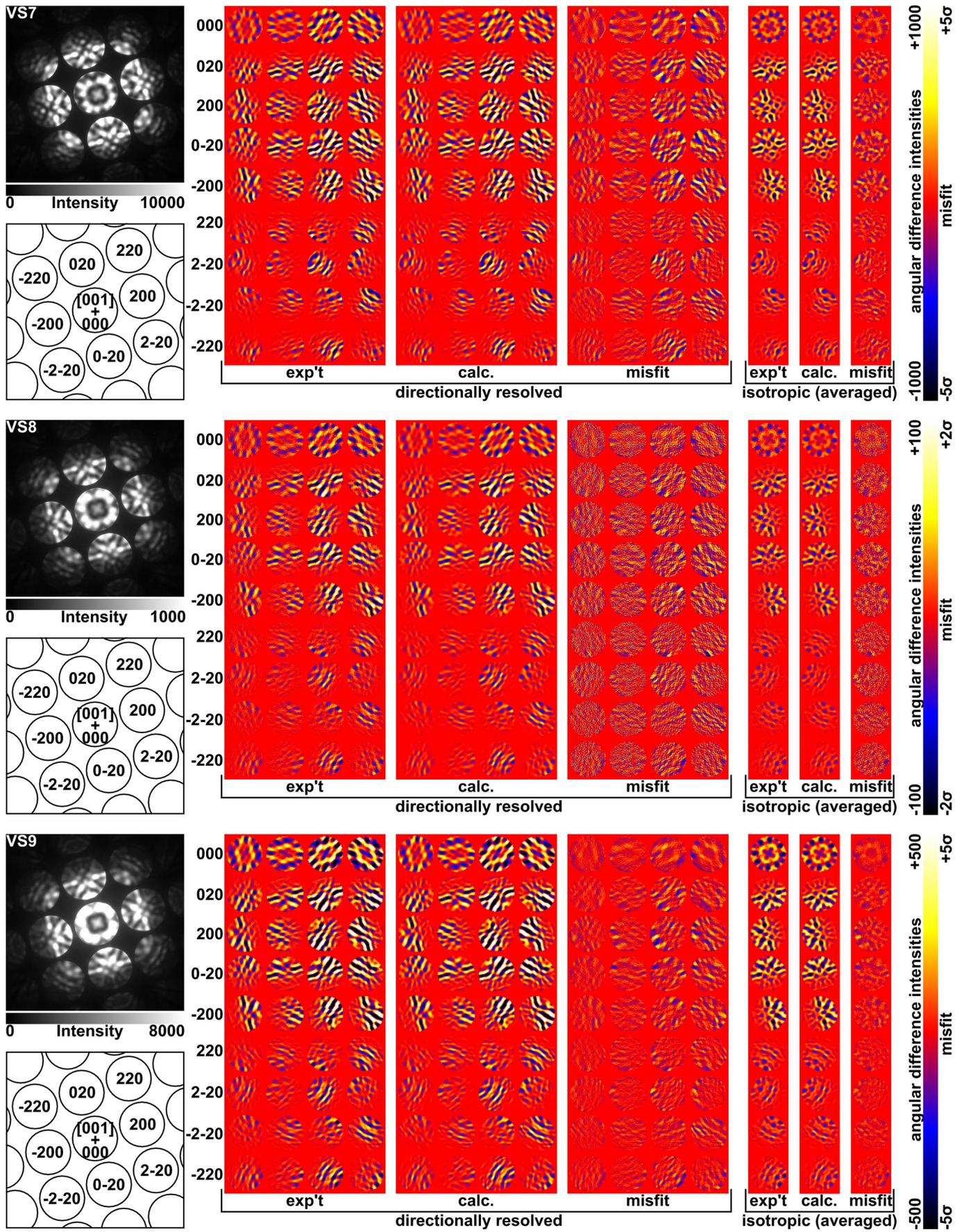

Fig. S3 (part 3/4): As per Fig. S3 (part 1/4), except for VS7-VS9.

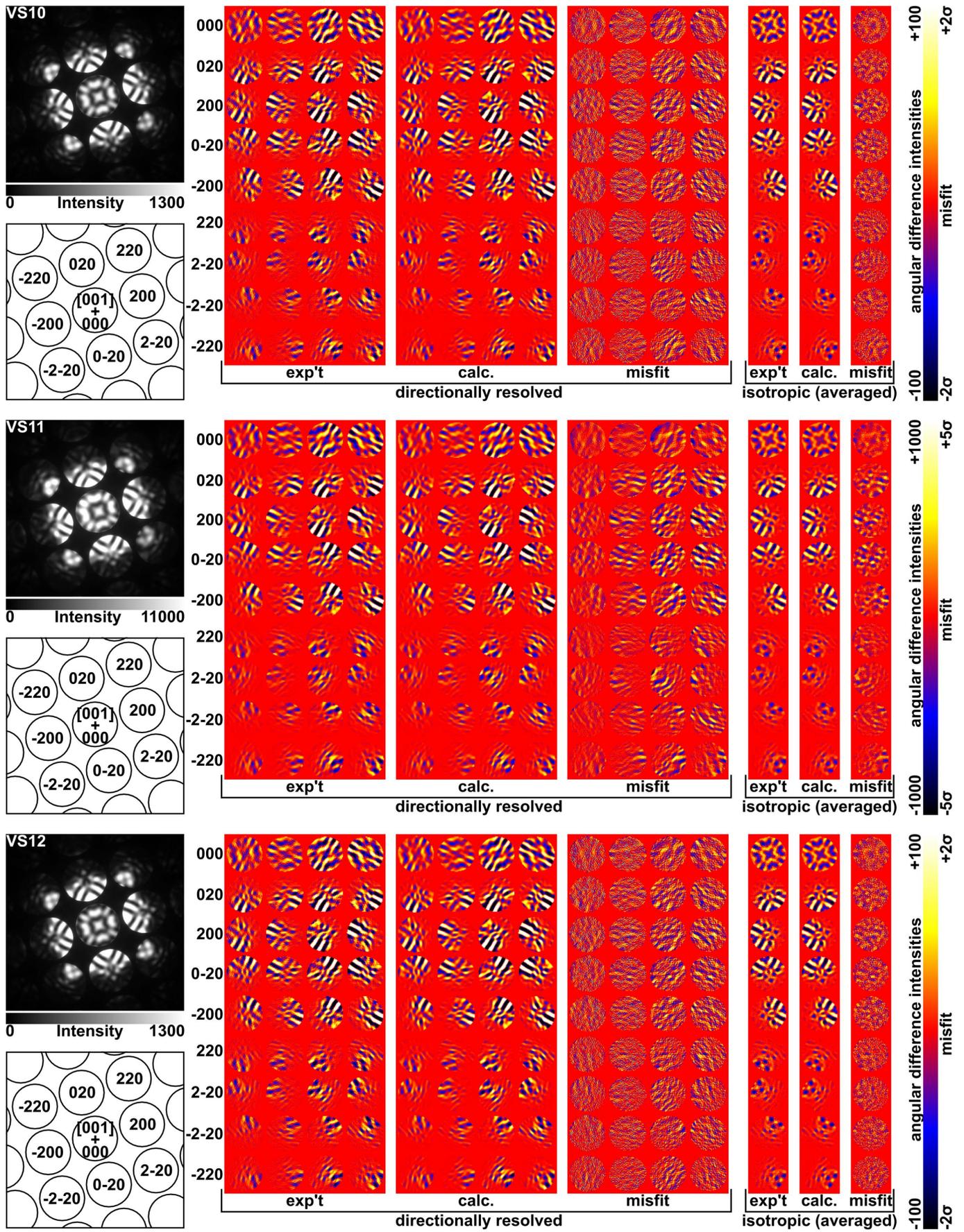

**Fig. S3 (part 4/4):** As per Fig. S3 (part 1/4), except for VS10-VS12. Note that VS10 is the example presented in Fig. 1e,g, where the isotropic (averaged) intensity differences have been realigned horizontally.

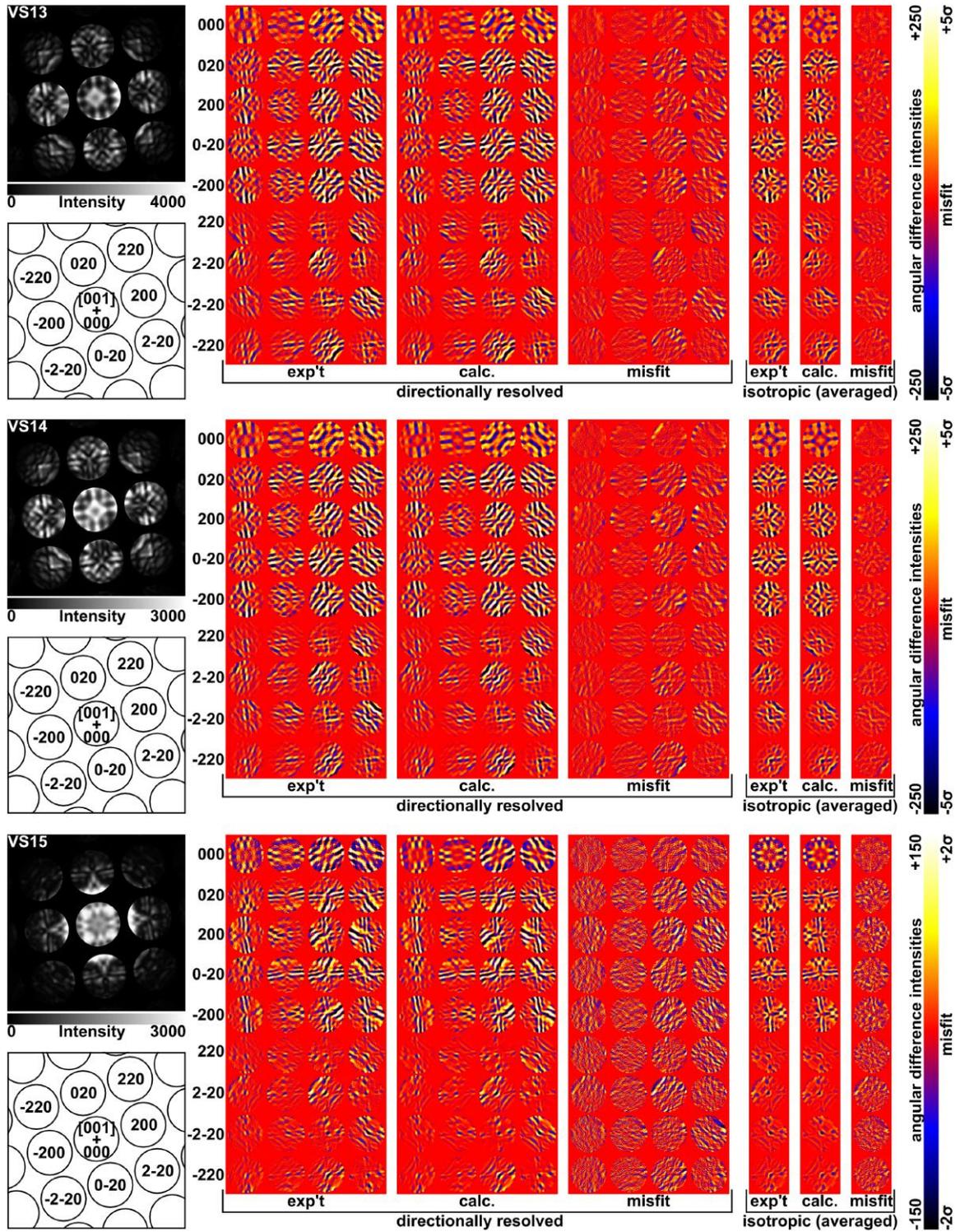

**Fig. S4 (part 1/3): The CBED patterns and QCBED pattern-matched directionally resolved angular difference intensities for VS13-VS15 are shown together with the isotropic angular difference intensities determined by averaging the directionally resolved components after QCBED was performed.** For each void state, the PSF-corrected[52] and noise reduced[53] pattern is shown in linear greyscale together with a matching schematic outline that indexes all the pattern-matched reflections and the location of the zone-axis orientation (+), which in the present cases is [001]. The angular difference intensities (determined as per the method in Extended Data Fig. 11) are all shown in false colour with the colour legend at right applying to all angular difference intensities shown for that void state. Note that the misfit maps are plots of $(I_i^{exp't} - I_i^{calc.})/\sigma_i$, where $I_i^{exp't}$ and $I_i^{calc.}$ are the experimental and calculated angular difference intensities in the $i^{th}$ pixel respectively, and $\sigma_i$ is the standard uncertainty in the $i^{th}$ pixel. The electron energy for these cases was 202.4 keV. The location from which each CBED pattern was collected is shown in Extended Data Fig. 3. While the present figure presents the output of just one pattern-matching refinement for each void state, the refinements were repeated many times for each CBED pattern (as described above and in Fig. S2 and in the main text). The averages and uncertainties of the key parameters determined from all refinements for each void state are summarised in Extended Data Fig. 6. Note that VS15 is the example presented in Extended Data Fig. 11 where the generation of the directionally resolved and isotropic (averaged) angular intensity differences is demonstrated.

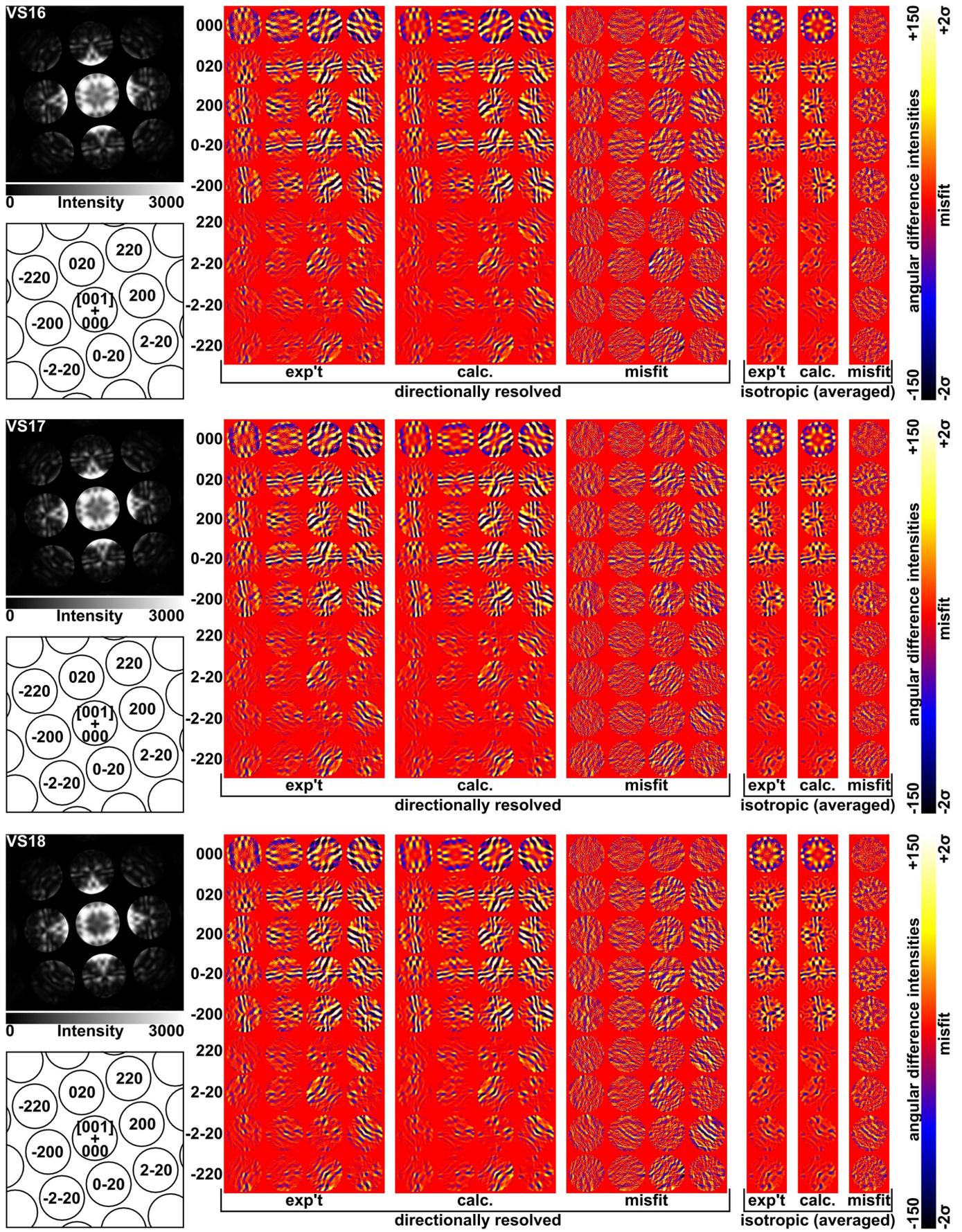

Fig. S4 (part 2/3): As per Fig. S4 (part 1/3), except for VS16-VS18.

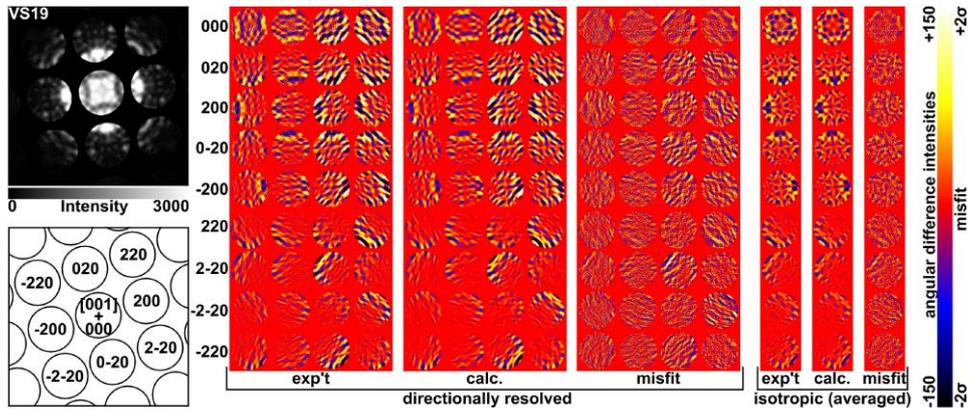

**Fig. S4 (part 3/3):** As per Fig. S4 (part 1/3), except for VS19.

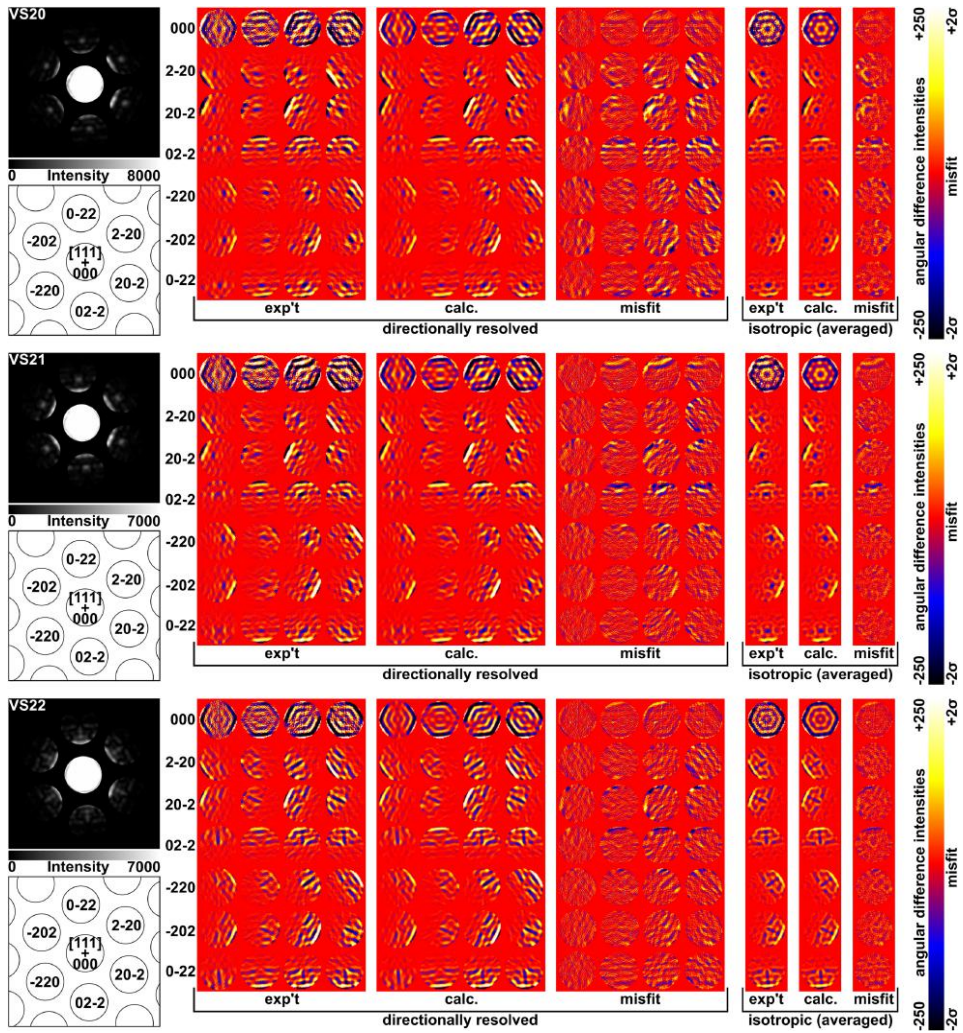

**Fig. S5 (part 1/2):** The CBED patterns and QCBED pattern-matched directionally resolved angular difference intensities for VS20-VS22 are shown together with the isotropic angular difference intensities determined by averaging the directionally resolved components after QCBED was performed. For each void state, the PSF-corrected[52] and noise reduced[53] pattern is shown in linear greyscale together with a matching schematic outline that indexes all the pattern-matched reflections and the location of the zone-axis orientation (+), which in the present cases is [111]. The angular difference intensities (determined as per the method in Extended Data Fig. 11) are all shown in false colour with the colour legend at right applying to all angular difference intensities shown for that void state. Note that the misfit maps are plots of $(I_i^{exp't} - I_i^{calc.})/\sigma_i$, where $I_i^{exp't}$ and $I_i^{calc.}$ are the experimental and calculated angular difference intensities in the $i^{th}$ pixel respectively, and $\sigma_i$ is the standard uncertainty in the $i^{th}$ pixel. The electron energy used for these cases was 202.4 keV. The location from which each CBED pattern was collected is shown in Extended Data Fig. 3. While the present figure presents the output of just one pattern-matching refinement for each void state, the refinements were repeated many times for each CBED pattern (as described above and in Fig. S2 and in the main text). The averages and uncertainties of the key parameters determined from all refinements for each void state are summarised in Extended Data Fig. 6.

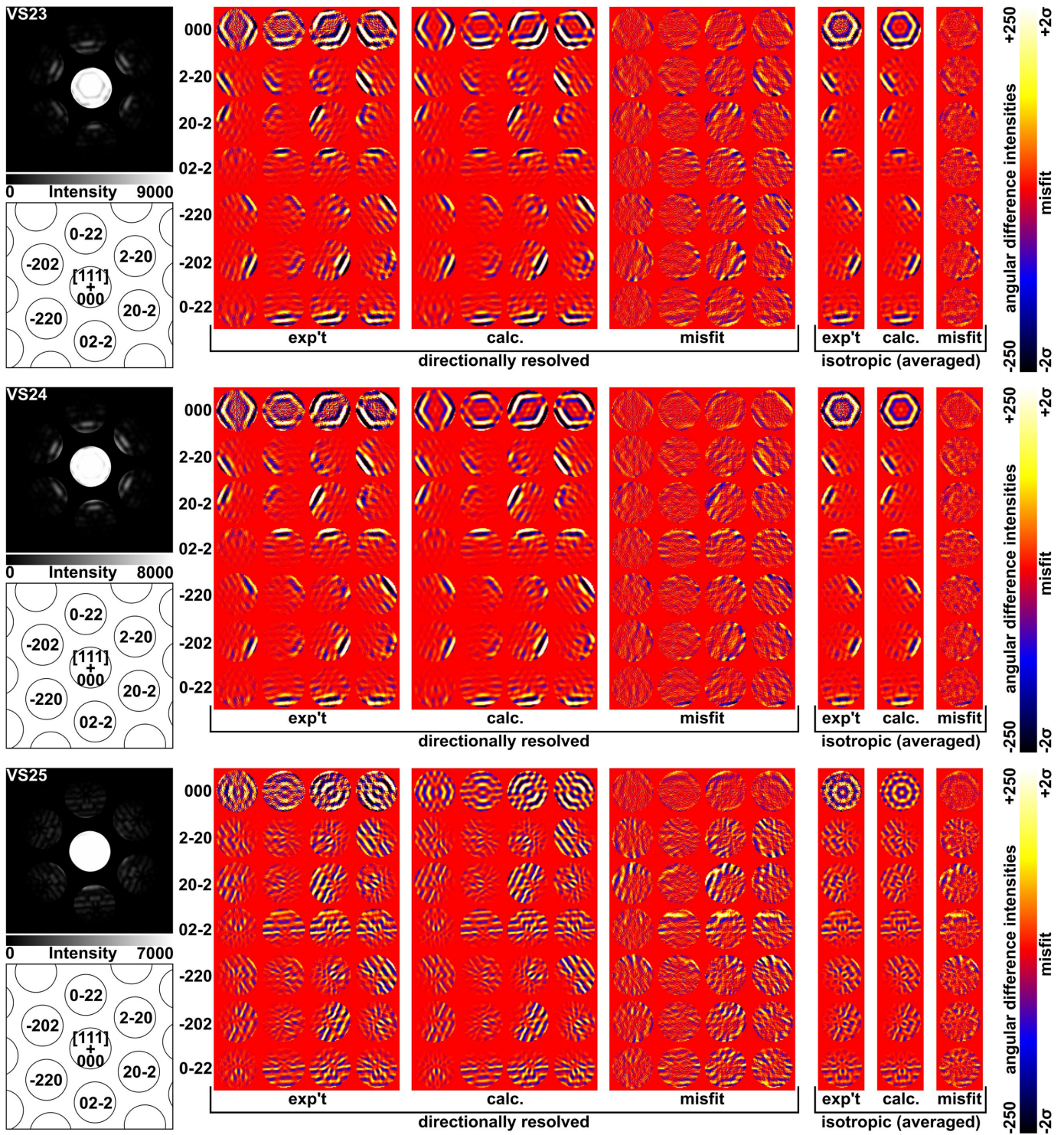

**Fig. S5 (part 2/2):** As per Fig. S5 (part 1/2), except for VS23-VS25.

## Matrix state refinements

The matrix state QCBED refinements were independent of one another because for any given state, none of the input for the next loop was dependent on the results from other matrix states. The refinement procedure is illustrated in Fig. S6 and was repeated 30 times for each void state, producing 600 sets of results for each state (12,000 in total for MS1-MS20).

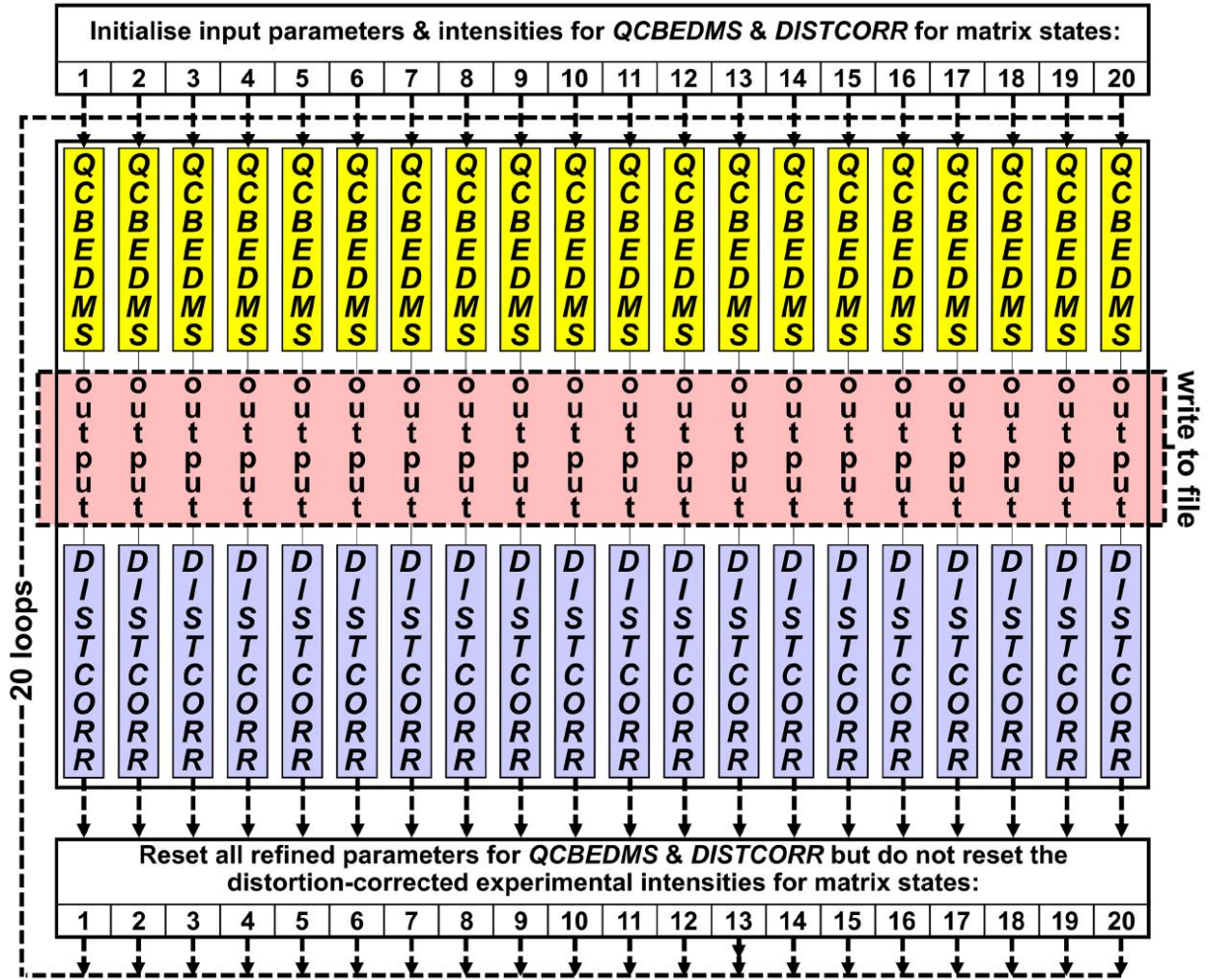

**Fig. S6: The architecture of QCBED refinements for MS1-MS20 does not involve any feedback of parameters averaged across matrix states, meaning that the matrix state refinements run independently of one another.** The refinement procedure is essentially the same as that for the void states (Fig. S2) except that no parameter averaging across matrix states is required for setting input parameters going into the next refinement loop. The 20 loops involving 20 independent matrix state refinements led to 400 sets of results for the entire refinement procedure illustrated above. This procedure was repeated 30 times to yield a total of 12,000 sets of results, i.e. 600 per matrix state.

The refined parameters were $H_{M1}$ – the thickness of the continuous matrix (Fig. 1d,h), $\{V_{hkl}\}_{M1}$ – the relevant set of structure factors (Fig. 1d,h), $C_{vac,M1}$ – the constant vacancy concentration in the probed volume (Extended Data Fig. 4d,e), the $B$ and $C$ coefficients of the parametrised phenomenological absorption model[67] (Eq. 5), and the incident orientation. Extended Data Fig. 7 presents the averages and uncertainties (from the 95% confidence interval) for $H_{M1}$, $C_{vac,M1}$ and $\{f_{hkl}\}_{M1}$ for all matrix states (where $f_{hkl}$ were converted from $V_{hkl}$ by the Mott-Bethe formula[33,34]), and graphs the refined structure factors as a fraction of the corresponding IAM[49] values.

Examples of QCBED pattern matching are given for all matrix states in Figs. S7–S9 and should be examined in conjunction with the more detailed description of QCBED pattern matching given in Extended Data Fig. 11 and the Methods.

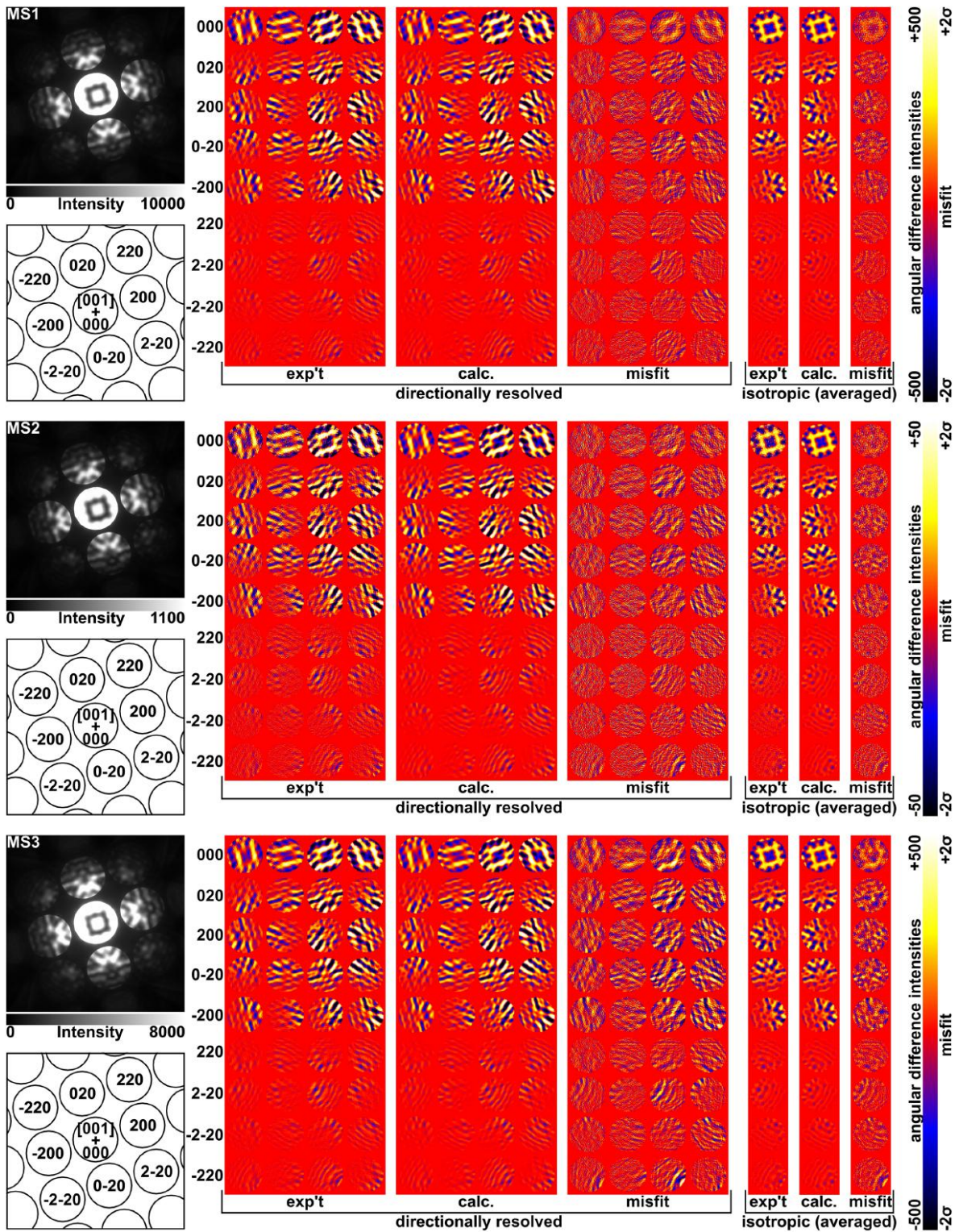

**Fig. S7 (part 1/3): The CBED patterns and QCBED pattern-matched directionally resolved angular difference intensities for MS1-MS3 are shown together with the isotropic angular difference intensities determined by averaging the directionally resolved components after QCBED was performed.** For each matrix state, the PSF-corrected[52] and noise reduced[53] pattern is shown in linear greyscale together with a matching schematic outline that indexes all the pattern-matched reflections and the location of the zone-axis orientation (+), which in the present cases is [001]. The angular difference intensities (determined as per the method in Extended Data Fig. 11) are all shown in false colour with the colour legend at right applying to all angular difference intensities shown for that matrix state. Note that the misfit maps are plots of $(I_i^{exp't} - I_i^{calc.})/\sigma_i$, where $I_i^{exp't}$ and $I_i^{calc.}$ are the experimental and calculated angular difference intensities in the $i^{th}$ pixel respectively, and $\sigma_i$ is the standard uncertainty in the $i^{th}$ pixel. The electron energy used for these cases was 160.3 keV. The location from which each CBED pattern was collected is shown in Extended Data Fig. 3. While the present figure presents the output of just one pattern-matching refinement for each matrix state, the refinements were repeated many times for each CBED pattern (as described above and in Fig. S6). The averages and uncertainties of the key parameters determined from all refinements for each matrix state are summarised in Extended Data Fig. 7.

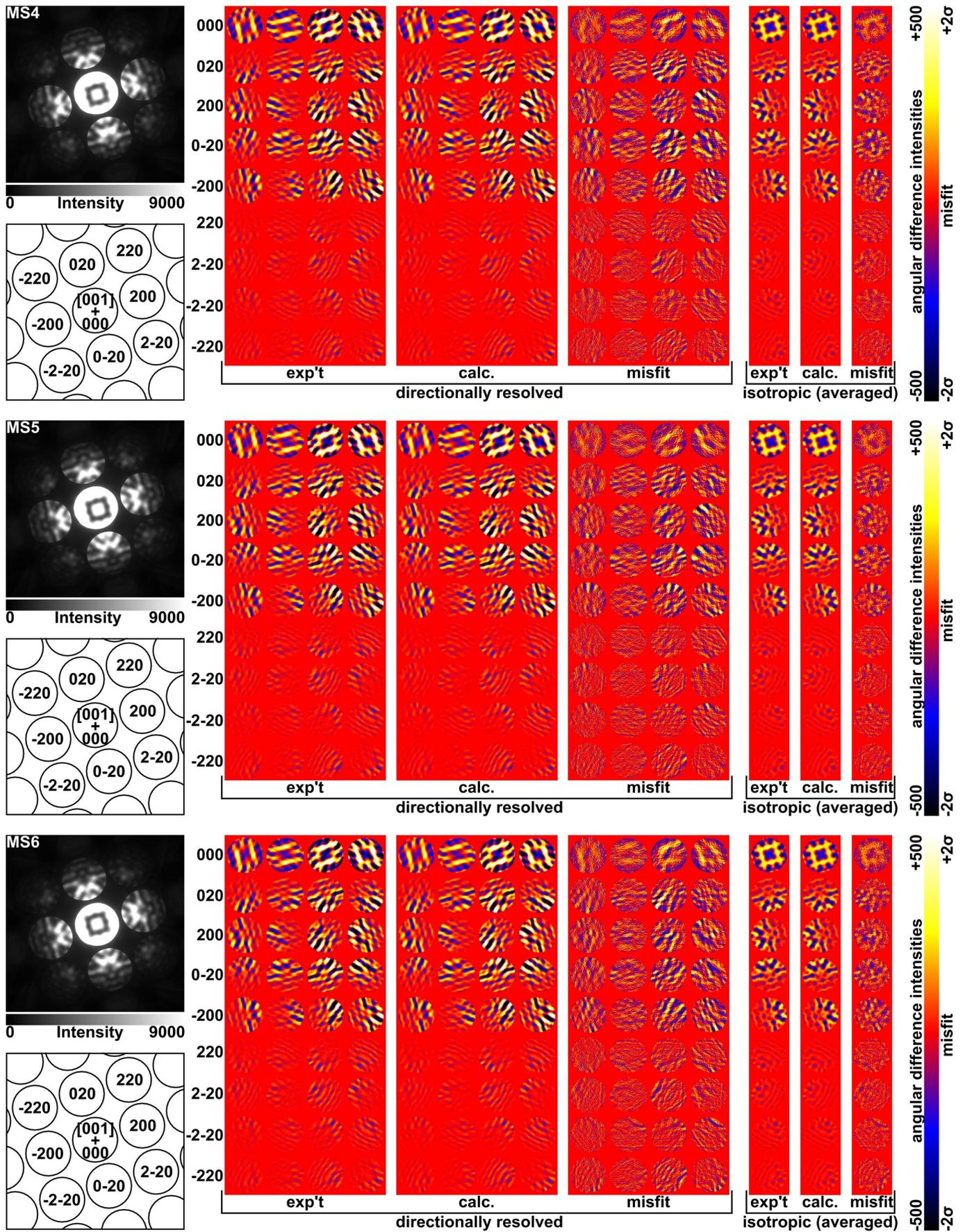

**Fig. S7 (part 2/3):** As per Fig. S7 (part 1/3), except for MS4-MS6. Note that MS6 is the example presented in Fig. 1f,h, where the isotropic (averaged) intensity differences have been realigned horizontally.

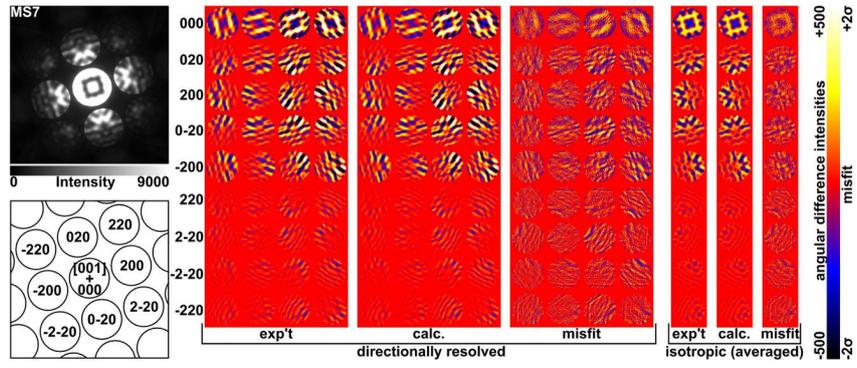

**Fig. S7 (part 3/3):** As per Fig. S7 (part 1/3), except for MS7.

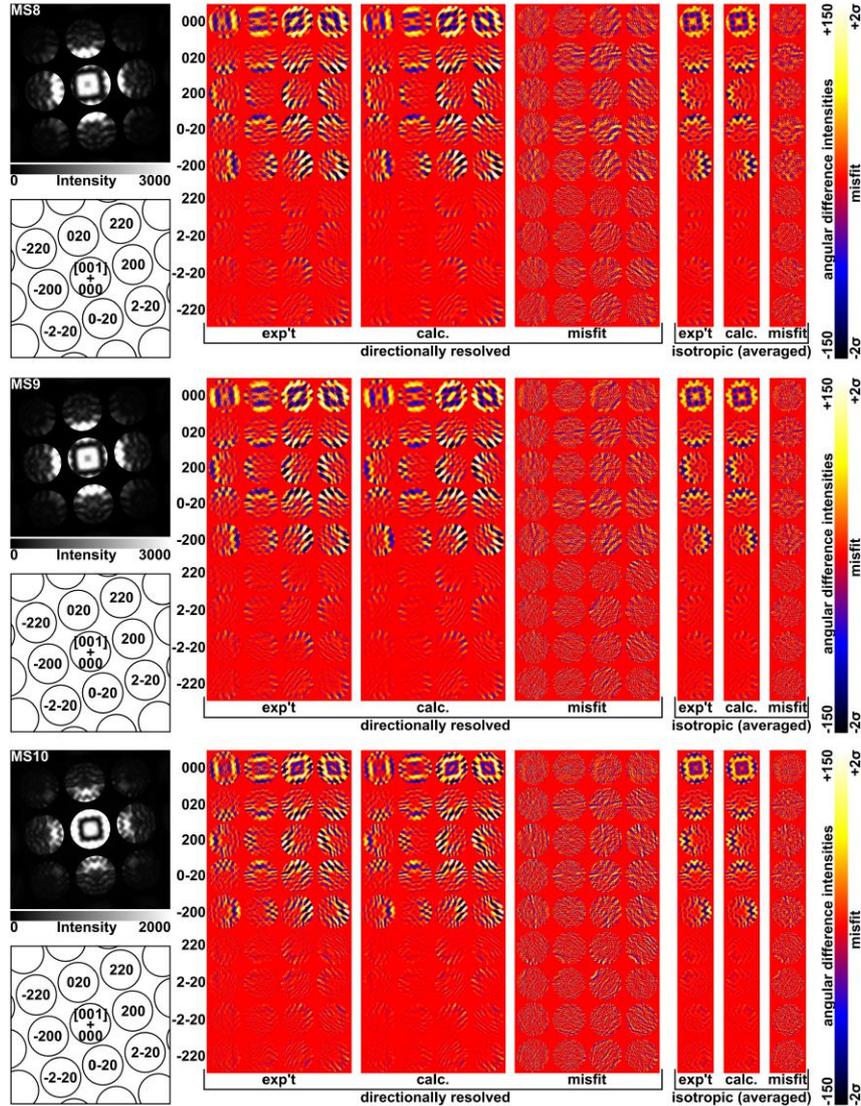

**Fig. S8 (part 1/3): The CBED pattern and QCBED pattern-matched directionally resolved angular difference intensities for MS8-MS10 are shown together with the isotropic angular difference intensities determined by averaging the directionally resolved components after QCBED was performed.** For each matrix state, the PSF-corrected[52] and noise reduced[53] pattern is shown in linear greyscale together with a matching schematic outline that indexes all the pattern-matched reflections and the location of the zone-axis orientation (+), which in the present case is [001]. The angular difference intensities (determined as per the method in Extended Data Fig. 11) are all shown in false colour with the colour legend at right applying to all angular difference intensities shown. Note that the misfit maps are plots of ($I_i^{exp't} - I_i^{calc.}$)/$\sigma_i$, where $I_i^{exp't}$ and $I_i^{calc.}$ are the experimental and calculated angular difference intensities in the i$^{th}$ pixel respectively, and $\sigma_i$ is the standard uncertainty in the i$^{th}$ pixel. The electron energy used was 202.4 keV. The location from which each CBED pattern was collected is shown in Extended Data Fig. 3. While this figure presents the output of just one pattern-matching refinement for each matrix state, the refinements were repeated many times for each CBED pattern (as described above and in Fig. S6). The averages and uncertainties of the key parameters determined from all refinements for each matrix state are summarised in Extended Data Fig. 7.

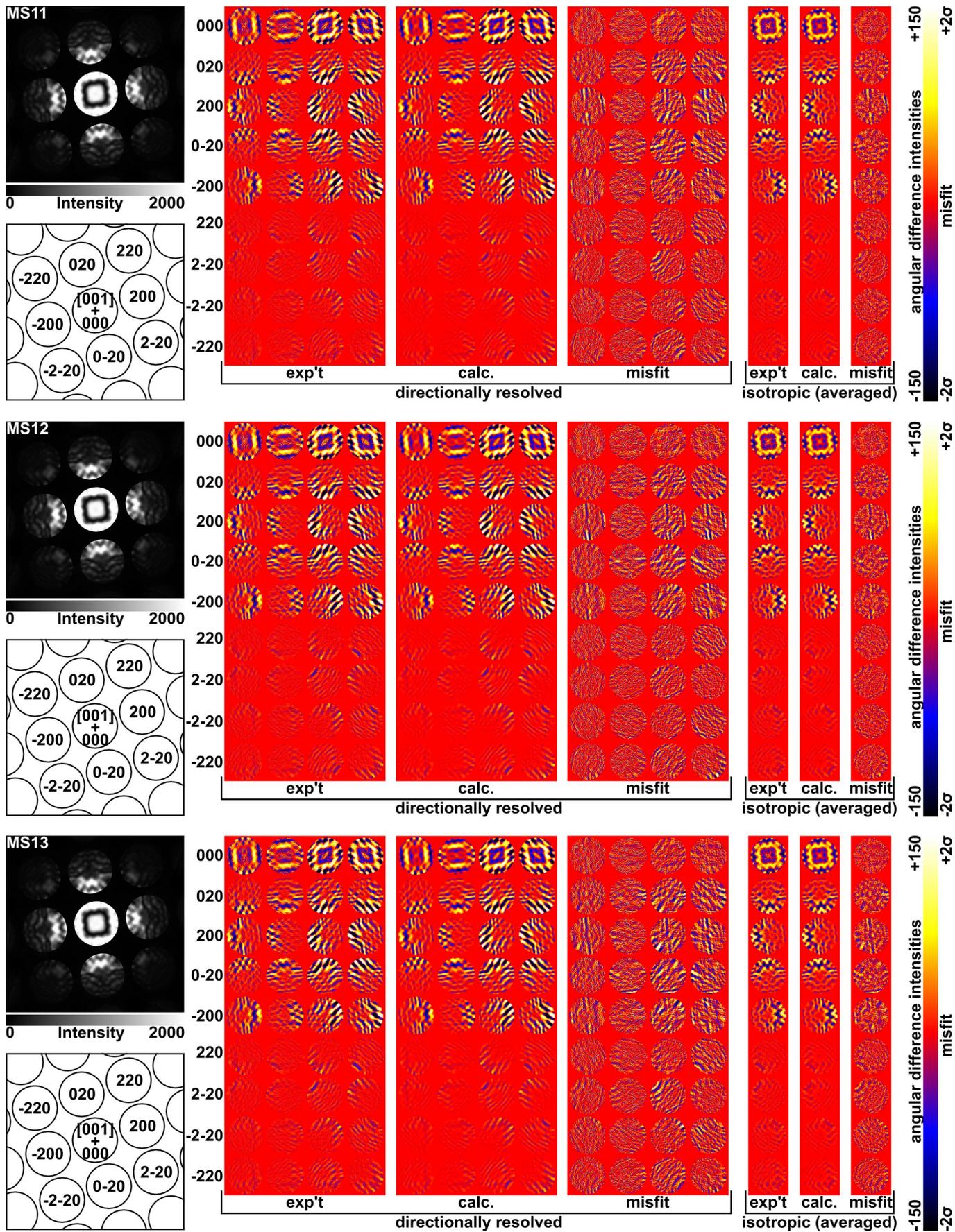

**Fig. S8 (part 2/3):** As per Fig. S8 (part 1/3), except for MS11-MS13.

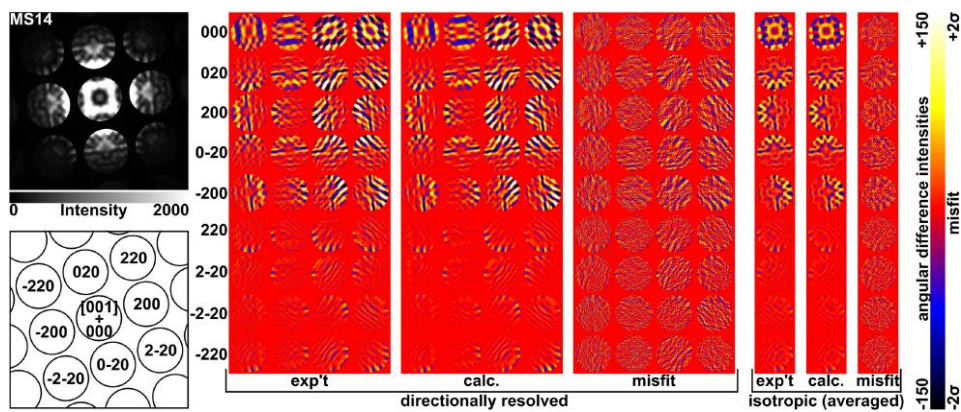

**Fig. S8 (part 3/3):** As per Fig. S8 (part 1/3), except for MS14.

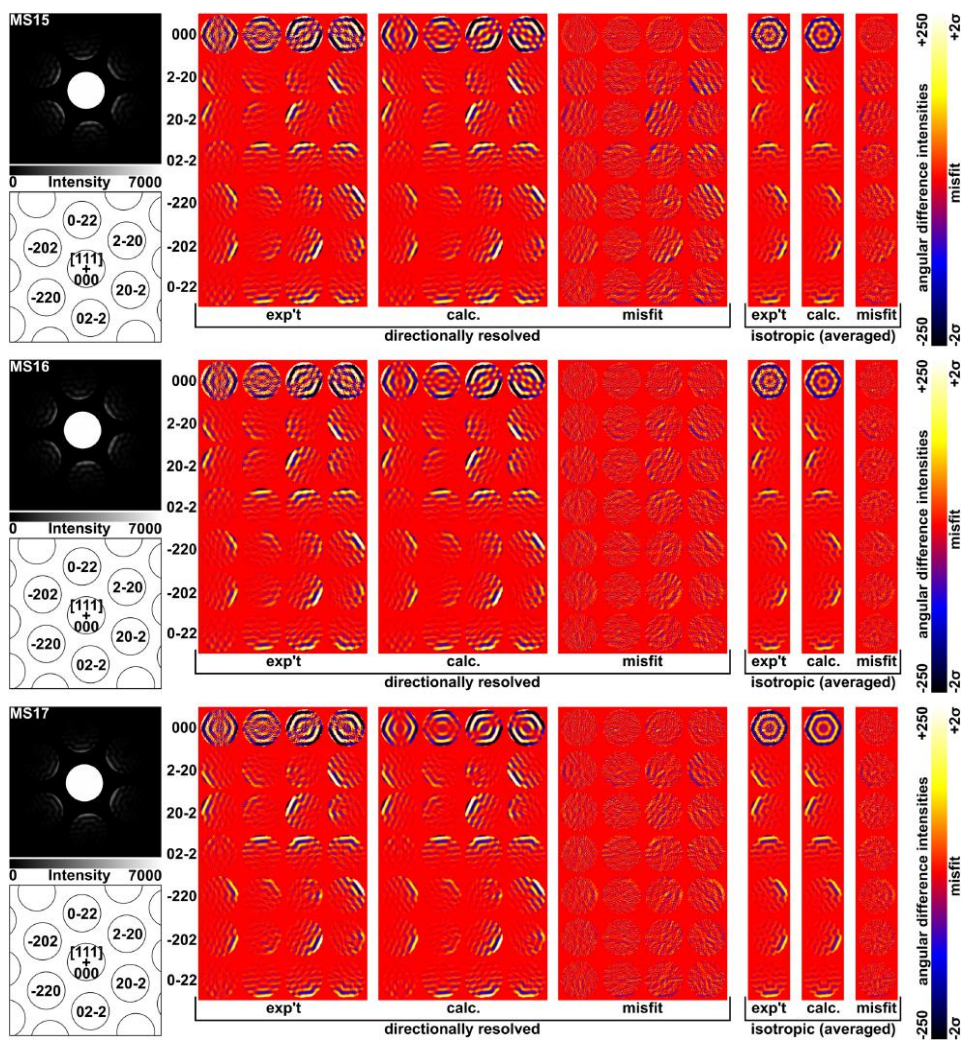

**Fig. S9 (part 1/2): The CBED patterns and QCBED pattern-matched directionally resolved angular difference intensities for MS15-MS17 are shown together with the isotropic angular difference intensities determined by averaging the directionally resolved components after QCBED was performed.** For each matrix state, the PSF-corrected[52] and noise reduced[53] pattern is shown in linear greyscale together with a matching schematic outline that indexes all of the pattern-matched reflections and the location of the zone-axis orientation (+), which in the present cases is [111]. The angular difference intensities (determined as per the method in Extended Data Fig. 11) are all shown in false colour with the colour legend at right applying to all angular difference intensities shown for that matrix state. Note that the misfit maps are plots of $(I_i^{\text{exp't}} - I_i^{\text{calc.}})/\sigma_i$, where $I_i^{\text{exp't}}$ and $I_i^{\text{calc.}}$ are the experimental and calculated angular difference intensities in the $i^{\text{th}}$ pixel respectively, and $\sigma_i$ is the standard uncertainty in the $i^{\text{th}}$ pixel. The electron energy used for these cases was 202.4 keV. The location from which each CBED pattern was collected is shown in Extended Data Fig. 3. While the present figure presents the output of just one pattern-matching refinement for each matrix state, the refinements were repeated many times for each CBED pattern (as described above and in Fig. S6). The averages and uncertainties of the key parameters determined from all refinements for each matrix state are summarised in Extended Data Fig. 7.

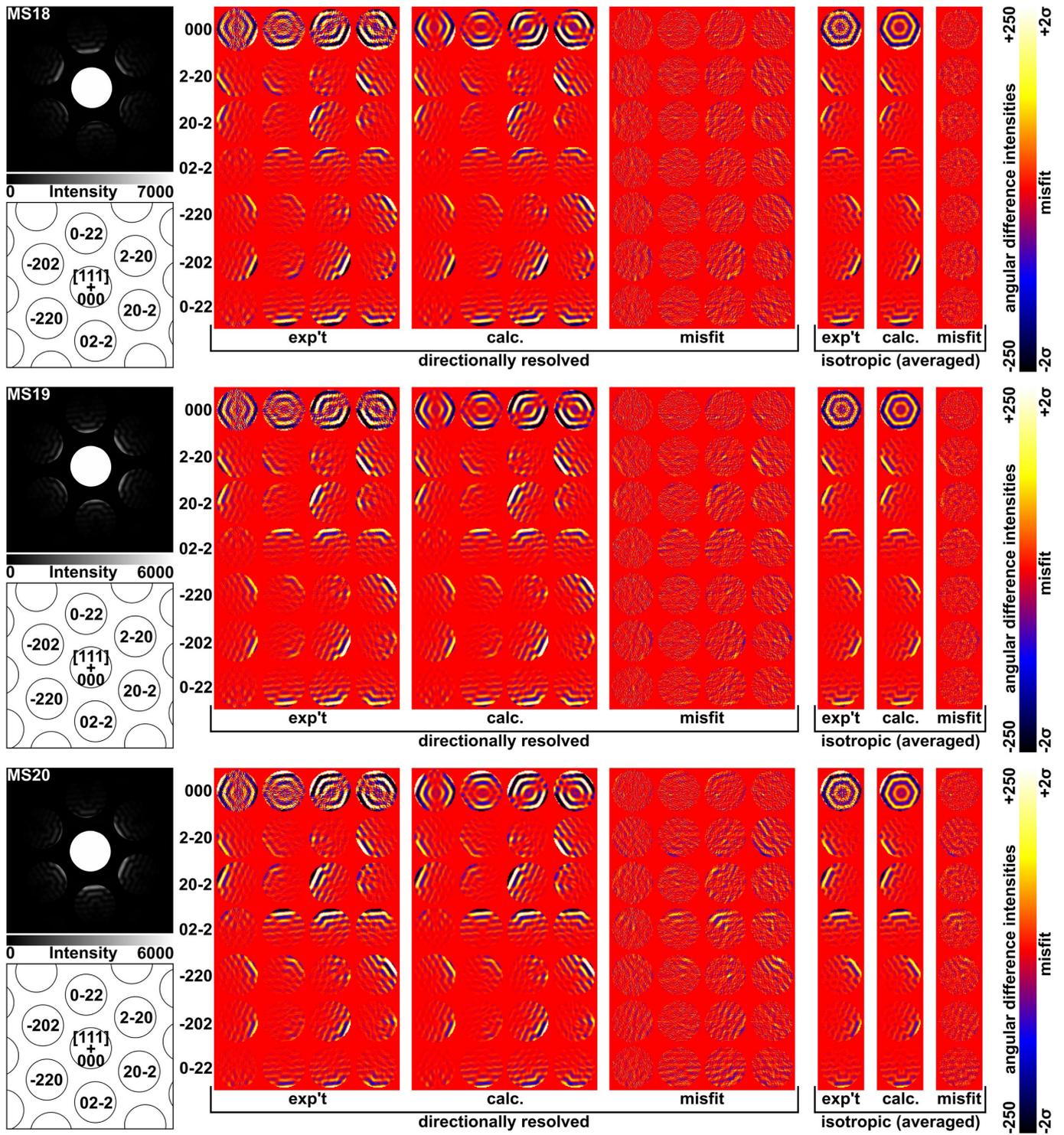

**Fig. S9 (part 2/2):** As per Fig. S9 (part 1/2), except for MS18-MS20.

For all 45 QCBED refinements, the initial parameter starting values were obtained from single applications of the procedures in Figs. S2 and S6 (void-state data and matrix-state data respectively). The *LCF* was the exception since the void-state refinements started from $LCF = 1$ or $LCF = 0$.

## The lattice contraction factor (*LCF*) results in context

Table S1 compares the present DFT and QCBED *LCF* determinations with results from the literature[44,76-96]. The variation is large across previous experimental and theoretical determinations and the present results represent a considerable reduction in uncertainty.

**Table S1: A summary (not exhaustive) of previous determinations of vacancy relaxation volumes or vacancy volumes in aluminium.** The top part of the table summarises previous experimental measurements, with the average and uncertainty of all previous work compared to the present QCBED result in the blue-shaded region of the table. Uncertainties span the 95% confidence interval of the mean. The bottom part of the table summarises previous theoretical determinations, with the average and uncertainty of all previous work compared to the present DFT result in the red-shaded region of the table. Here, $V_r$ is the relaxation volume of the lattice due to a vacancy, $V_v$ is the volume of a vacancy, and $V_a$ is the volume of an atom. Note that the lattice contraction factor (*LCF*) measured by QCBED is $V_r/V_a$ and equivalently, $LCF = 1 - V_v/V_a$.

| $V_r/V_a$ (i.e. *LCF*) = 1 - $V_v/V_a$ | $V_v/V_a$ | reference/comments |
|---|---|---|
| 0.43 | 0.57 | Ref.[44] |
| 0.38±0.02 | 0.62±0.02 | Ref.[76] |
| 0.46 | 0.54 | Ref.[77] |
| 0.40±0.06 | 0.60±0.06 | In Ref.[77] using the results of Ref.[78]. |
| 0.40±0.02 | 0.60±0.02 | In Ref.[79] reinterpreting Ref.[76] by removing the effect of divacancies. |
| 0.45 | 0.55 | In Ref.[79] reinterpreting Ref.[80]. |
| 0.32±0.10 | 0.68±0.10 | In Ref.[79] reinterpreting Ref.[81]. |
| -0.23 | 1.23 | In Ref.[82], citing Ref.[83], who used Ref.[84]. |
| 0.05 | 0.95 | In Ref.[85], quoting Ref.[86]. |
| 0.3±0.2 | 0.7±0.2 | Mean of previous experiments (95% CI) |
| 0.33±0.03 | 0.67±0.03 | QCBED - this work (95% CI) |
| 0.26 | 0.74 | |
| 0.25 | 0.75 | Ref.[87] |
| 0.27 | 0.73 | |
| 0.27 | 0.73 | |
| 0.23 | 0.77 | Ref.[88] |
| 0 | 1.00 | |
| 0 | 1.00 | In Ref.[88] using Ref.[89]. |
| 0.53 | 0.47 | |
| -0.69 | 1.69 | In Ref.[88] using Ref.[90]. |
| -1.50 | 2.50 | |
| 0 | 1.00 | In Ref.[88] using Ref.[91]. |
| 0.481 | 0.519 | Ref.[92] |
| 0.33±0.05 | 0.67±0.05 | Ref.[85] |
| 0.33 | 0.67 | Ref.[93], using GGA |
| 0.33 | 0.67 | Ref.[93], using LDA |
| 0.37 | 0.63 | Ref.[94] |
| 0.40 | 0.60 | Ref.[95] |
| 0.49 | 0.51 | |
| 0.87 | 0.13 | Ref.[96] |
| 0.50 | 0.50 | |
| 0.2±0.2 | 0.8±0.2 | Mean of previous theoretical modelling (95% CI) |
| 0.309 ± 0.006 | 0.691±0.006 | DFT - this work (95% CI) |

In this context, the effect of the temperature discrepancy between the experimental determination at 295 K and the DFT result at 0 K is negligible. A zeroth-order correction to the DFT result involves scaling $V/V_0$ from 0 K to 295 K by the cube of the ratio of the 295 K defect-free lattice parameter[48], $a_{295\ K} = 4.0490$ Å, to the 0 K defect-free DFT lattice parameter[20], $a_{DFT} = 4.0407$ Å, i.e.

$(a_{295\,K}/a_{DFT})^3 = 1.0062$. At most, the DFT-determined *LCF* would increase by this ratio when elevating the result to 295 K, i.e. $LCF_{DFT,\,295\,K} = 0.311\pm0.006$. Compared with $LCF_{DFT,\,0\,K} = 0.309\pm0.006$, the correction is insignificant. We note that the slight change is in the direction of the present QCBED determination of $LCF_{QCBED} = 0.33\pm0.03$ (corresponding to 295 K).

Bonding electron density plots from DFT

All DFT bonding electron density, $\Delta\rho(\mathbf{r})$, plots (Fig. 4 and Extended Data Figs. 9 and 10) were computed from the output structure factors from the *Wien2k* calculations, for $(\sin\theta)/\lambda \leq 2.0$ Å$^{-1}$. The relevant effective Debye-Waller parameters, $B_{Al,eff}$, (listed in the captions to Fig. 4 and Extended Data Fig. 9) were applied to the 0 K DFT structure factors to elevate them to 295 K (the temperature of the QCBED work). This is trivial in the case of a monatomic crystal. Doyle and Turner IAM[49] structure factors computed with the same values of $B_{Al,\,eff}$, were subtracted from their DFT counterparts and the Fourier sum produced the $\Delta\rho(\mathbf{r})$ plots in Extended Data Figs. 9a,e,i,m,q and 10a-c. Each IAM structure factor was determined using a cubic spline interpolation of the Doyle and Turner IAM scattering factors[49,71] as opposed to a parametrised sum of Gaussians fitted to the Doyle and Turner scattering factors[71-74], the former approach being considerably more accurate[71-74].

With the 3D plots of $\Delta\rho(\mathbf{r})$ in hand, site averaging was carried out by translating each site in a particular 3D $\Delta\rho(\mathbf{r})$ plot to the origin and summing voxels followed by dividing by the number of sites. This resulted in the sub-cells shown in Extended Data Fig. 9b,f,j,n,r. The subsequent projection of the site-averaged $\Delta\rho(\mathbf{r})$ as shown, gave rise to the plots in Extended Data Fig. 9c,d,g,h,k,l,o,p,s,t, and Fig. 4a,b. The production of Fig. 4c and Extended Data Fig. 10 did not involve site averaging as the intention was to show the topological distortion of the projected $\Delta\rho(\mathbf{r})$ due to the vacancy (black cross) in the slices bounding it in the 108-site supercell.